\documentclass[11pt, letterpaper]{article}

\usepackage[left=1in, right=1in, top=1in, bottom=1in]{geometry}
\usepackage[utf8]{inputenc}
\usepackage[T1]{fontenc}
\usepackage{bm}
\usepackage{type1cm}
\usepackage{lettrine}
\usepackage{amsmath,amssymb,amsthm}
\usepackage{moreverb}
\usepackage{mathtools}
\usepackage{amsmath}
\usepackage{amssymb}
\usepackage{algorithmic}
\usepackage{graphics}
\usepackage{graphicx}
\usepackage{subfigure}
\usepackage{caption}
\usepackage{extarrows}
\usepackage{color}
\usepackage{framed}
\usepackage{wrapfig}
\usepackage{bm}
\usepackage{mathrsfs}
\usepackage{mathabx}
\usepackage{multirow}
\usepackage{longtable}
\usepackage{hyperref}
\usepackage{paralist}
\usepackage{indentfirst}
\usepackage{relsize}
\usepackage{extarrows}
\usepackage{upgreek}
\usepackage{bm}
\usepackage{mwe}
\usepackage[dvipsnames,table]{xcolor}
\usepackage{booktabs}
\usepackage{authblk}
\usepackage{lettrine}
\usepackage{type1cm}
\usepackage{threeparttable}
\usepackage[sort&compress,numbers]{natbib}
\usepackage[figurename=Figure]{caption}
\usepackage[normalem]{ulem}
\usepackage{makecell}
\usepackage{overpic}
\usepackage{ragged2e}

\graphicspath{ {./Figure/} }
\usepackage[font=footnotesize,labelfont=bf]{caption}

\providecommand{\keywords}[1]{\textbf{\textit{Keywords: }} #1}

\hypersetup{
bookmarks=true,
bookmarksopen=true,
bookmarksnumbered=true,
unicode=false,
pdftoolbar=true,
pdfmenubar=true,
pdffitwindow=false,
pdfstartview={FitH},
pdftitle={My title},
pdfauthor={Author},
pdfsubject={Subject},
pdfcreator={Creator},
pdfproducer={Producer},
pdfkeywords={keywords},
pdfnewwindow=true,
colorlinks=true,
linkcolor=blue,
citecolor=blue,
filecolor=blue,
urlcolor=blue
}

\begin{document}

\title{\textbf{Generalizable Neural Downscaling of Earth System Model Wind Fields via Continuous Dynamics Modeling}}

\author[1,$^\dagger$]{Chenxi Yu}
\author[1,$^\dagger$]{Jianan Wei}
\author[1,$^\dagger$]{Hanlin Kong}
\author[2]{Hao Sun}
\author[3]{Bian He}
\author[1,$^*$]{Wenguan Wang}

\affil[1]{\small State Key Lab of Brain-Machine Intelligence, Zhejiang University}
\affil[2]{\small Gaoling School of Artificial Intelligence, Renmin University of China} 
\affil[3]{\small State Key Laboratory of Earth System Numerical Modeling and Application, Institute of Atmospheric Physics, Chinese Academy of Sciences \vspace{18pt}}

\affil[*]{Corresponding author}
\affil[$\dagger$]{Equal contribution}

\date{}

\maketitle

\normalsize

\vspace{-18pt} 
\begin{abstract}
	\small
    Accurate high-resolution wind field simulations are critical for resolving fine-scale atmospheric dynamics, yet the simulation of wind fields in Earth System Models (ESMs) remains limited by coarse spatial resolution and systematic biases.
    To address this, data-driven downscaling techniques have been widely used to enhance coarse-resolution ESM outputs.
    However, existing methods are typically tied to fixed discretizations, limiting generalization across models with different native resolutions.
    Here we formulate global near-surface wind downscaling as an operator-learning problem on continuous atmospheric state fields and develop a downscaling neural operator that maps coarse-scale fields to fine-scale counterparts across heterogeneous discretizations.
    The operator learning-based neural downscaling framework outperforms dominant baselines, recovers fine-scale physical structures, and generalizes to previously unseen ESMs and future climate scenarios without retraining, while preserving long-term wind projection trends.
    These findings establish a generalizable paradigm for high-resolution climate downscaling across diverse simulation outputs and future scenarios.
\end{abstract}



\newpage

\section{Introduction}

Rising global energy consumption renders the development of renewable energy sources increasingly urgent~\cite{zhang2020new}. Wind energy is a pivotal renewable resource in future global energy portfolios~\cite{leung2012Wind}, and wind speed, as a primary component of atmospheric circulation, directly governs energy exchange across the surface-lower atmosphere interface~\cite{kim2015Recent}, modulates global hydrological cycling~\cite{mcvicar2012Global}, atmospheric visibility~\cite{zhang2020Decreasing} and air quality~\cite{pryor2012Analyses}. Consequently, long-term accurate projections of global wind speed are of paramount importance. They are essential not only for the reliable assessment and cost-effective development of global wind energy resources, but also for environmental quality management and climate change adaptation~\cite{zha2022Effects,lyu2026Reconstruction}. Numerical Earth System Models (ESMs), especially the widely used Coupled Model Intercomparison Project Phase 6 (CMIP6)~\cite{Eyring2016overview} multi-model ensemble, are the primary tools for simulating and projecting these dynamics. However, the coarse spatial resolution of ESMs limits the representation of extreme features on regional scales~\cite{Daines2016ModelBased}, and systematic biases further compromise reliability~\cite{Maraun2016Bias}. Although the CMIP6 High Resolution Model Intercomparison Project (HighResMIP) provides high-resolution simulations up to 0.25$^\circ\times$0.25$^\circ$~\cite{haarsma2016High}, it requires substantial computational resources and is not economically feasible for real-time predictions. Therefore, effective downscaling and bias correction methods are essential to translate raw ESM outputs into actionable information for global wind energy projections, enabling more accurate, efficient predictions.

Conventional ESM downscaling methods mainly include dynamical and statistical downscaling~\cite{tang2016Statistical}. Dynamical downscaling methods~\cite{Xu2019Dynamical,Tapiador2020Regional} are implemented using regional climate models~\cite{rummukainen2010Stateofthearta}, which explicitly solve the governing equations of atmospheric dynamics and thermodynamics to simulate fine-scale wind fields, but are computationally expensive, thereby limiting their application at the global scale or for long-term ensembles. Statistical downscaling methods~\cite{Schoof2013Statistical,Reder2025Estimating} are more computationally efficient but lack physical consistency and suffer from limited transferability and poor generalization in unseen climate scenarios.

Recent advances~\cite{Vandal2017DeepSD,sun2024Deepa,Rampal2024Enhancing,soares2024Highresolution} in deep learning have opened new avenues for data-driven ESM downscaling. In contrast to conventional statistical methods~\cite{Schoof2013Statistical,Reder2025Estimating} that depend on handcrafted predictors or simple linear relationships, neural networks can learn complex, nonlinear mappings directly from coarse-scale ESM simulations to fine-scale observational data. Early studies~\cite{Vandal2017DeepSD,Neelesh2022high,Zhong2024investigating,AmirHossein2025PrecipNet} mainly rely on convolutional neural networks or attention-based models, while more recent studies~\cite{Glawion2025global,Oyama2023deep,Hess2025fast,Schmidt2025agenerative,Ignacio2025dynamical,Hess2022Physically} have adopted generative models to learn conditional distributions and produce stochastic downscaled realizations with high fidelity.

However, current deep learning methods for ESM downscaling remain limited in their ability to generalize across climate models~\cite{Hernanz2024limitations,Bano-Medina2022Downscaling,Hess2025fast}. In practice, ESM downscaling is often applied to multi-model ensembles~\cite{Ignacio2025dynamical,Bano-Medina2022Downscaling,Yoshikane2023downscaling,Yuan2024HiCPC}, yet different ESMs differ substantially in their native spatial resolutions~\cite{nguyen2024Selecting,Eyring2016overview}, making it difficult to define a unified refinement strategy that is applicable across scales. More importantly, ESM simulations are generated under different initial conditions and model formulations, including differences in physical parameterizations and external forcings~\cite{zanchettin2022Effects,Eyring2016overview}, which result in model-specific climatologies, variability, and bias structures. Consequently, the coarse-to-fine mapping learned from one model may not transfer to another, leaving most deep learning methods model-specific and limiting their applicability to unseen models with different spatial scales or forcing scenarios.

Beyond conventional neural networks that are typically restricted to fixed input-output discretizations, neural operator learning~\cite{Azizzadenesheli2024neural,kovachki2023neural} has emerged as a powerful paradigm for modeling mappings between continuous physical fields. Rather than learning grid-specific transformations, neural operators capture mappings between underlying functions that can be evaluated across different discretizations. This property is also critical for developing a flexible and generalizable climate downscaling framework~\cite{yang2024fourier,Sinha2025on}. From the perspective of operator learning, the low-resolution (LR) data ensemble is treated as functions defined over the spatial domain that characterize coarse, large-scale atmospheric states, and the high-resolution (HR) data ensemble as functions representing fine-scale atmospheric variability. Climate downscaling can thus be formulated as an operator learning problem, where the goal is to learn a mapping from coarse-scale state functions to fine-scale representations. Such an operator is expected to be agnostic to specific climate models and future scenarios, while capturing the underlying dynamical relationships across spatial resolutions.

Motivated by this formulation, we develop a neural downscaling framework based on operator learning to address the challenge of cross-model and cross-scenario generalization in global surface wind downscaling for ESMs. 
(i) We formulate global wind field downscaling as a function-to-function mapping between continuous atmospheric state fields.
(ii) Accordingly, a U-shaped downscaling neural operator is designed to approximate the theoretical operator that maps coarse-scale wind fields to fine-scale counterparts.
(iii) Trained solely on target datasets with multi-scale paired samples, the neural operator learns to model the continuous dynamical mappings between underlying state functions, without relying on model-specific characteristics and enabling direct application to bias-corrected, low-resolution outputs from diverse climate simulations at inference time, without retraining.
(iv) We conduct comprehensive experiments to evaluate the downscaling performance of our proposed neural framework across three different ESM simulations and three future scenarios, demonstrating its ability to recover essential physical structures while preserving statistical consistency across spatial scales.
(v) The framework supports efficient inference through a single forward pass of the network, enabling HR downscaled wind fields ($\sim$25 km) to be generated over large spatial domains with minimal computational cost and avoiding the substantial overhead associated with iterative generative modeling methods.

\section{Results}

\subsection{State mapping for continuous downscaling}

\subsubsection{Problem formulation}\label{subsubsec:problem_formulation}

Conventional methods~\cite{Wood2004hydrologic,Vandal2017DeepSD} typically formulate downscaling as a mapping between fixed discretizations of gridded fields, \textit{i.e.}, $\bm{b}_{\text{HR}} = f(\bm{a}_{\text{LR}})$, where $\bm{a}_{\text{LR}}\in\mathbb{R}^{d_a}$ and $\bm{b}_{\text{HR}}\in\mathbb{R}^{d_b}$ are the LR and HR gridded fields, respectively, and $f$ is the downscaling function.
Since the input and output dimensions ($d_a$ and $d_b$) are explicitly tied to the underlying grid, these methods are not directly transferable across simulations with different native resolutions.

We reformulate climate downscaling as an operator-learning problem defined on function spaces. Instead of treating the inputs and outputs of the downscaling task as gridded fields, we represent them as continuous atmospheric state fields, \textit{i.e.}, functions defined over a spatial domain $\Omega \subset \mathbb{R}^p$. Specifically, let $\bm{a}: \Omega \rightarrow \mathbb{R}^{d_a}, \bm{a} \in \mathcal{A}(\Omega;\mathbb{R}^{d_a})$, denote the LR input atmospheric state field, which primarily encodes large-scale, coarse-grained atmospheric dynamics resolved by the driving climate model, and $\bm{b}: \Omega \rightarrow \mathbb{R}^{d_b}, \bm{b} \in \mathcal{B}(\Omega;\mathbb{R}^{d_b})$, denote the corresponding HR target atmospheric state field, which contains finer-scale and mesoscale dynamical structures absent from the LR representation. Here, $\mathcal{A}$ and $\mathcal{B}$ are separable Banach spaces of vector-valued functions defined on $\Omega$, following the standard operator-learning formulation~\cite{kovachki2023neural}.
We then define a theoretical downscaling operator $G$ that maps the LR state field $\bm{a}$ to the HR state field $\bm{b} = G(\bm{a})$. 
Accordingly, the downscaling operator $G$ is formulated as a mapping between function spaces:
\begin{equation}
    G: \mathcal{A}(\Omega;\mathbb{R}^{d_a}) \rightarrow \mathcal{B}(\Omega;\mathbb{R}^{d_b}).
\end{equation}

\begin{figure*}[!t]
    \centering
    \begin{overpic}[width=\textwidth,percent]{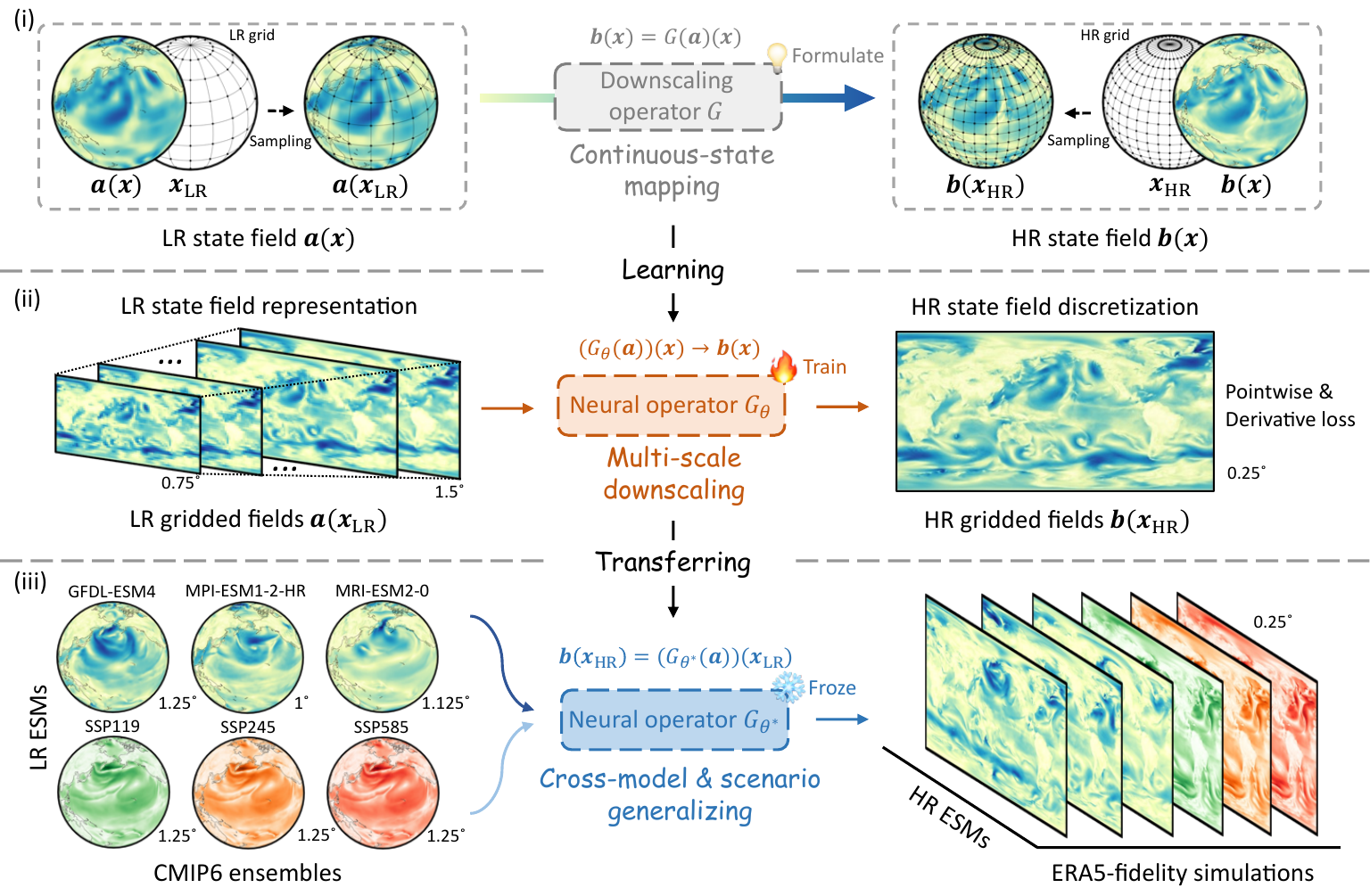}
        \tiny
        \put(89,38.5){
            \textbf{Eq.~\ref{eq:loss_func}}
        }
    \end{overpic}

    \caption{\textbf{Overview of our work.}
    (i) Climate downscaling is formulated as a mapping between continuous state fields $\bm{a}$ and $\bm{b}$ defined over the domain $\Omega$, such that $\bm{b} = G(\bm{a})$, where $G$ denotes the theoretical downscaling operator.
    (ii) A neural operator $G_{\boldsymbol{\theta}}$ is trained to approximate the underlying downscaling operator $G$. The HR gridded fields are first progressively coarsened to construct a hierarchy of LR gridded fields, which are used to represent the LR state field $\bm{a}$ during training; the neural operator then learns the LR-to-HR function mapping such that $\bm{b}(\bm{x}) \approx (G_{\boldsymbol{\theta}}(\bm{a}))(\bm{x})$, and the predicted HR state field is discretized on the target HR grids to compute the loss against the HR reference values for optimizing $G_{\boldsymbol{\theta}}$.
    (iii) When applying the pretrained neural operator $G_{\boldsymbol{\theta}^*}$ to bias-corrected LR outputs $\bm{a(x_{\textbf{LR}})}$ from ESMs, which are interpreted as representation of the underlying LR state field, it directly generates HR gridded fields $\bm{b(x_{\textbf{HR}})}$ discretized on the target HR grids.
    }
    \label{fig:framework}
\end{figure*}

In practice, the continuous state fields $\bm{a}$ and $\bm{b}$ are not directly available. Instead, we only have access to their discretized evaluations at a finite set of spatial locations. For $M$ sampling pairs $\{(\bm{a}(\bm{x}_{\text{LR}}^i), \bm{b}(\bm{x}_{\text{HR}}^i))\}_{i=1}^M$, let $\bm{x}_{\text{LR}}^i = \{x_1^i, \ldots, x_{N_i}^i\} \subset \Omega$ denote the set of observation points for the $i$-th sample with $N_i$ points. The corresponding discrete representations of the input state field $\bm{a}(\bm{x}_{\text{LR}}^i)$ are given by:
\begin{equation}
    \bm{a}(\bm{x}_{\text{LR}}^i) = \{\bm{a}(x_1^i), \ldots, \bm{a}(x_{N_i}^i)\}, \quad i=1,\ldots,M,
\end{equation}
and similarly for the output state field $\bm{b}(\bm{x}_{\text{HR}}^i)$:
\begin{equation}
    \bm{b}(\bm{x}_{\text{HR}}^i) = \{\bm{b}(x_1^i), \ldots, \bm{b}(x_{N_i}^i)\}, \quad i=1,\ldots,M,
\end{equation}

As the analytical form of the theoretical downscaling operator $G$ is unknown, we adopt a data-driven parameterization of $G$ via a neural operator $G_{\boldsymbol{\theta}}$:
\begin{equation}
    G_{\boldsymbol{\theta}}:\mathcal{A}(\Omega;\mathbb{R}^{d_a}) \times \Theta \to \mathcal{B}(\Omega;\mathbb{R}^{d_b}), 
    \quad \boldsymbol{\theta} \in \Theta,
\end{equation}
where $\boldsymbol{\theta}$ denotes the learnable parameters. The training dataset is constructed from $M$ sampling pairs $\{(\bm{a}(\bm{x}_{\text{LR}}^i), \bm{b}(\bm{x}_{\text{HR}}^i))\}_{i=1}^M$, corresponding to the discrete representations of input-output function pairs, from which the neural operator $G_{\boldsymbol{\theta}}$ is inferred.

The learned neural operator $G_{\boldsymbol{\theta}}$ approximates the downscaling operator $G$ that maps between continuous atmospheric state fields.
In this formulation, the gridded fields are treated as discrete representations of the underlying state fields defined over the domain $\Omega$, which allows the state fields to be flexibly queried at any location $\bm{x} \subset \Omega$ under different sampling discretizations, thereby enabling the neural operator to accommodate inputs and outputs defined on different grids.

\subsubsection{Learning continuous state mappings}\label{subsubsec:learning_continuous_state_mappings}

\begin{table}[!t]
    \centering
    \setlength{\tabcolsep}{8pt}
    \caption{\textbf{Continuous state field mapping performance comparison across the neural downscaling framework and SD baselines.}
    }
    \small
    \label{tab:state_mapping_comparison}
    \begin{tabular}{lccccccccc}
        \toprule
        \multirow{2}{*}{Methods} 
        & \multirow{2}{*}{\shortstack{Scale \\ factor}} 
        & \multirow{2}{*}{\shortstack{Input \\ grid}} 
        & \multicolumn{2}{c}{MAE} 
        & \multicolumn{2}{c}{RMSE} 
        & \multicolumn{2}{c}{\textit{95}$^{\text{th}}$ PE} 
        & \textit{5}$^{\text{th}}$ PE \\
        \cmidrule(lr){4-5} \cmidrule(lr){6-7} \cmidrule(lr){8-9} \cmidrule(lr){10-10}
        & & & $UV$ & Speed 
        & $UV$ & Speed 
        & $UV$ & Speed 
        & $UV$ \\
        \midrule
        SD (Interp.) & \multirow{3}{*}{$\times$3} & \multirow{3}{*}{$0.75^\circ$} 
            & 0.142 & 0.161 
            & 0.313 & 0.367 
            & 0.137 & 0.190 
            & 0.140 \\
        SD (Clim.)  &    &      
            & 0.113 & 0.126 
            & 0.244 & 0.290 
            & 0.107 & 0.137
            & 0.111 \\
        Ours          &    &      
            & \textbf{0.046} & 0.048 
            & \textbf{0.079} & \textbf{0.083} 
            & \textbf{0.028} & \textbf{0.032} 
            & \textbf{0.028} \\
        \midrule
        SD (Interp.) & \multirow{3}{*}{$\times$4} & \multirow{3}{*}{$1^\circ$} 
            & 0.210 & 0.237 
            & 0.411 & 0.475 
            & 0.200 & 0.279 
            & 0.207 \\
        SD (Clim.)  &    &      
            & 0.169 & 0.190 
            & 0.322 & 0.377 
            & 0.158 & 0.204 
            & 0.165 \\
        Ours          &    &      
            & \textbf{0.065} & \textbf{0.067} 
            & \textbf{0.107} & \textbf{0.112} 
            & \textbf{0.029} & \textbf{0.031} 
            & \textbf{0.029} \\
        \midrule
        SD (Interp.) & \multirow{3}{*}{$\times$5} & \multirow{3}{*}{$1.25^\circ$} 
            & 0.261 & 0.295 
            & 0.488 & 0.560 
            & 0.249 & 0.348 
            & 0.257 \\
        SD (Clim.)  &    &      
            & 0.213 & 0.239 
            & 0.387 & 0.450 
            & 0.197 & 0.254 
            & 0.206 \\
        Ours          &    &      
            & \textbf{0.091} & \textbf{0.095} 
            & \textbf{0.145} & \textbf{0.152} 
            & \textbf{0.038} & \textbf{0.043} 
            & \textbf{0.039} \\
        \midrule
        SD (Interp.) & \multirow{3}{*}{$\times$6} & \multirow{3}{*}{$1.5^\circ$} 
            & 0.318 & 0.360 
            & 0.566 & 0.648 
            & 0.305 & 0.430 
            & 0.319 \\
        SD (Clim.)  &    &      
            & 0.263 & 0.295 
            & 0.455 & 0.527 
            & 0.244 & 0.317 
            & 0.258 \\
        Ours          &    &      
            & \textbf{0.114} & \textbf{0.118} 
            & \textbf{0.179} & \textbf{0.187} 
            & \textbf{0.041} & \textbf{0.043} 
            & \textbf{0.041} \\
        \bottomrule
    \end{tabular}
    \newline
    
    \footnotesize 
    \justifying
    \noindent The global mean absolute error (MAE) and root mean square error (RMSE) for wind components (zonal $U$ and meridional $V$, fourth column and sixth column) and wind speed (fifth column and seventh column) with respect to HR ERA5. 
    The global absolute error in \textit{95}$^{\text{th}}$ wind component percentile (eighth column), \textit{95}$^{\text{th}}$ wind speed percentile (ninth column), and \textit{5}$^{\text{th}}$ wind component percentile (tenth column) with respect to HR ERA5. 
    All metrics are reported for downscaling scale factors of $\times$3, $\times$4, $\times$5, and $\times$6, corresponding to input grid spacings of 0.75$^\circ$, 1$^\circ$, 1.25$^\circ$, and 1.5$^\circ$, respectively. 
    The best performance highlighted in \textbf{bold}.
\end{table}

We develop a neural downscaling framework (Fig.~\ref{fig:framework}) from the downscaling problem formulation, in which the neural operator $G_{\boldsymbol{\theta}}$ is trained to approximate the theoretical downscaling operator $G$, enabling the mapping from LR atmospheric state fields to the HR state fields.

During training, we adopt a multi-scale self-supervised learning strategy, where each data pair $(\bm{a}(\bm{x}_{\mathrm{LR}}), \bm{b}(\bm{x}_{\mathrm{HR}}))$ is constructed by progressively coarsening the HR reanalysis field $\bm{b}(\bm{x}_{\mathrm{HR}})$ to multiple lower resolutions with varying grid spacings. These coarsened gridded fields are treated as discrete representations of the underlying LR state field $\bm{a}$, and are used as inputs to train the neural operator to recover the corresponding HR target.
Given these LR representations, the neural operator $G_{\boldsymbol{\theta}}$ estimates the HR state field, whose predictions are further discretized on the target HR grids. The estimated HR state field is compared against the reference HR reanalysis field sampled at the finest available granularity for optimizing the neural operator.

During inference, ESM simulation outputs are first preprocessed through simple bias correction at their native resolutions. The bias-corrected LR ESM simulations are then interpreted as discrete representations $\bm{a}(\bm{x}_{\text{LR}})$ of the underlying LR state fields and are provided as inputs to the pretrained neural operator $G_{\boldsymbol{\theta}^*}$, generating the HR gridded fields $\bm{b}(\bm{x}_{\text{HR}})$ discretized on the target HR grids.

The capability of the neural operator to map between state fields is a key property of our neural downscaling framework.
Training and evaluation are conducted exclusively on ERA5~\cite{Hersbach2020era5} reanalysis to isolate the operator's intrinsic capacity for learning continuous state field mappings, independent of any ESM influence. 
The framework is compared against two spatial disaggregation-based downscaling baselines, namely Spatial Disaggregation with interpolation, SD (Interp.), and Spatial Disaggregation with climatology adjustment, SD (Clim.), which correspond to the Bias Correction and Spatial Disaggregation (BCSD)~\cite{Wood2004hydrologic} methods without explicit prior bias correction. These baselines are chosen for their continuous downscaling capability.

Table~\ref{tab:state_mapping_comparison} shows that our framework outperforms the SD baselines across all downscaling factors. It achieves the lowest MAE and RMSE for both wind components and speed at every input resolution, with the performance gap increasing as the downscaling factor grows, demonstrating a strong capability to recover fine-scale state fields from continuous coarse-scale inputs. The framework yields the smallest errors at both the \textit{95}$^{\text{th}}$ and \textit{5}$^{\text{th}}$ percentiles across all scales, with an enhanced capability to correct biases in extreme wind conditions and reliably capture tail behaviours of the state fields. We report additional evaluation results for the state mapping capability of our framework, computing the global MAE distribution (Fig.~\ref{fig:a_state_mapping_qualitative}) and radial averaged power spectral density (RAPSD, Fig.~\ref{fig:a_state_mapping_rapsd}) discrepancies of the downscaled fields across all spatial scales.

\subsection{CMIP6 modeling of ERA5}\label{subsec:cmip6_modeling_of_era5}

We employ our neural downscaling framework to reconstruct the HR near-surface wind fields from LR CMIP6~\cite{Eyring2016overview} ESM simulations. Its performance is compared against two popular continuous downscaling methods~\cite{nahar2017rankbased,delimamoraes2024Climate,xu2019Application}: BCSD with interpolation, BCSD (Interp.), and BCSD with climatology adjustment, BCSD (Clim.)~\cite{Wood2004hydrologic}. Here, we present results using GFDL-ESM4~\cite{Dunne2020GFDL} simulations as LR inputs, with ERA5~\cite{Hersbach2020era5} serving as the reference. Notably, the native resolution of GFDL-ESM4 corresponds to downscaling scale factors unseen during training, making it a representative test case for evaluating the generalization of our framework to previously unseen discretization schemes. We evaluate the framework in terms of its ability to recover spatial patterns, correct systematic biases, and quantify uncertainty in the generated wind fields.

\begin{table}[!t]
    \setlength{\tabcolsep}{2.1pt}
    \centering
    \caption{
        \textbf{Statistical comparison in CMIP6~\cite{Eyring2016overview} modeling of ERA5 performance across the neural downscaling framework and BCSD baselines.}
    }
    \small
    \label{tab:cmip6_modeling_comparison}
    \begin{tabular}{lccccccccccccc}
    \toprule
    \multirow{2}{*}{Methods}
    & \multicolumn{1}{c}{\makecell[c]{\shortstack{MAE \\ (pool)}}}
    & \multicolumn{2}{c}{\makecell[c]{PCC}}
    & \multicolumn{4}{c}{\makecell[c]{ME}} 
    & \multicolumn{4}{c}{\makecell[c]{\textit{95}$^{\text{th}}$ PE}} 
    & \multicolumn{2}{c}{\makecell[c]{\textit{5}$^{\text{th}}$ PE}} \\
    \cmidrule(lr){2-2}\cmidrule(lr){3-4}\cmidrule(lr){5-8}\cmidrule(lr){9-12}\cmidrule(lr){13-14}
    & $UV$ 
    & $UV$ & Speed
    & $UV$ & \% & Speed & \%
    & $UV$ & \% & Speed & \%
    & $UV$ & \% \\
    \midrule
    BCSD (Interp.)  
        & 0.114 
        & 0.993 & 0.985 
        & 0.249 & 55.50 
        & 0.306 & 55.24 
        & 0.432 & 54.98 
        & 0.523 & 56.74 
        & 0.422 & 55.10  \\
    BCSD (Clim.)
        & 0.099
        & \textbf{0.997} & 0.992
        & \textbf{0.192} & \textbf{65.77} 
        & 0.254 & 62.87 
        & 0.396 & 58.70 
        & 0.453 & 62.51 
        & 0.386 & 58.96  \\
    Ours      
        & \textbf{0.048}
        & 0.996 & \textbf{0.996} 
        & 0.195 & 65.15 
        & \textbf{0.184} & \textbf{73.08} 
        & \textbf{0.322} & \textbf{66.47} 
        & \textbf{0.309} & \textbf{74.42} 
        & \textbf{0.296} & \textbf{68.48} \\
    \bottomrule
    \end{tabular}
    \newline

    \footnotesize 
    \justifying
    \noindent The pooled global mean absolute error, MAE (pool), for wind components (second column) with respect to LR bias-corrected GFDL-ESM4~\cite{Dunne2020GFDL} simulations. 
    The Pearson correlation coefficient, PCC, for wind components (third column) and wind speed (fourth column) with respect to ERA5.
    The global long-term mean error, ME, for wind components (zonal $U$ and meridional $V$, fifth column) and wind speed (seventh column) with respect to ERA5. 
    The global absolute error in \textit{95}$^{\text{th}}$ wind component percentile (\textit{95}$^{\text{th}}$ PE, ninth column), \textit{95}$^{\text{th}}$ wind speed percentile (eleventh column), and global absolute error in \textit{5}$^{\text{th}}$ wind component percentile (\textit{5}$^{\text{th}}$ PE, thirteenth column) with respect to ERA5.
    The best performance highlighted in \textbf{bold}. The percentage improvement over raw ESM simulations is reported in the parentheses.
\end{table}

\subsubsection{Recovery of spatial patterns in downscaled wind fields}\label{subsubsec:recovery_of_spatial_patterns_in_downscaled_wind_fields}

\begin{figure*}[!h]
    \centering
    \begin{overpic}[width=\textwidth,percent]{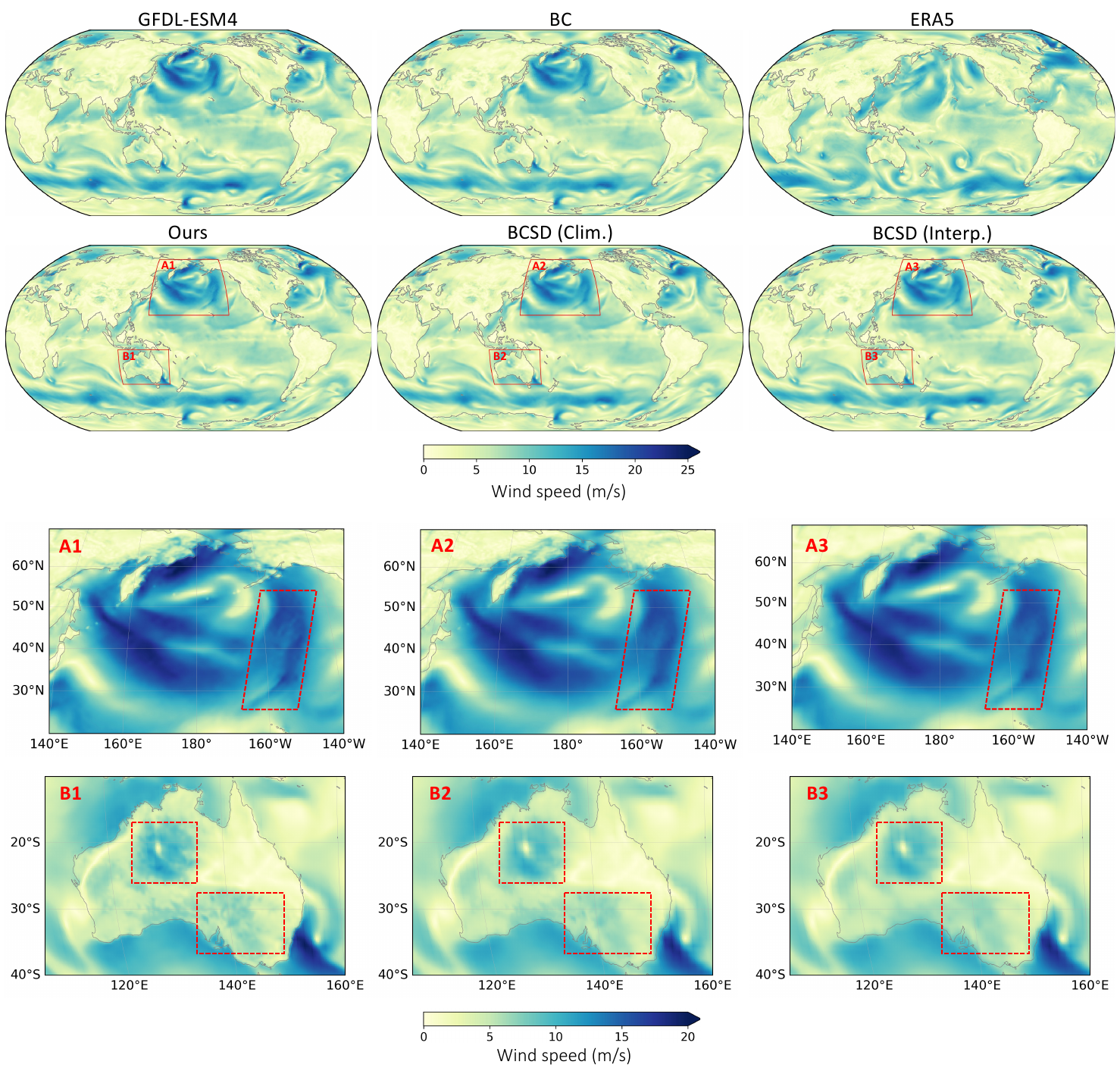}
        \put(1,95){\scriptsize{$\textbf{a}$}}
        \put(34,95){\scriptsize{$\textbf{b}$}}
        \put(67,95){\scriptsize{$\textbf{c}$}}
        \put(1,75){\scriptsize{$\textbf{d}$}}
        \put(34,75){\scriptsize{$\textbf{e}$}}
        \put(67,75){\scriptsize{$\textbf{f}$}}
        \put(1,49){\scriptsize{$\textbf{g}$}}
        \put(1,27){\scriptsize{$\textbf{h}$}}
    \end{overpic}
    \caption{
        \textbf{Qualitative comparison of daily near-surface wind fields.}
        \textbf{a}, Daily near-surface wind field from raw GFDL-ESM4~\cite{Dunne2020GFDL} simulation.
        \textbf{b}, Bias-corrected (BC) wind field on the native resolution.
        \textbf{c}, Near-surface wind field from the target ERA5 reanalysis.
        \textbf{d,e,f}, Downscaled fields from our neural operator (\textbf{d}), BCSD (Clim.) (\textbf{e}), and BCSD (Interp.) (\textbf{f}), respectively. The local regions highlighted by red boxes and labeled \textcolor{red!90!black}{\textbf{A}} and \textcolor{red!90!black}{\textbf{B}} in these panels are enlarged and shown in \textbf{g} and \textbf{h}, respectively.
    }
    \label{fig:qualitative}
\end{figure*}

All ESM simulations are first bias-corrected during inference, for both our neural downscaling framework and BCSD.
Under this setting, the main challenge is to transform CMIP6 simulations into their HR counterparts.
This requires accurate and computationally efficient spatial disaggregation.
For qualitative comparison, we observe that the raw GFDL-ESM4~\cite{Dunne2020GFDL} simulation (Fig.~\ref{fig:qualitative}a) and its bias-corrected counterpart (Fig.~\ref{fig:qualitative}b) exhibit a high degree of consistency in large-scale wind field structures. The bias correction procedure preserves the intrinsic characteristics of the ESM simulations, while effectively adjusting wind patterns in regions with complex topography, such as mountainous areas, as well as over extensive water bodies. Compared to the HR ERA5 (Fig.~\ref{fig:qualitative}c), the near-surface wind fields from the ESM simulations differ substantially, even for the same day.

The downscaled fields produced by our framework (Fig.~\ref{fig:qualitative}d) exhibit comparable realism to the ERA5 wind fields (Fig.~\ref{fig:qualitative}c), capturing fine-scale structures with similar physical fidelity. This is in stark contrast to BCSD (Clim.) (Fig.~\ref{fig:qualitative}e) and BCSD (Interp.) (Fig.~\ref{fig:qualitative}f), which fail to reproduce such fine-scale details. 
We present regional downscaling results over two representative areas: the North Pacific Ocean in the Northern Hemisphere (Fig.~\ref{fig:qualitative}g) and the Australian region in the Southern Hemisphere (Fig.~\ref{fig:qualitative}h). In these regions, our downscaled wind fields exhibit visually clearer, more natural, and richer local structures compared with BCSD. These fine-scale features cannot be recovered through methods based solely on long-term climatology or direct spatial interpolation. In the highly dynamic marine environment of the Pacific Ocean, local wind structures are dominated by transient atmospheric processes rather than stationary climatological patterns. Consequently, BCSD that rely on climatological bias adjustment or direct interpolation produce nearly indistinguishable wind fields over the ocean, failing to introduce additional dynamic details. The framework is able to generate previously unresolved fine-scale vortex-like structures, reflecting the intrinsic variability of the regional wind field. A similar conclusion holds for the Australian region over land: in areas lacking stable long-term climatological patterns, BCSD (Clim.) becomes ineffective and fails to recover fine-scale dynamic wind field structures, whereas our framework successfully generates such details. Similar qualitative comparisons are reported for the downscaled wind fields of MPI-ESM1-2-HR~\cite{Mauritsen2019developments} and MRI-ESM2-0~\cite{Seiji2019meteorological} simulations (Fig.~\ref{fig:a_spatial_patterns_qualitative}).

\begin{figure*}[!t]
    \centering
    \begin{overpic}[width=\textwidth,percent]{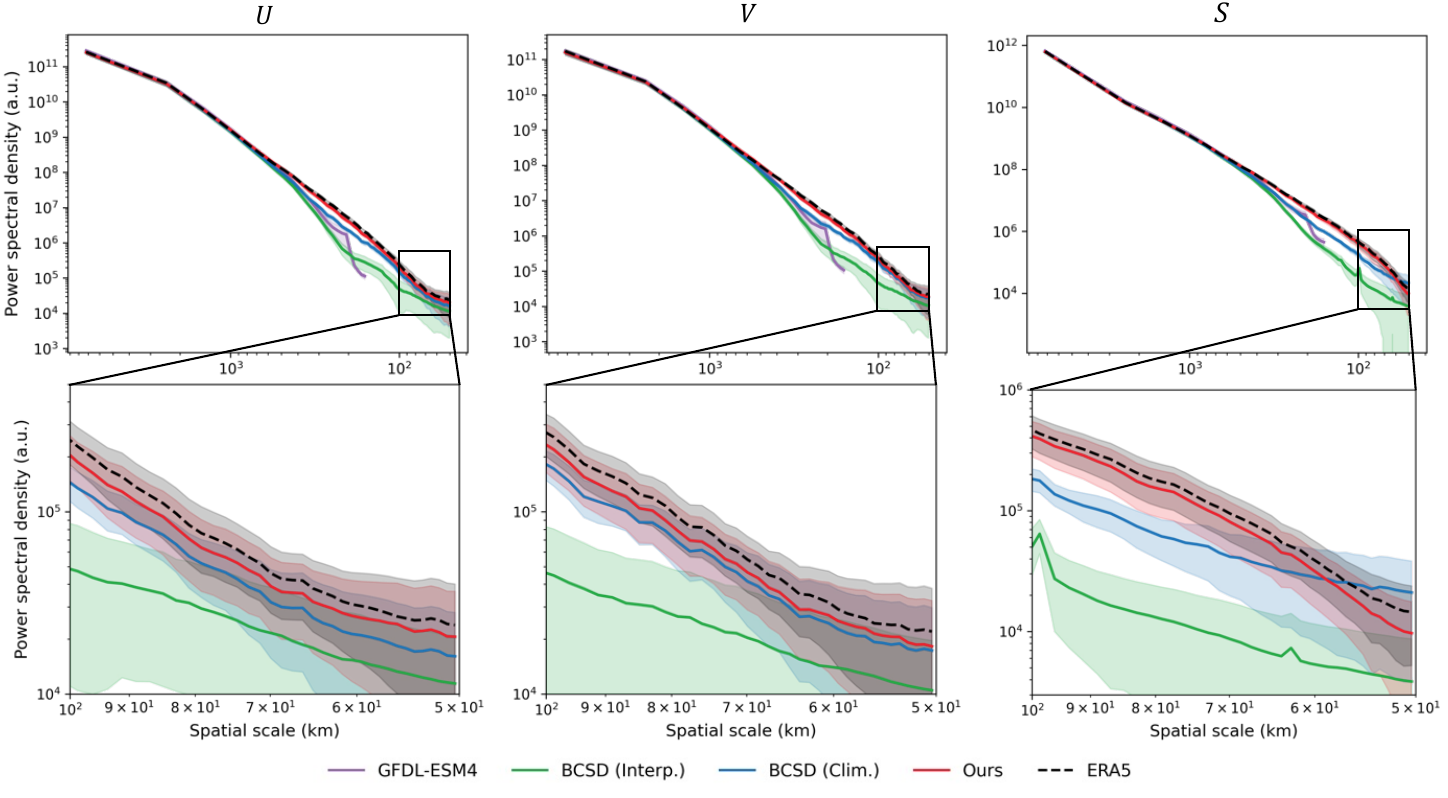}
    \end{overpic}
    \caption{
        \textbf{RAPSD comparison of neural downscaling framework and BCSD baselines.}
        Radial averaged power spectral density (RAPSD) of the zonal wind component $U$, meridional wind component $V$, and wind speed $S$ from the downscaled fields. The lower panel shows the enlarged views of the small-scale regime (spatial scales $< 100$ km). The shaded region denotes $\pm 1$ standard deviation over all time steps during the evaluation period.
    }
    \label{fig:rapsd}
\end{figure*}

We analyze the radial averaged power spectral density (RAPSD)~\cite{Ding2025Superresolution} of the downscaled wind fields to assess the capability to recover the fine-scale dynamical variability of the ESM simulations (Fig.~\ref{fig:rapsd}). The native GFDL-ESM4 wind fields, with a spatial resolution of approximately 1$^\circ\times$1.25$^\circ$, exhibit a pronounced loss of spectral power at small spatial scales relative to the 0.25$^\circ$ downscaled results and ERA5, indicating an absence of high-frequency variability. For BCSD (Interp.), power deficits persist across the intermediate and small spatial scales. Incorporating climatological high-frequency biases partially improves spectral power at these scales; however, the scale-dependent decay of the power spectrum remains different from that of ERA5. The framework reproduces the wind power spectra across a wide range of spatial scales, closely matching both the spectral magnitude and the scale-dependent decay observed in ERA5, with the recovery of small-scale spectral power indicating an effective reconstruction of small-scale dynamical variability that is absent in the original coarse-resolution model output. We apply RAPSD analysis to the downscaled wind fields of MPI-ESM1-2-HR and MRI-ESM2-0, showing that our framework produces spectral characteristics more consistent with ERA5 in terms of energy distribution across scales than BCSD (Fig.~\ref{fig:a_spatial_patterns_rapsd}).

We report the MAE (Table~\ref{tab:cmip6_modeling_comparison}) between the LR ESM simulations and the downscaled fields after spatial pooling back to the original low resolution to assess the capability of different methods to preserve the large-scale characteristics of the ESM simulations. The results show that our framework exhibits richer fine-scale visual details and achieves the lowest pooled MAE among all compared methods, indicating that the original large-scale features of the ESM simulations are preserved to the greatest extent. Beyond spatial consistency, we also assess the temporal consistency of the downscaled wind fields by comparing their temporal autocorrelation with that of ERA5 (Fig.~\ref{fig:a_temporal_consistency}). The framework produces downscaled fields whose temporal autocorrelation closely matches ERA5, outperforming BCSD methods.

\subsubsection{Correction of climatological and distributional biases}\label{subsubsec:correction_of_climatological_and_distributional_biases}

As ESM simulations and ERA5 reanalysis are not paired at the sample level~\cite{Hess2025fast,Aich2025conditional}, we assess the bias correction capability of the downscaling methods in terms of long-term statistical consistency. Specifically, we assess the capability of the downscaled wind fields to reproduce the climatological characteristics observed in ERA5.

\begin{figure*}[!t]
    \centering
    \begin{overpic}[width=\textwidth]{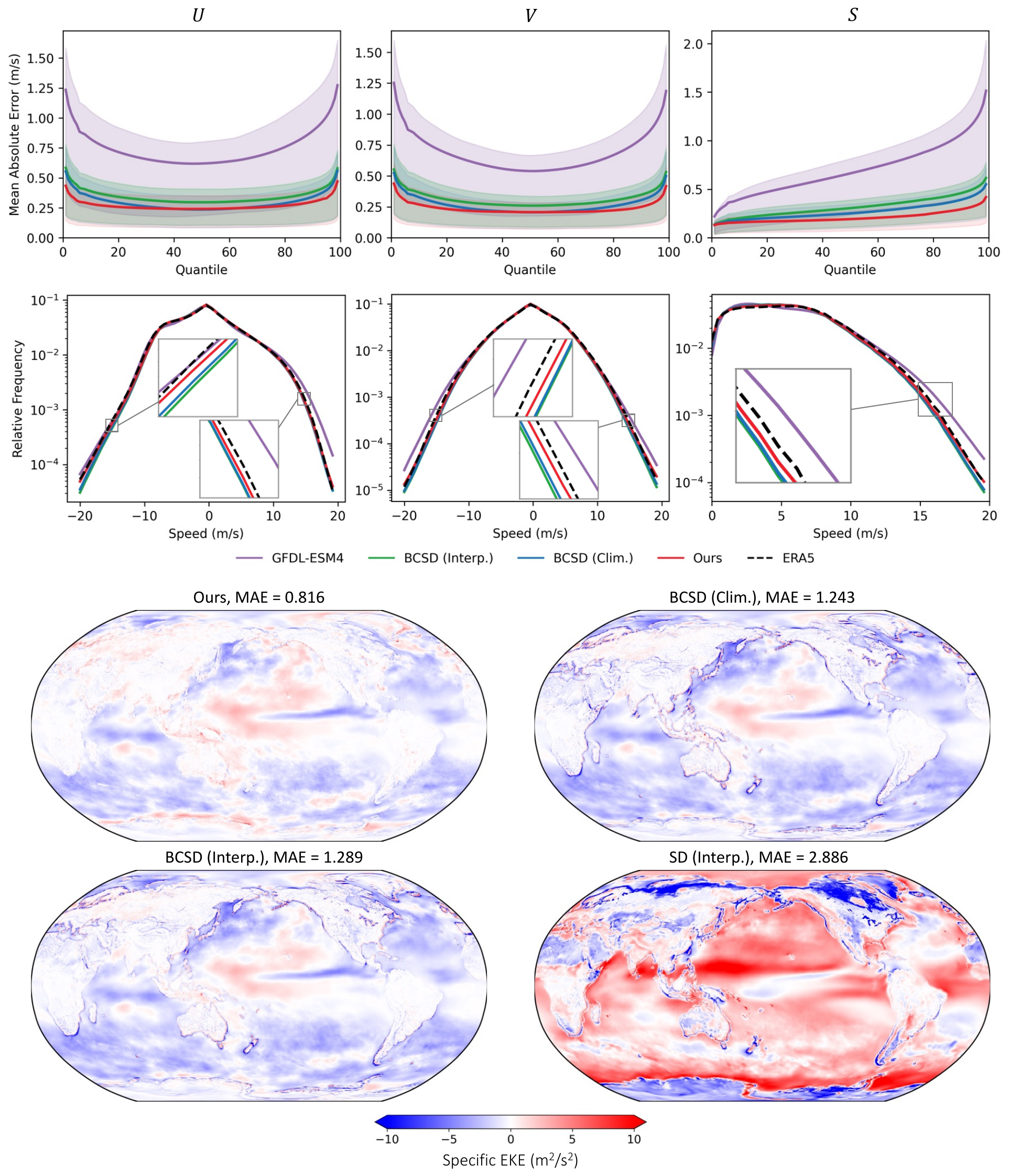}
        \put(1,98){\scriptsize{$\textbf{a}$}}
        \put(1,76){\scriptsize{$\textbf{b}$}}
        \put(1,50){\scriptsize{$\textbf{c}$}}
    \end{overpic}
    \caption{
        \textbf{Bias correction capability comparison of neural downscaling framework and BCSD baselines.}
        \textbf{a}, MAE distribution across speed percentiles for global near-surface wind components and wind speed relative to ERA5, with the shaded region representing the interquartile range of errors (25$^\text{th}$--75$^\text{th}$ percentile).
        \textbf{b}, The relative frequency of near-surface wind components and wind speed with respect to ERA5.
        \textbf{c}, The global distribution of mean EKE bias.
    }
    \label{fig:bias_correction}
\end{figure*}

For the MAE at each wind speed percentile (Fig.~\ref{fig:bias_correction}a), our framework outperforms both the raw GFDL-ESM4 simulations and BCSD for the near-surface wind components. Relative to GFDL-ESM4, the bias-corrected wind fields largely reduce quantile biases evaluated at each grid point along the temporal dimension. We quantify these improvements by reporting the error magnitudes within the percentile ranges associated with high wind speeds (Table~\ref{tab:cmip6_modeling_comparison}), enabling a direct comparison of performance in dynamically active regimes. At the \textit{95}$^{\text{th}}$ percentile, our framework achieves MAEs of 0.322~m~s$^{-1}$ for the wind components and 0.309~m~s$^{-1}$ for wind speed, representing performance improvements of 66.47\% and 74.42\%, respectively, relative to the raw GFDL-ESM4 simulations, which exhibit errors of 0.96~m~s$^{-1}$ and 1.21~m~s$^{-1}$ without bias correction. At the \textit{5}$^{\text{th}}$ percentile, our framework attains a MAE of 0.296~m~s$^{-1}$ for the wind components, corresponding to a 68.48\% improvement over GFDL-ESM4 with error of 0.94~m~s$^{-1}$, with performance gains surpassing those achieved by BCSD. Fig.~\ref{fig:a_bias_correction_q95} presents the global distribution of MAE at the \textit{95}$^{\text{th}}$ percentile wind speed, where the elevated errors are largely attributed to the downscaling discontinuities introduced by BCSD at land-sea boundaries.

We directly assess climatological bias by comparing errors in the global long-term mean of downscaled wind components and wind speed (Table~\ref{tab:cmip6_modeling_comparison}). Among the baselines, BCSD (Clim.) yields the lowest error for the wind components, with a MAE of 0.192~m~s$^{-1}$, corresponding to a 65.77\% improvement over GFDL-ESM4. Our framework achieves a comparable climatological MAE of 0.195~m~s$^{-1}$, representing a 65.15\% improvement. Despite not explicitly targeting climatological mean correction, the proposed framework effectively reduces long-term mean biases through learning from historical ERA5 data, achieving performance comparable to BCSD (Clim.)---which is specifically designed for climatology-adjustment bias correction---while retaining the capacity to capture spatial and temporal variability beyond the mean state. The spatial Pearson correlation coefficient (PCC) of the climatology further confirms this competitive performance. Consistent with these findings, the bias correction performances of the downscaled MPI-ESM1-2-HR and MRI-ESM2-0 simulations are reported in Table~\ref{tab:a_bias_correction_1}.

In terms of relative frequency (Fig.~\ref{fig:bias_correction}b), the raw GFDL-ESM4 simulations show substantial deviations from ERA5, characterized by a strong underestimation of calm and moderate winds and a slight overestimation of strong winds. The prior bias correction markedly reduces these discrepancies, although systematic distortions persist, with a tendency to overestimate low wind speeds and underestimate high wind speeds. For wind components, near-zero wind conditions are intrinsically unstable because small directional fluctuations can induce large relative variations when wind magnitudes approach zero. Although BCSD partially preserves weak winds, this sensitivity leads to pronounced overestimation in the weak-wind regime. We mitigate this instability by favouring calm conditions for winds below 1~m~s$^{-1}$, thereby improving the overall fidelity of wind component downscaling but relatively overestimating calm wind speeds. For wind speed, BCSD methods exhibit considerably weaker bias correction capability than our framework, particularly across the mid-to-high wind-speed range, where systematic biases remain evident. These improvements are more pronounced for the downscaled fields of MPI-ESM1-2-HR and MRI-ESM2-0 simulations (Fig.~\ref{fig:a_bias_correction_freqency}).

We decompose the total specific kinetic energy of global wind fields into mean kinetic energy (MKE) and eddy kinetic energy (EKE)~\cite{Kang2017Evaluation}. Both our framework and BCSD (Clim.) restore much of the missing MKE, yielding distributions that closely follow the ERA5. BCSD (Clim.) achieves a MAE of 0.855 m$^2$ s$^{-2}$ for MKE, corresponding to a 66.03\% reduction in error relative to raw GFDL-ESM4, while our framework attains a comparable MAE of 0.856 m$^2$ s$^{-2}$, reducing the error by 65.98\%. Despite these substantial improvements, small residual MKE biases persist for both methods, reflecting the influence of underlying systematic errors that cannot be fully removed. The differences become more pronounced for the EKE fields (Fig.~\ref{fig:bias_correction}c). BCSD (Clim.) achieves a MAE of 1.243 m$^2$ s$^{-2}$ for EKE, corresponding to a 60.17\% reduction in error relative to GFDL-ESM4; however, it exhibits pronounced biases near land-sea transition zones, where sharp surface heterogeneity leads to degraded performance. The framework attains a lower MAE of 0.816 m$^2$ s$^{-2}$, reducing the error by 71.73\% relative to GFDL-ESM4, and demonstrates a superior ability to recover the fine-scale turbulent kinetic energy associated with small-scale fluctuations. The quantitative analysis of the specific kinetic energy decomposition is presented in Table~\ref{tab:a_bias_correction_2}.

Fig.~\ref{fig:a_bias_correction_double} further demonstrates the capability of our framework to correct wind field divergence over the tropical Pacific Ocean, reflecting the mitigation of the Double Intertropical Convergence Zone (Double-ITCZ) bias~\cite{Zhang2019double,Liu2002Double} inherent in ESMs.

\subsubsection{Predictive uncertainty in spatial downscaling}\label{subsubsec:predictive_uncertainty_in_spatial_downscaling}

As fine-scale wind variability cannot be uniquely determined from the coarse-scale fields, multiple physically consistent solutions are admissible, indicating that wind field downscaling constitutes an intrinsically ill-posed and uncertain problem~\cite{daust2025Capturinga}.

\begin{figure*}[!t]
    \centering
    \begin{overpic}[width=\textwidth]{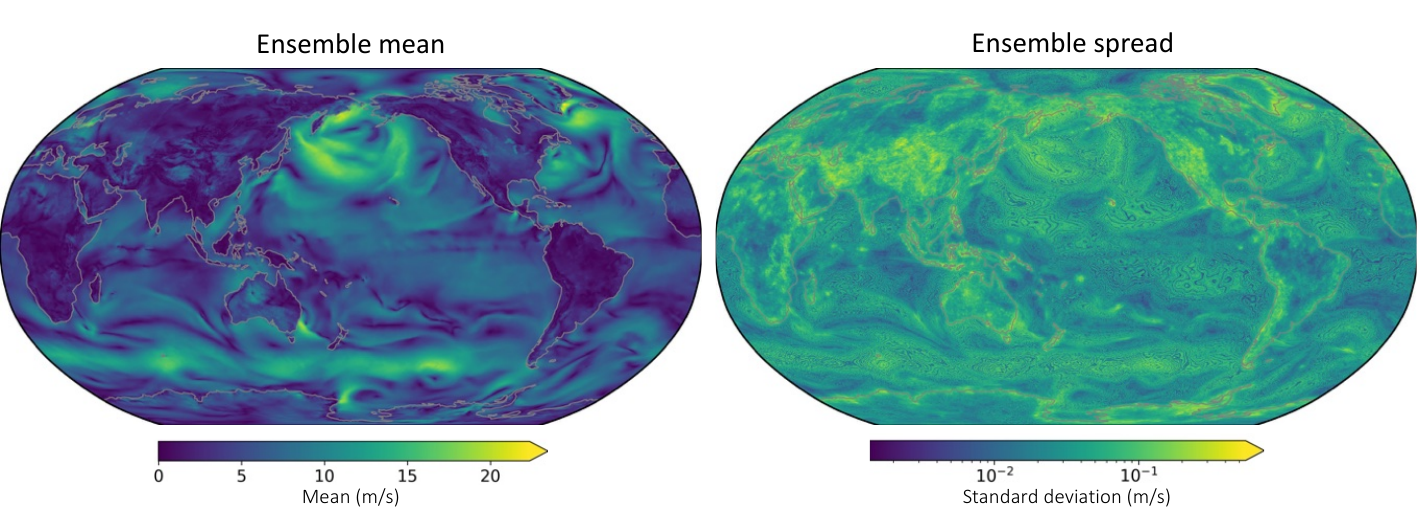}
        \put(1,35){\scriptsize{$\textbf{a}$}}
        \put(51,35){\scriptsize{$\textbf{b}$}}
    \end{overpic}
    \caption{
        \textbf{Predictive uncertainty of the neural downscaling framework for daily near-surface wind fields from the GFDL-ESM4 simulations.}
        \textbf{a,b}, The ensemble mean (\textbf{a}) and standard deviation (\textbf{b}) of the downscaled fields from our framework with stochastic forward passes for the near-surface wind speed, respectively.
    }
    \label{fig:uncertainty}
\end{figure*}

The neural operator in our framework is deterministic at inference, ensuring full reproducibility for given inputs and fixed model parameters. At the same time, the framework naturally supports uncertainty quantification in the downscaling process. In practice, this is achieved by generating an ensemble of predictions through stochastic forward passes with dropout layers activated~\cite{Gal2016Dropout}, which enables efficient estimation of predictive uncertainty associated with the downscaling results.

We produce 100 ensemble members for single-day near-surface wind field from GFDL-ESM4. The corresponding ensemble mean and standard deviation are presented in Fig.~\ref{fig:uncertainty}. The ensemble mean largely converges toward a smoothly varying field comparable to direct interpolation, reflecting the stability of large-scale components across stochastic realizations. The ensemble spread is primarily localized in regions where the model introduces additional fine-scale structures. This pattern is within our expectation, as such fine-scale variability is not uniquely determined by coarse-resolution inputs and thus carries inherently higher uncertainty. 

To quantify whether the generated ensemble captures the spatial distribution of downscaling errors, we conduct an additional experiment using ERA5 data degraded to the native resolution of GFDL-ESM4 as inputs. The coarsened ERA5 fields are downscaled with the pretrained neural operator. We compare the long-term spatial distribution of downscaling errors against the corresponding ensemble spread. The two exhibit a strong spatial correspondence, with correlation coefficients approximately 0.9 (Fig.~\ref{fig:a_uncertainty}a,b). This strong agreement indicates that the ensemble spread reliably reflects the spatial structure of prediction errors, suggesting that the estimated uncertainty is meaningful rather than merely stochastic.

We construct an ensemble baseline for comparison by producing a downscaled ensemble from BCSD (Clim.). For each downscaled realization, we randomly sample 100 climatological bias fields from the day-of-year-specific climatological adjustment dataset with 366 days and further generate a downscaled ensemble with 100 members. We subsequently evaluate the probabilistic performance using the Continuous Ranked Probability Score (CRPS) and the Continuous Ranked Probability Skill Score (CRPSS), thereby assessing both absolute ensemble quality and skill relative to the BCSD (Clim.) baseline. The resulting ensemble derived from our framework outperforms the BCSD (Clim.) ensemble in terms of probabilistic downscaling performance (Fig.~\ref{fig:a_uncertainty}c,d).

\subsection{Generalization to future climate scenarios}\label{subsec:generalization_to_future_climate_scenarios}

We apply the neural downscaling framework to multiple future Shared Socioeconomic Pathway (SSP)~\cite{Riahi2017Shared} scenarios with different radiative forcing trajectories to assess its zero-shot generalization capability under previously unseen climate conditions. 
Specifically, the framework is evaluated using simulations from GFDL-ESM4 under the SSP1-1.9 (SSP119), SSP2-4.5 (SSP245), and SSP5-8.5 (SSP585) scenarios, which represent low, intermediate, and high radiative forcing pathways by the end of the 21st century, respectively.

For SSP245, which is closest to the present-day trajectory of anthropogenic climate forcing~\cite{Bevacqua2025year}, we perform a systematic evaluation over the 2015--2024 period, during which ERA5 ground truth data are available. 
The results show that our framework generalizes effectively to the SSP245 future scenario, improving deficiencies in the GFDL-ESM4 near-surface wind fields relative to BCSD methods. 
Consistent with the historical evaluation, the improvements manifest in realistic spatial patterns and largely reduced systematic biases (Table~\ref{tab:a_future_1} and Table~\ref{tab:a_future_2}). 

\begin{figure*}[!t]
    \centering
    \begin{overpic}[width=\textwidth]{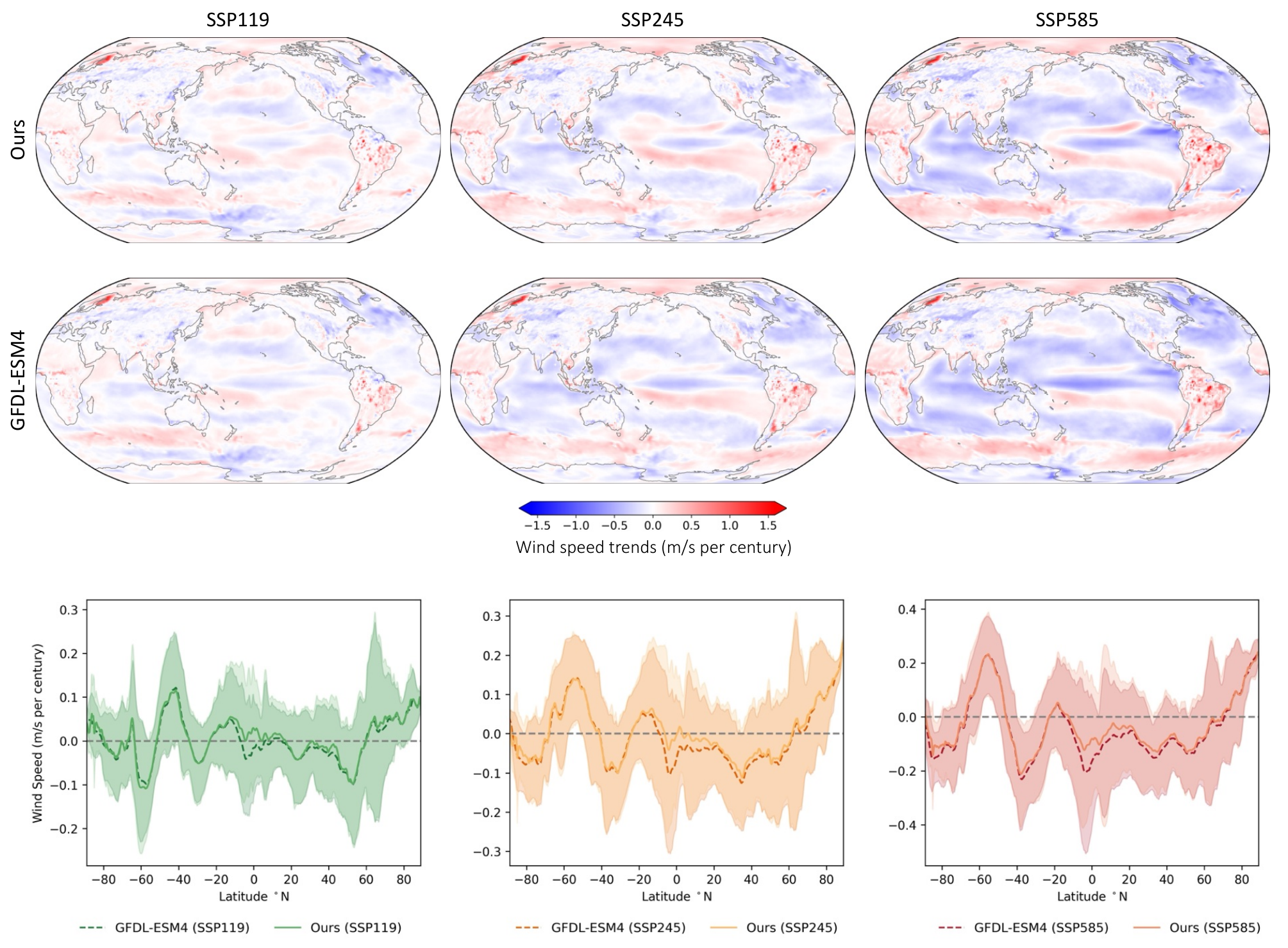}
        \put(2,72){\scriptsize{$\textbf{a}$}}
        \put(2,53){\scriptsize{$\textbf{b}$}}
        \put(2,28){\scriptsize{$\textbf{c}$}}
    \end{overpic}
    \caption{
        \textbf{Future climate scenario projections downscaled by the neural downscaling framework.}
        \textbf{a,b}, Spatial patterns of future global near-surface wind speed trends produced by our framework (\textbf{a}) and the original GFDL-ESM4~\cite{Dunne2020GFDL} simulations (\textbf{b}) under the SSP119, SSP245, and SSP585 scenarios.
        \textbf{c}, Latitudinally averaged wind speed trends of our framework and the original GFDL-ESM4 simulations over the global domains for the SSP119, SSP245, and SSP585 scenarios, from left to right.
    }
    \label{fig:future}
\end{figure*}

To demonstrate that our framework does not merely reproduce historical climate characteristics, but instead learns a generalizable state mapping through continuous dynamical modeling, we apply our framework to late-21st-century wind field projections under the three different future SSP scenarios. 
We then compare the long-term trends in the original GFDL-ESM4 simulations with those obtained from the downscaled fields produced by our framework (Fig.~\ref{fig:future}). 
For all three future scenarios, the spatial patterns of global near-surface wind speed trends produced by the downscaled results (Fig.~\ref{fig:future}a) closely resemble those of the original GFDL-ESM4 simulations (Fig.~\ref{fig:future}b), with the primary difference arising from the enhanced spatial resolution in the downscaled fields. 
Importantly, the large-scale structure and magnitude of the projected trends are well preserved, indicating that our framework does not distort the underlying climate change signal. 

We further quantify the agreement between the downscaled fields and the original GFDL-ESM4 simulations by computing latitudinally averaged wind speed trends over the global domains (Fig.~\ref{fig:future}c). 
The resulting profiles exhibit a high degree of consistency across all scenarios, with only minor discrepancies primarily confined to the equatorial oceanic region. 
To further elucidate the spatial structure of future wind speed changes, we conduct a regional analysis by separately examining land and ocean domains across different latitude bands (Fig.~\ref{fig:a_future_gen}).
From the simulation results of GFDL-ESM4, we find that land areas of the Northern Hemisphere generally exhibit a weakening trend in near-surface wind speed, while pronounced wind speed intensification emerges at high northern latitudes, particularly over the Arctic Ocean. A comparison across scenarios further reveals that the magnitude of these wind speed changes increases from SSP119 to SSP585, consistent with the progressively stronger radiative forcing and in agreement with previous studies~\cite{Shen2022Evaluation,Liu2024Rapid}.

\section{Discussion}

We propose a neural downscaling framework that enables global downscaling of near-surface wind fields, while generalizing across different climate models and future scenarios. The framework learns a coarse-to-fine atmospheric state mapping solely from ERA5 reanalysis~\cite{Hersbach2020era5} via multi-scale self-supervised learning and integrates this learned mapping with prior bias correction to produce global HR wind fields applicable to arbitrary ESM simulations. Across comprehensive evaluations covering both historical and future projection experiments, the framework consistently outperforms the widely used BCSD methods~\cite{Wood2004hydrologic} in terms of recovering spatial patterns, correcting systematic biases, preserving future projection trends and quantifying uncertainty. These results demonstrate that learning a state mapping through a downscaling neural operator provides a promising and flexible pathway for generalizable downscaling of ESM simulations, enabling a single trained model to be applied across diverse climate model ensembles without retraining.

A key implication of our framework is its strong generalizability across both climate models and future scenarios. This generalization capability is enabled by two central designs. First, the operator-learning formulation represents downscaling as a mapping between coarse- and fine-scale continuous atmospheric state fields, rather than as a mapping between discrete vectors defined on a specific pair of input-output grids. This means that the model is trained to capture the underlying functional relationship between atmospheric states across scales, instead of learning resolution-specific or model-specific statistical correspondences. As a result, the learned mapping is less dependent on a particular discretization and can be transferred more naturally to simulations produced by different ESMs. Second, the model is trained in a self-supervised manner using only ERA5 reanalysis, where HR fields are progressively coarsened to multiple lower resolutions to construct synthetic coarse-fine pairs. Through this process, the model is exposed to a wide range of cross-scale transformations and learns the general dynamical relationships that link large-scale atmospheric structure to fine-scale variability (Table~\ref{tab:a_ablation_2} and Fig.~\ref{fig:a_ablation}). Since this training strategy does not rely on paired ERA5-ESM samples from any individual climate model, the learned operator is not anchored to the biases or statistical characteristics of a specific ESM.

The framework improves the fidelity of near-surface wind fields even for relatively computationally expensive ESM simulations such as GFDL-ESM4~\cite{Dunne2020GFDL}, while maintaining flexibility and computational efficiency. Increasing the native resolution of GFDL-ESM4 from approximately $1^{\circ}$ to $0.25^{\circ}$ through dynamical simulation would incur a computational cost of about $\sim$4.4 days per simulated year using 1,728 CPU cores~\cite{Lin2024GFDL}. In contrast, the downscaling framework operates directly from the original $1^{\circ}$ GFDL-ESM4 output at a substantially lower cost. Generating the $1^{\circ}$ simulations requires $\sim$2.7 hours per simulated year~\cite{Lin2024GFDL}, followed by an additional inference cost of only $\sim$92 seconds on a single NVIDIA H20 GPU, or $\sim$55 minutes on a single CPU core (on average, with $\sim$54 seconds overhead for data loading and preprocessing). Despite this several-orders-of-magnitude reduction in computational expense, the resulting downscaled fields achieve fidelity comparable to or higher than that of dynamically simulated HR outputs, particularly in terms of bias reduction, highlighting the efficiency and practical value of the framework for large-scale climate applications.

The predictive uncertainty in our framework is obtained via stochastic forward passes enabled by dropout, which produce ensemble downscaled realizations; however, these estimates should be interpreted as conservative, as they primarily capture epistemic uncertainty associated with model parameters and do not explicitly account for aleatoric uncertainty arising from intrinsic climate variability and observational limitations in the reference data~\cite{kendall2017What}. Future work could address this limitation by developing native probabilistic variants of the neural operator for more comprehensive uncertainty quantification~\cite{wu2022Quantifying}, which is particularly important for risk-sensitive applications and ensemble-based climate assessments~\cite{lafferty2023Downscaling}. In addition, this study focuses on the general downscaling of near-surface wind fields within an operator learning framework and does not consider other atmospheric variables. Extending the framework to additional variables, such as precipitation, as well as to multivariate and multi-level atmospheric fields, would provide a more comprehensive evaluation of its generality and robustness, while enabling a more complete representation of boundary-layer dynamics and vertically coupled atmospheric processes.

\section{Methods}

\subsection{Data}\label{subsec41}

\noindent\textbf{ERA5 reanalysis.}
We employ the state-of-the-art ERA5~\cite{Hersbach2020era5} reanalysis dataset provided by the European Centre for Medium-Range Weather Forecasts (ECMWF), and in this study we restrict the analysis to the period 1960--2024 at a horizontal resolution of 0.25$^\circ$ with 720$\times$1440 grid points. The atmospheric variables used are 10 m zonal wind ($u_{10}$) and 10 m meridional wind ($v_{10}$). ERA5 is used to train the neural operator and regarded as the reference for all evaluations.

\noindent\textbf{CMIP6 ESMs.}
We use climate model simulations from three CMIP6~\cite{Eyring2016overview} ESMs with varying levels of complexity and spatial resolutions:

\begin{itemize}
    \item GFDL-ESM4~\cite{Dunne2020GFDL}, developed by NOAA's Geophysical Fluid Dynamics Laboratory, uses a 1$^\circ$$\times$1.25$^\circ$ horizontal grid and includes advanced representations of aerosols, atmospheric chemistry, and ocean biogeochemistry, making it well-suited for studying coupled carbon-climate-chemistry interactions.
    \item MPI-ESM1-2-HR~\cite{Mauritsen2019developments}, from the Max Planck Institute for Meteorology, employs a 0.9375$^\circ$ horizontal resolution for both atmosphere and ocean components and is designed to capture long-term climate variability through coupled atmosphere-ocean processes.
    \item MRI-ESM2-0~\cite{Seiji2019meteorological}, developed by the Meteorological Research Institute of Japan, features a finer 1.125$^\circ$ grid and incorporates detailed cloud microphysics and aerosol-cloud interaction schemes, making it particularly relevant for regional climate analysis in East Asia.
\end{itemize}

For all models, historical ESM simulations from 1960 to 2014 are used. Future projections are based on late-21st-century simulations (2070--2099) under the SSP119, SSP245, and SSP585 scenarios. In addition, simulations from 2015 to 2024 under SSP245 are included to enable evaluation against ERA5 during a future period with available observational ground truth. The analyzed variables are near-surface zonal ($uas$) and meridional ($vas$) winds.
The neural operator is trained using daily data over the 1960--1999 period. 
For evaluation, daily fields from 2005--2014 are used for the historical period, while daily data from 2015--2024 are employed for the future period where ERA5 ground truth is available. 
For climatological analyses of wind speed trends, daily data are aggregated over 1985--2014 to characterize the historical climatology, and over 2070--2099 for future scenario projections.


\noindent\textbf{Data preprocessing.} 
The training dataset is constructed from ERA5 reanalysis fields that are progressively coarsened via direct interpolation to form paired samples. 
For evaluation, all climate model simulations are bias corrected at their native resolutions using quantile delta mapping (QDM)~\cite{Douglas2013bias} with 500 quantile bins. 
This procedure aligns the statistical distributions of the model-simulated variables with those of the ERA5 reference, thereby reducing systematic biases in the model inputs prior to downscaling. 
The original ERA5 reanalysis data are used as ground truth for all evaluations.

\subsection{Downscaling neural operator}\label{subsec42}

Neural operators provide a practical mechanism for realizing resolution-independent downscaling by learning mappings directly between continuous function spaces. 
In this work, we adopt the neural operator paradigm as the backbone for approximating the theoretical downscaling operator $G$, which maps a coarse-scale atmospheric state field $\bm{a}\in\mathcal{A}(\Omega;\mathbb{R}^{d_a})$ to a fine-scale target field $\bm{b}\in\mathcal{B}(\Omega;\mathbb{R}^{d_b})$ defined over the same physical domain $\Omega$. 
Following the operator learning framework introduced in Section~\ref{subsubsec:learning_continuous_state_mappings}, this mapping is approximated by a parameterized neural operator $G_{\boldsymbol{\theta}}$ with learnable parameters $\boldsymbol{\theta}\in\Theta$, which enables the modeling of nonlocal interactions inherent in atmospheric processes while supporting flexible discretization. 

The general representation of a neural operator can be written in integral form as:
\begin{equation}\label{eq:integral_operator}
    \bm{b}(\bm{x}) = G_{\boldsymbol{\theta}}(\bm{a})(\bm{x}) 
    = \int_{\Omega} \kappa_{\boldsymbol{\theta}}(\bm{x},\bm{y})\, \bm{a}(\bm{y})\, d\bm{y},
\end{equation}
where the parameterizable kernel $\kappa_{\boldsymbol{\theta}}(\bm{x},\bm{y})$ characterizes how information at location $\bm{y}\in\Omega$ influences the output at location $\bm{x}\in\Omega$. This formulation naturally captures long-range and multi-scale dependencies that are essential for representing transport- and circulation-driven processes in the atmospheric system. In practice, neural operator architectures do not explicitly construct the kernel $\kappa_{\boldsymbol{\theta}}$, but instead learn an implicit parameterization of the operator through structured neural representations.

\begin{figure*}[!t]
    \centering
    \begin{overpic}[width=\textwidth]{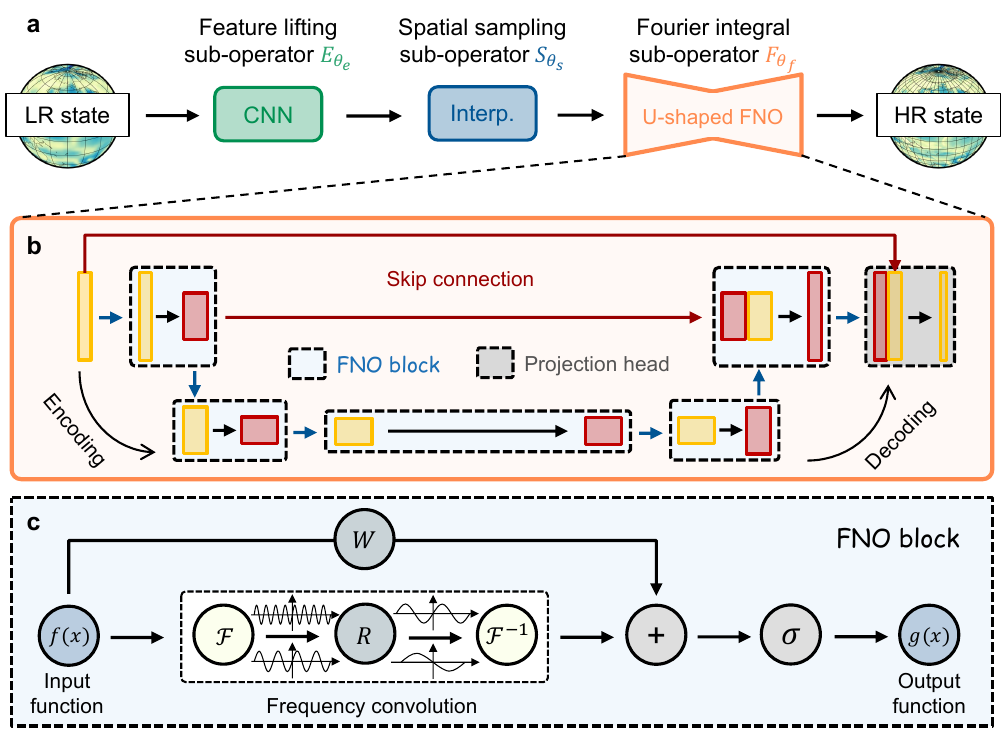}
    \end{overpic}
    \caption{
        \textbf{Overview architecture of the downscaling neural operator.}
        \textbf{a}, The overall architecture of the downscaling neural operator.
        \textbf{b}, The detailed architecture of the Fourier integral operator $F_{\boldsymbol{\theta}_3}$, \textit{i.e.}, the U-shaped FNO, which consists of an encoder-decoder architecture with skip-connections in which convolutional layers are replaced by FNO~\cite{Li2021Fourier} blocks.
        \textbf{c}, The detailed architecture of single FNO block.
    }
    \label{fig:neural_operator}
\end{figure*}

To realize the neural operator $G_{\boldsymbol{\theta}}$ for generalizable climate downscaling, we explicitly account for the fact that, unlike conventional operator learning problems, the input and output functions of the downscaling operator represent atmospheric states at different spatial scales with different discretization. Accordingly, the downscaling neural operator is designed to naturally accommodate resolution enhancement between the input and output representations, and is implemented using a three-stage architecture that extends the basic integral operator formulation with two additional operator components (Fig.~\ref{fig:neural_operator}):
\begin{equation}
    G_{\boldsymbol{\theta}} := F_{\boldsymbol{\theta}_f} \circ S_{\boldsymbol{\theta}_s} \circ E_{\boldsymbol{\theta}_e},
\end{equation}
where $E_{\boldsymbol{\theta}_e}$ is the Feature lifting sub-operator, $S_{\boldsymbol{\theta}_s}$ is the Spatial sampling sub-operator, $F_{\boldsymbol{\theta}_f}$ is the Fourier integral sub-operator, and $\circ$ denotes the function composition. These sub-operators are defined as follows:

\begin{itemize}
    \item \textbf{Feature lifting sub-operator} $E_{\boldsymbol{\theta}_e}\!:\!\mathcal{A}(\Omega;\mathbb{R}^{d_a}) \to \mathcal{L}(\Omega;\mathbb{R}^{d_l})$ maps the coarse-scale state field function $\bm{a}(\bm{x}) \in \mathcal{A}(\Omega;\mathbb{R}^{d_a})$ into a high-dimensional latent function $\bm{l}(\bm{x}) \in \mathcal{L}(\Omega;\mathbb{R}^{d_l})$. The sub-operator is implemented as a convolutional neural network, which is translation-invariant and learns a compact and informative latent function that captures both large-scale circulation patterns and localized spatial gradients.
    \item \textbf{Spatial sampling sub-operator} $S_{\boldsymbol{\theta}_s}\!:\!\mathcal{L}(\Omega;\mathbb{R}^{d_l}) \to \mathcal{L}(\Omega;\mathbb{R}^{d_b})$ disaggregates the coarse-scale latent function representation $\bm{l}(\bm{x}) \in \mathcal{L}(\Omega;\mathbb{R}^{d_l})$ into a fine-scale latent function representation $\bm{l}'(\bm{x}) \in \mathcal{L}(\Omega;\mathbb{R}^{d_b})$ by a parameter-free interpolation, \textit{e.g.}, bicubic interpolation. It aligns the discretization strategy of the coarse-scale latent function with the target fine-scale state field function $\bm{b}(\bm{x}) \in \mathcal{B}(\Omega;\mathbb{R}^{d_b})$.
    \item \textbf{Fourier integral sub-operator} $F_{\boldsymbol{\theta}_f}\!:\!\mathcal{L}(\Omega;\mathbb{R}^{d_b}) \to \mathcal{B}(\Omega;\mathbb{R}^{d_b})$ constructs the fine-scale state field function $\bm{b}(\bm{x}) \in \mathcal{B}(\Omega;\mathbb{R}^{d_b})$ from the fine-scale latent function representation $\bm{l}'(\bm{x}) \in \mathcal{L}(\Omega;\mathbb{R}^{d_b})$ by a U-shaped Fourier neural operator (U-FNO)~\cite{Li2021Fourier}. The sub-operator is implemented based on the integral formulation in Eq.~\ref{eq:integral_operator}, using an encoder-decoder architecture in which convolutional layers are replaced by Fourier neural operator blocks.  
\end{itemize}


The parameters $\boldsymbol{\theta}$ are learned by minimizing the expected discrepancy between the predicted fine-scale state field $\hat{\bm{b}}(\bm{x}_{\text{HR}}) = G_{\boldsymbol{\theta}}(\bm{a}(\bm{x}_{\text{LR}}))(\bm{x}_{\text{HR}})$ and the target fine-scale state field $\bm{b}(\bm{x}_{\text{HR}})$ over the training pairs $(\bm{a}(\bm{x}_{\text{LR}}), \bm{b}(\bm{x}_{\text{HR}}))$:
\begin{equation}
    \min_{\boldsymbol{\theta}} \;
    \mathbb{E}_{(\bm{a},\bm{b})}
    \left[
    \lVert
    G_{\boldsymbol{\theta}}(\bm{a}(\bm{x}_{\text{LR}}))(\bm{x}_{\text{HR}}) - \bm{b}(\bm{x}_{\text{HR}})
    \rVert_{1}
    \right].
\end{equation}

To further promote multi-scale consistency during model optimization, we additionally impose constraints on the first-order spatial derivatives (\textit{i.e.}, gradients) of the near-surface wind fields. The derivative-based loss penalizes discrepancies in local gradient structures between the predicted and target fields, encouraging the neural operator to better preserve fine-scale spatial variability and dynamical features beyond pointwise amplitude errors. The optimization problem for derivatives is formulated as:
\begin{equation}
    \min_{\boldsymbol{\theta}} \;
    \mathbb{E}_{(\bm{a},\bm{b})}
    \left[
    \lVert
    \nabla G_{\boldsymbol{\theta}}(\bm{a}(\bm{x}_{\text{LR}}))(\bm{x}_{\text{HR}})-\nabla \bm{b}(\bm{x}_{\text{HR}})
    \rVert_{1}
    \right],
\end{equation}
where $\nabla$ is the gradient operator, and the gradient is computed using the Sobel operator. We finally combine the two optimization problems into a single optimization problem:
\begin{equation}\label{eq:loss_func}
    \min_{\boldsymbol{\theta}} \;
    \mathbb{E}_{(\bm{a},\bm{b})}
    \left[
    \alpha_1 \lVert
    \hat{\bm{b}}(\bm{x}_{\text{HR}}) - \bm{b}(\bm{x}_{\text{HR}})
    \rVert_{1}
    +
    \alpha_2 \lVert
    \nabla \hat{\bm{b}}(\bm{x}_{\text{HR}}) - \nabla \bm{b}(\bm{x}_{\text{HR}})
    \rVert_{1}
    \right],
\end{equation}
where $\alpha_1$ and $\alpha_2$ are the loss weights for the pointwise and derivative-based loss terms, respectively. We set $\alpha_1 = 1$ and $\alpha_2 = 0.1$ in our experiments (Table~\ref{tab:a_ablation_1}).

\subsection{Bias correction and spatial disaggregation baseline}

We take BCSD~\cite{Wood2004hydrologic} as the main baseline methodes for comparison. It is implemented by first bias correcting the model simulation to the ERA5 reanalysis at its coarse-scale native resolution, and then spatially disaggregating the bias corrected simulation to the target fine-scale resolution.

The prior bias correction is implemented using QDM~\cite{Douglas2013bias}, which is the same as the preprocessing step used in our neural downscaling framework. 
For spatial disaggregation, \textit{i.e.}, upscaling the spatial resolution of the model simulation, we employ two commonly used baseline strategies: interpolation (Interp.) and climatology adjustment (Clim.). 

Interpolation is implemented using a simple bilinear interpolation to upscale the LR bias corrected model simulation to the target fine-scale resolution. 

The climatology adjustment method is more involved and is implemented by adding climatological biases estimated from the historical overlap period between the model simulations and ERA5 reanalysis. 
Specifically, day-of-year-specific mean climatological differences are computed at each grid point between the interpolated fields derived from coarsened ERA5 reanalysis data and the original HR ERA5 fields, and are subsequently added to the interpolated fields at the corresponding day of year to obtain the final downscaled fields. 
The climatological bias is computed as:
\begin{equation}\label{eq:clim_bias}
    \Delta \bm{b}_{\mathrm{clim}}(\bm{x}, d)
    =
    \frac{1}{Y}
    \sum_{y=1}^{Y}
    \bm{b}_{\mathrm{ERA5}}(\bm{x}, d, y)
    -
    \frac{1}{Y}
    \sum_{y=1}^{Y}
    \tilde{\bm{b}}_{\mathrm{ERA5}}(\bm{x}, d, y),
\end{equation}
where $Y$ is the total number of years of the historical overlap period, $\bm{b}_{\mathrm{ERA5}}(\bm{x}, d, y)$ denotes the original HR ERA5 reanalysis field in year $y$, $\tilde{\bm{b}}_{\mathrm{ERA5}}(\bm{x}, d, y)$ denotes the HR field obtained by interpolating the coarsened ERA5 reanalysis to the target grid in year $y$, and $d=\mathrm{DOY}(t)$ represents the corresponding day of year (DOY) for the time step $t$.
The climatology adjustment is formulated as:
\begin{equation}
    \hat{\bm{b}}_{\mathrm{Clim}}(\bm{x}, t) = 
    \tilde{\bm{b}}(\bm{x}, t)
    +
    \Delta \bm{b}_{\mathrm{clim}}(\bm{x}, d),
\end{equation}
where $\tilde{\bm{b}}(\bm{x}, t)$ represents the HR field obtained by interpolating the LR, bias-corrected model simulation to the target grid, and $\Delta \bm{b}_{\mathrm{clim}}(\bm{x}, d)$ denotes the day-of-year-specific climatological bias.

In this work, a 40-year period from 1960 to 2000 is used to estimate the day-of-year-specific climatological biases, which is chosen to coincide with the training period of the neural operator and is sufficiently long to provide stable estimates of the climatological mean biases. 
For BCSD (Clim.), the day-of-year-specific climatological biases are computed solely based on the ERA5 reanalysis, and the degree of spatial coarsening is flexibly specified. 
As a result, BCSD naturally supports cross-model application and downscaling across arbitrary spatial resolutions without requiring model-specific retraining, making it a suitable and widely adopted baseline for comparison.

\subsection{Experimental setup}

\subsubsection{Network architecture and hyperparameters}\label{subsec:network_architecture_and_hyperparameters}

\noindent\textbf{Feature lifting sub-operator $E_{\boldsymbol{\theta}_e}$.}
The module employs a modified convolutional neural network to construct a high-dimensional latent function representation. The encoder begins with a 3$\times$3 convolutional head that projects the input channels into a 64-dimensional feature space. This is followed by a stack of 32 residual blocks, each composed of two 3$\times$3 convolutions with a ReLU activation inserted after the first convolution. Each residual branch is scaled by a factor of 0.1 to stabilize optimization. All blocks operate with a constant width of 64 channels. After passing through the residual stack, a final 3$\times$3 convolution forms the encoder tail and produces the latent feature map used by the subsequent stages of the operator.

\noindent\textbf{Spatial sampling sub-operator $S_{\boldsymbol{\theta}_s}$.}
To match the size of the encoded features to the target output discretization, $S_{\boldsymbol{\theta}_s}$ applies an explicit spatial interpolation step before $F_{\boldsymbol{\theta}_f}$. The interpolation is parameterized by a scaling factor that is randomly sampled between 3 and 6 during training, whereas at inference time the scaling factor is arbitrarily specified to match the desired target resolution. Bicubic interpolation is used by default, providing a good balance between smoothness and structural fidelity.

\noindent\textbf{Fourier integral sub-operator $F_{\boldsymbol{\theta}_f}$.}
After sampling the features to the target output discretization, the features are processed by a two-layer U-FNO to construct the fine-scale state field function. 
The network follows an encoder-decoder architecture composed of FNO~\cite{Li2021Fourier} blocks, each of which combines a spectral convolution with a pointwise residual pathway. The spectral component performs truncated Fourier transforms with a fixed number of retained modes in each spatial dimension and applies learned complex-valued linear transformations to capture global, nonlocal interactions, while the residual pathway employs $1\times1$ convolutions together with interpolation to preserve local information and support resolution changes.

The U-FNO begins with a stem projection that lifts the input features to a hidden width of 64 channels. 
The encoder consists of two FNO blocks that progressively reduce the spatial resolution to one-half and one-quarter of the target grid, while increasing the channel dimensions from 64 to 128 and then to 256, retaining $(64,64)$ and $(32,32)$ Fourier modes in the zonal and meridional directions, respectively. 
A bottleneck FNO block operates at the coarsest scale with 256 channels and retains $(32,32)$ Fourier modes. 
The decoder mirrors this hierarchy by successively restoring the spatial resolution to one-half and full resolution, reducing the channel dimensions from 256 to 128 and finally to 64, and incorporates skip connections from the corresponding encoder stages to recover fine-scale spatial details; the final decoding stage retains $(96,96)$ Fourier modes to enhance spectral expressiveness at the target resolution. 
The final output is produced through pointwise projection layers that map the 64-channel representation to the desired output dimension.

The output resolution is controlled by the target grid specified at inference time, enabling flexible and continuous resolution upscaling, even for resolutions that are not encountered during training.

\subsubsection{Evaluation metrics}

\noindent\textbf{Radial averaged power spectral density.}
We assess the spectral consistency between the downscaled fields and the fine-scale ERA5 reanalysis using the RAPSD~\cite{Ding2025Superresolution}. Specifically, the two-dimensional Fourier transform is first applied to each spatial field to obtain its spectral representation. The resulting power spectrum is then converted from Cartesian wavenumber space to radial wavenumber space by averaging spectral energy over concentric annuli with equal spatial wavelength. The RAPSD is defined as:
\begin{equation}
    \mathrm{RAPSD}(\lambda) = \frac{1}{N_\lambda} \sum_{k \in \mathcal{A}(\lambda)} |\mathcal{F}(k)|^2,
\end{equation}
where $\mathcal{F}(k)$ denotes the Fourier coefficient at wavenumber $k$, $\mathcal{A}(\lambda)$ represents the set of wavenumbers whose magnitude corresponds to wavelength $\lambda$, and $N_\lambda$ is the number of spectral components within the corresponding annulus.

\noindent\textbf{Specific kinetic energy decomposition.}
We decompose the specific kinetic energy into MKE and EKE components~\cite{Kang2017Evaluation}. 
The MKE characterizes the strength of the climatological mean flow, whereas the EKE represents the variance associated with higher-frequency fluctuations around the mean state.
Unless otherwise specified, the term kinetic energy (including MKE and EKE) used throughout this manuscript refers to the horizontal specific kinetic energy, \textit{i.e.}, kinetic energy per unit mass.

Let $u(\bm{x},t)$ and $v(\bm{x},t)$ denote the near-surface zonal and meridional wind components, respectively, and $\overline{(\cdot)}$ denote a long-term temporal mean taken over the analysis period. 
The climatological mean kinetic energy $\mathrm{MKE}(\bm{x})$ is defined as:
\begin{equation}
    \mathrm{MKE}(\bm{x}) = \frac{1}{2}\left( \overline{u}^2(\bm{x}) + \overline{v}^2(\bm{x}) \right),
\end{equation}
while the instantaneous turbulent kinetic energy $\mathrm{EKE}(\bm{x},t)$ is computed from the velocity fluctuations as:
\begin{equation}
    \mathrm{EKE}(\bm{x},t) = \frac{1}{2}\left( (u(\bm{x},t)-\overline{u}(\bm{x}))^2 + (v(\bm{x},t)-\overline{v}(\bm{x}))^2 \right),
\end{equation}
and the total specific kinetic energy $\mathrm{KE}(\bm{x},t)$ is approximately conserved, \textit{i.e.},
\begin{equation}
    \mathrm{KE}(\bm{x},t) = \frac{1}{2}(u^2(\bm{x},t) + v^2(\bm{x},t))
    \approx \mathrm{MKE}(\bm{x}) + \mathrm{EKE}(\bm{x},t).
\end{equation}

\noindent\textbf{Future wind speed trends.}
We apply the framework to the future wind field projections from the GFDL-ESM4 model under the three different future SSP scenarios (SSP119, SSP245, and SSP585) during the period of 2070--2099 to compute the future wind speed climatology $\overline{S}(\bm{x}, t_{2070-2099})$. For historical wind speed climatology $\overline{S}(\bm{x}, t_{1985-2014})$, we compute the long-term temporal mean of the wind speed over the period from 1985 to 2014. We then compute the future wind speed trends $\Delta S(\bm{x})$ by subtracting the historical wind speed climatology from the future wind speed climatology:
\begin{equation}\label{eq:future_wind_speed_trends}
    \Delta S(\bm{x}) = \overline{S}(\bm{x}, t_{2070-2099}) - \overline{S}(\bm{x}, t_{1985-2014}),
\end{equation}
where $\overline{(\cdot)}$ denotes the long-term temporal mean taken over the analysis period.
The future wind speed trends projected by the GFDL-ESM4 simulations under three different future SSP scenarios are calculated using the same procedure in Eq.~\ref{eq:future_wind_speed_trends}.

\noindent\textbf{Probabilistic scores.}
We employ the Continuous Ranked Probability Score (CRPS) and the Continuous Ranked Probability Skill Score (CRPSS) to evaluate the probabilistic performance of the downscaled ensembles.
The CRPS for a predictive cumulative distribution function $F(x)$ and an observation $y$ is defined as:
\begin{equation}
    \mathrm{CRPS}(F,y) = \int_{-\infty}^{+\infty} \left( F(x) - \mathbb{I}(x \ge y) \right)^2 dx ,
\end{equation}
where $\mathbb{I}(x \ge y)$ is the indicator function. A lower CRPS indicates a more accurate probabilistic prediction. The CRPSS is defined relative to a reference predictions as:
\begin{equation}
    \mathrm{CRPSS} = 1 - \frac{\mathrm{CRPS}_{\text{model}}}{\mathrm{CRPS}_{\text{ref}}},
\end{equation}
where $\mathrm{CRPS}_{\text{model}}$ and $\mathrm{CRPS}_{\text{ref}}$ denote the CRPS of our framework and BCSD (Clim.) ensemble predictions, respectively. A positive CRPSS indicates that our framework outperforms BCSD (Clim.) baseline.

\section*{Author contributions} 
W.W. conceived the research idea and designed the study. W.W. C.Y. J.W. and H.K. developed the neural downscaling framework. C.Y. J.W. and H.K. implemented the baseline models, and conducted experiments on cross-model and cross-scenario generalization. C.Y. J.W. H.K. and B.H. analyzed the data and interpreted the results. W.W. C.Y. J.W. and H.K. wrote the initial draft of the manuscript. W.W. H.S. and B.H. provided critical revisions and refined the technical presentation of the manuscript. W.W. supervised the overall project and coordinated the research activities. All authors discussed the results and approved the final manuscript.

\section*{Corresponding author} 
Wenguan Wang (\url{wenguanwang@zju.edu.cn}).

\section*{Competing interests}
The authors declare no competing interests.

\clearpage
\section*{Supplementary information}

\setcounter{section}{0}
\setcounter{subsection}{0}
\setcounter{subsubsection}{0}
\renewcommand\thesection{\Alph{section}}
\renewcommand\thesubsection{\thesection.\arabic{subsection}}
\renewcommand\thesubsubsection{\thesubsection.\arabic{subsubsection}}
\renewcommand\theHsection{supp.\arabic{section}}
\renewcommand\theHsubsection{supp.\arabic{section}.\arabic{subsection}}
\renewcommand\theHsubsubsection{supp.\arabic{section}.\arabic{subsection}.\arabic{subsubsection}}
\setcounter{figure}{0}
\renewcommand{\thefigure}{S.\arabic{figure}}
\renewcommand{\theHfigure}{supp.\arabic{figure}}
\renewcommand{\figurename}{Figure}
\setcounter{table}{0}
\renewcommand{\thetable}{S.\arabic{table}}
\renewcommand{\theHtable}{supp.\arabic{table}}
\renewcommand{\tablename}{Table}
\graphicspath{ {./Figure/} {./Figure_supp/} }

\section{Additional evaluation results}
\subsection{State mapping capability}\label{subsec:state_mapping_capability}

\begin{figure*}[!h]
    \centering
    \begin{overpic}[width=\textwidth]{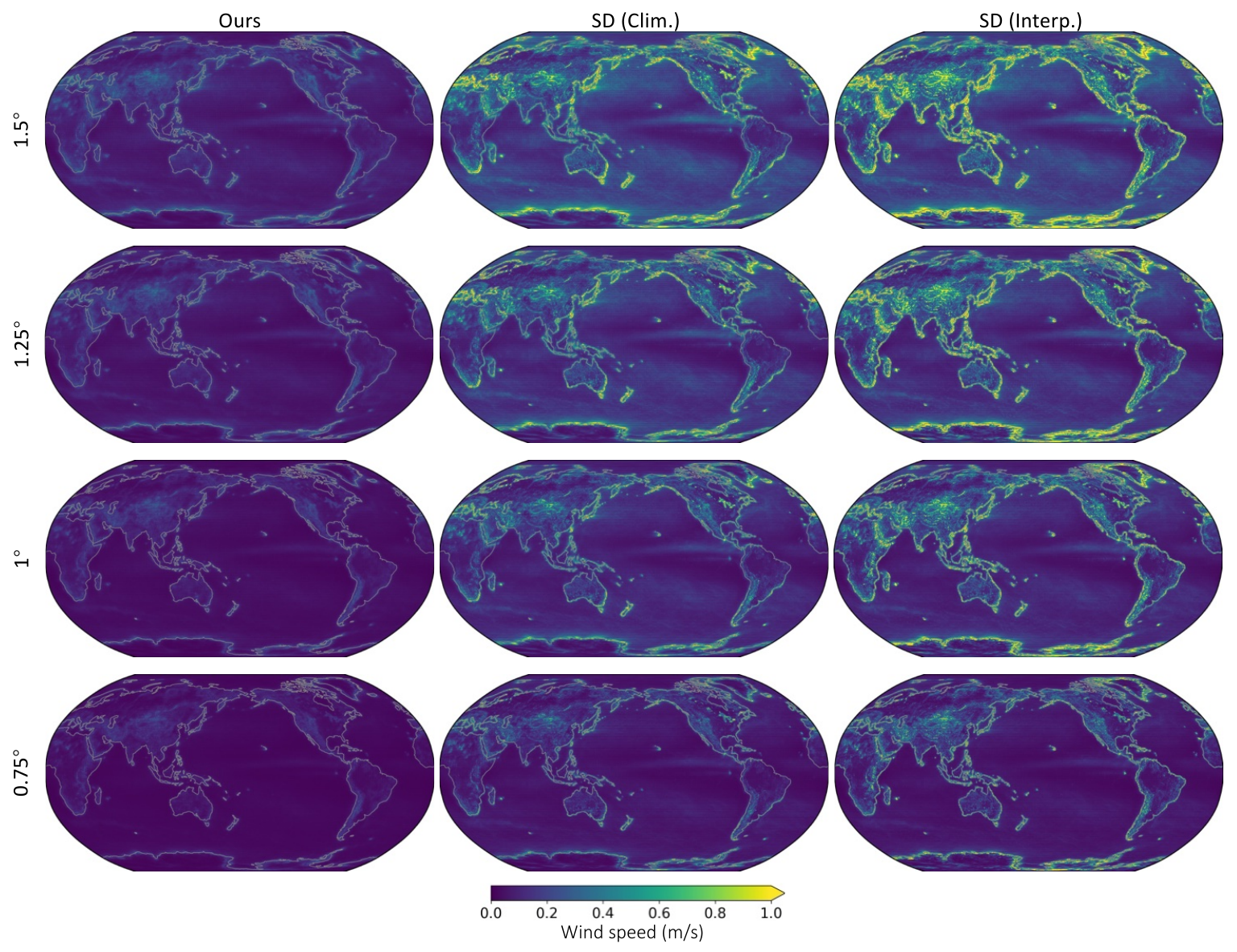}
    \end{overpic}
    \caption{
        \textbf{Global MAE distribution comparison for state mapping across the neural downscaling framework and SD baselines.} 
        From left to right, MAE of our framework, SD (Clim.), and SD (Interp.), respectively.
        From top to bottom, MAE of the downscaled fields from coarsened ERA5 reanalysis data with spatial resolution of 1.5$^\circ\times$1.5$^\circ$, 1.25$^\circ\times$1.25$^\circ$, 1$^\circ\times$1$^\circ$, and 0.75$^\circ\times$0.75$^\circ$ to HR ERA5 reanalysis with spatial resolution of 0.25$^\circ$.
    }
    \label{fig:a_state_mapping_qualitative}
\end{figure*}

We compare the global mean absolute error (MAE) distributions of our neural downscaling framework against Spatial Disaggregation (SD) baselines for state mapping of low-resolution (LR) ERA5~\cite{Hersbach2020era5} reanalysis data (Fig.~\ref{fig:a_state_mapping_qualitative}). The results demonstrate that high-resolution (HR) fields downscaled by our framework consistently exhibit lower errors relative to the fine-scale ERA5 reference, outperforming the SD baselines.

\begin{figure*}[!t]
    \centering
    \begin{overpic}[width=\textwidth]{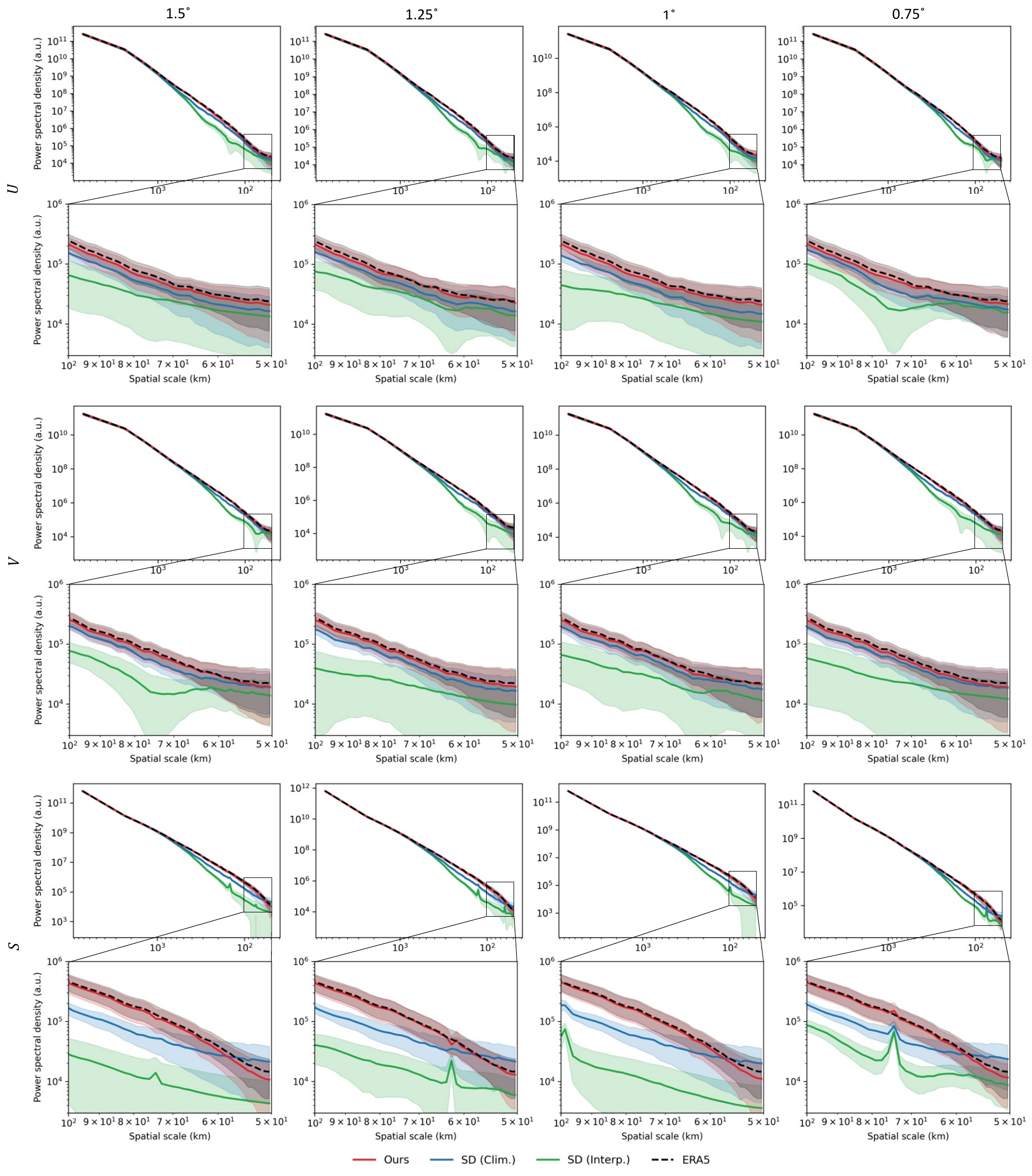}
        \put(1,98){\scriptsize{$\textbf{a}$}}
        \put(1,66){\scriptsize{$\textbf{b}$}}
        \put(1,34){\scriptsize{$\textbf{c}$}}
    \end{overpic}
    \caption{
        \textbf{RAPSD comparison for state mapping across the neural downscaling framework and SD baselines.} The RAPSD of the zonal wind component $U$ (\textbf{a}), meridional wind component $V$ (\textbf{b}), and wind speed $S$ (\textbf{c}) from coarsened ERA5 reanalysis data with spatial resolution of 1.5$^\circ\times$1.5$^\circ$, 1.25$^\circ\times$1.25$^\circ$, 1$^\circ\times$1$^\circ$, and 0.75$^\circ\times$0.75$^\circ$ to HR ERA5 reanalysis with spatial resolution of 0.25$^\circ$ for our framework and SD baselines. The lower panel shows the enlarged views of the small-scale regime (spatial scales $<100$ km). The shaded region denotes $\pm 1$ standard deviation over all time steps during the evaluation period.
    }
    \label{fig:a_state_mapping_rapsd}
\end{figure*}

To evaluate the spatial fidelity of the downscaled state mapping outputs, we analyze the radially averaged power spectral density (RAPSD) of the downscaled fields produced by our framework and SD baselines (Fig.~\ref{fig:a_state_mapping_rapsd}). The results show that our framework achieves closer spectral agreement with the fine-scale ERA5 reference across all spatial scales. This indicates that our framework more faithfully preserves the dynamical characteristics of the wind fields in the state mapping process.

\clearpage

\subsection{Spatial downscaling of different ESMs}\label{app:spatial_downscaling_different_esms}

\begin{figure*}[!h]
    \centering
    \begin{overpic}[width=\textwidth]{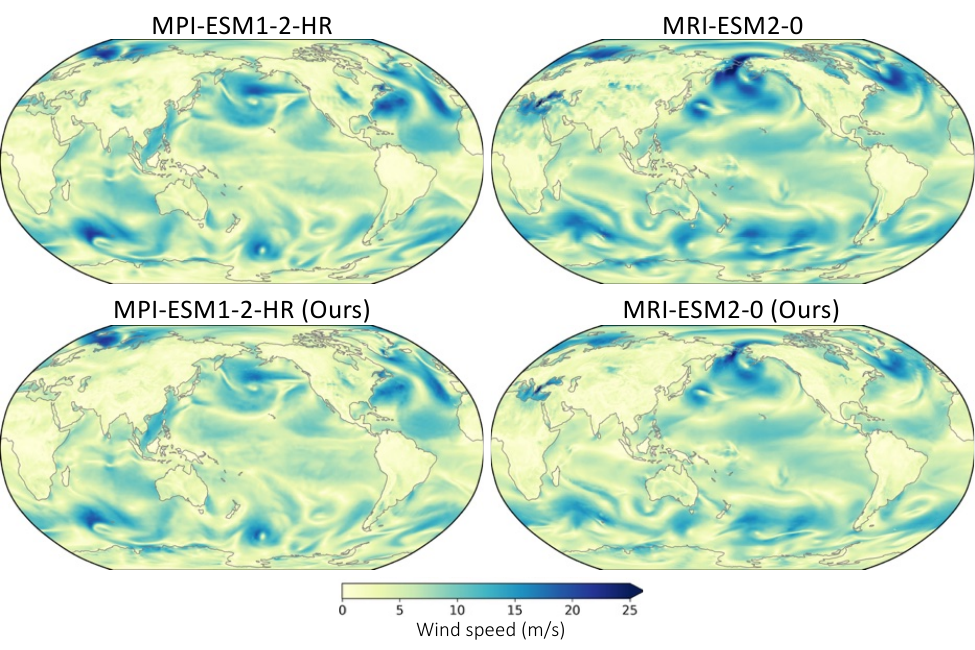}
        \put(0,64){\scriptsize{$\textbf{a}$}}
        \put(50,64){\scriptsize{$\textbf{b}$}}
        \put(0,36){\scriptsize{$\textbf{c}$}}
        \put(50,36){\scriptsize{$\textbf{d}$}}
    \end{overpic}
    \caption{
        \textbf{Qualitative comparison of raw ESM simulations and downscaled fields for the same date.}
        \textbf{a,b}, The raw coarse-scale ESM simulations for the MPI-ESM1-2-HR~\cite{Mauritsen2019developments} (\textbf{a}), and MRI-ESM2-0~\cite{Seiji2019meteorological} (\textbf{b}) simulations, respectively.
        \textbf{c,d}, The downscaled fields from our framework for the MPI-ESM1-2-HR (\textbf{c}), and MRI-ESM2-0 (\textbf{d}) simulations.
    }
    \label{fig:a_spatial_patterns_qualitative}
\end{figure*}

\subsubsection{Spatial patterns}\label{app:spatial_patterns}

We qualitatively compare the near-surface wind fields from the raw MPI-ESM1-2-HR~\cite{Mauritsen2019developments} and MRI-ESM2-0~\cite{Seiji2019meteorological} simulations, as well as their corresponding downscaled fields for the same date (Fig.~\ref{fig:a_spatial_patterns_qualitative}). The results show that our framework is able to produce realistic HR wind fields from both ESM simulations, despite their differing native spatial resolutions. Notably, the large-scale structures of the ESM simulations differ substantially from those of the HR ERA5 fields, and also vary among the ESMs themselves on the same date.

\begin{figure*}[!h]
    \centering
    \begin{overpic}[width=\textwidth]{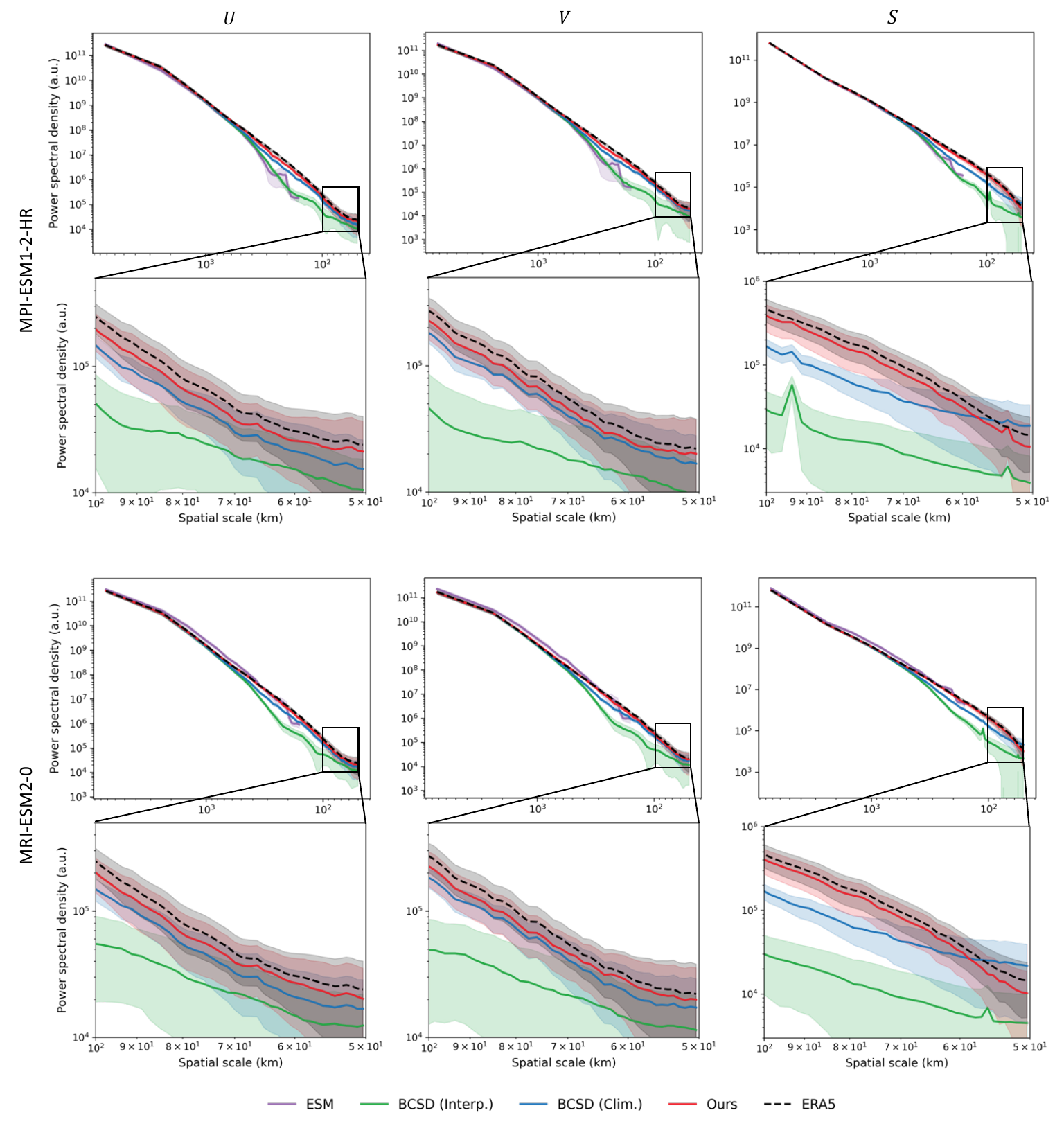}
        \put(2,99){\scriptsize{$\textbf{a}$}}
        \put(2,50){\scriptsize{$\textbf{b}$}}
    \end{overpic}
    \caption{
        \textbf{RAPSD comparison of neural downscaling framework and BCSD baselines across different ESMs.}
        \textbf{a}, RAPSD of the zonal wind component $U$, meridional wind component $V$, and wind speed $S$ for the MPI-ESM1-2-HR~\cite{Mauritsen2019developments} simulation.
        \textbf{b}, RAPSD of $U$, $V$, and $S$ for the MRI-ESM2-0~\cite{Seiji2019meteorological} simulation.
        The lower panel shows the enlarged views of the small-scale regime (spatial scales $< 100$ km). The shaded region denotes $\pm 1$ standard deviation over all time steps during the evaluation period.
    }
    \label{fig:a_spatial_patterns_rapsd}
\end{figure*}

Fig.~\ref{fig:a_spatial_patterns_rapsd} presents the RAPSD of the zonal wind component $U$, meridional wind component $V$, and wind speed $S$ for the downscaled MPI-ESM1-2-HR and MRI-ESM2-0 fields. The spectral characteristics of the downscaled fields are broadly consistent with those of the original ESM simulations, demonstrating that our framework effectively preserves the spectral properties of wind fields across different ESM simulations.

\clearpage

\subsubsection{Bias correction}\label{app:bias_correction}

\begin{figure*}[!h]
    \centering
    \begin{overpic}[width=\textwidth]{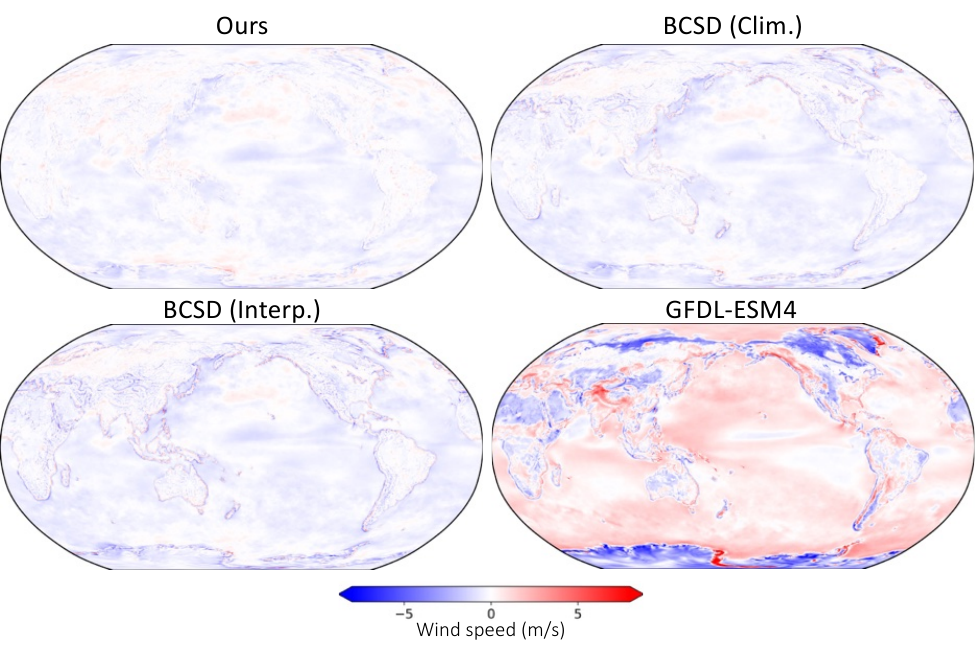}
        \put(0,64){\scriptsize{$\textbf{a}$}}
        \put(50,64){\scriptsize{$\textbf{b}$}}
        \put(0,36){\scriptsize{$\textbf{c}$}}
        \put(50,36){\scriptsize{$\textbf{d}$}}
    \end{overpic}
    \caption{
        \textbf{Global distribution of the \textit{95}$^{\text{th}}$ percentile wind speed error relative to ERA5 for the neural downscaling framework, BCSD baselines, and GFDL-ESM4~\cite{Dunne2020GFDL} simulations.}
        The global distribution of the \textit{95}$^{\text{th}}$ percentile wind speed error relative to ERA5 for our framework (\textbf{a}), BCSD (Clim.) (\textbf{b}), BCSD (Interp.) (\textbf{c}), and GFDL-ESM4 simulations (\textbf{d}), respectively.
    }
    \label{fig:a_bias_correction_q95}
\end{figure*}

We present the global distribution of the \textit{95}$^{\text{th}}$ percentile wind speed error relative to ERA5 for the downscaled fields from our framework (Fig.~\ref{fig:a_bias_correction_q95}a), the BCSD baselines (Fig.~\ref{fig:a_bias_correction_q95}b,c) and raw GFDL-ESM4 simulations (Fig.~\ref{fig:a_bias_correction_q95}d). The GFDL-ESM4 simulations exhibit a pronounced contrast between land and ocean domains. This discrepancy is substantially reduced in the downscaled fields produced by both our framework and BCSD. For BCSD methods, the most prominent residual discrepancies occur in coastal and land-sea transition regions, where strong surface heterogeneity leads to pronounced mismatches between land and ocean wind regimes.



\begin{figure*}[!t]
    \centering
    \begin{overpic}[width=\textwidth]{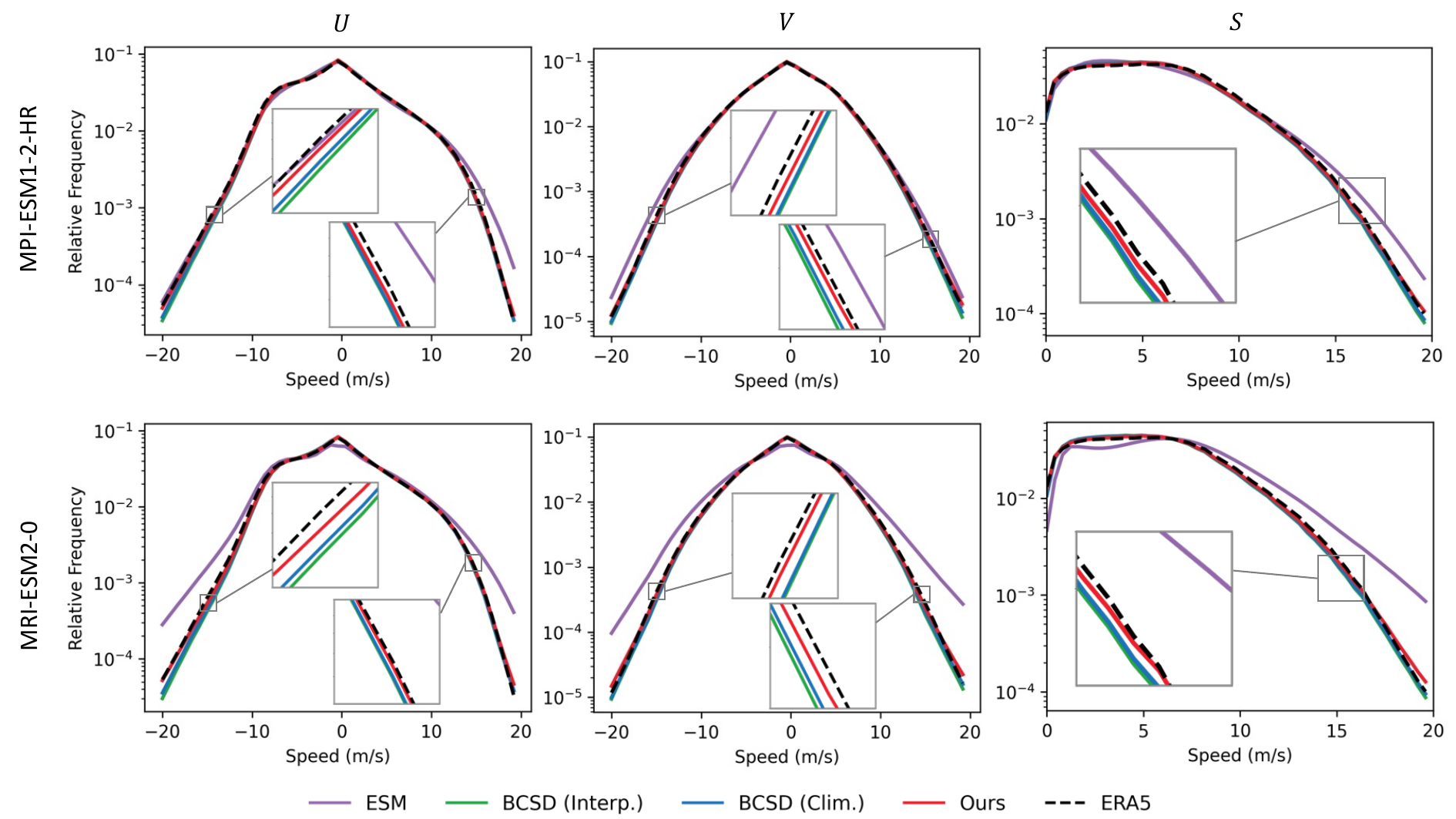}
        \put(3,53){\scriptsize{$\textbf{a}$}}
        \put(3,28){\scriptsize{$\textbf{b}$}}
    \end{overpic}
    \caption{
        \textbf{Relative frequency distributions of near-surface wind fields downscaled by the neural downscaling framework and BCSD baselines across different ESMs.}
        \textbf{a}, The relative frequency distribution of the near-surface wind fields for MPI-ESM1-2-HR~\cite{Mauritsen2019developments} simulations and their downscaled fields with respect to ERA5.
        \textbf{b}, The relative frequency distribution of the near-surface wind fields for MRI-ESM2-0~\cite{Seiji2019meteorological} simulations and their downscaled fields.
    }
    \label{fig:a_bias_correction_freqency}
\end{figure*}

We compare the relative frequency distributions of near-surface wind fields, including wind components and wind speed (Fig.~\ref{fig:a_bias_correction_freqency}), with respect to ERA5 for the MPI-ESM1-2-HR and MRI-ESM2-0 simulations and their downscaled fields obtained from our framework and BCSD baselines. The results show that the framework more effectively improves the relative frequency distributions across the full wind speed range and substantially reduces the discrepancies present in the ESM simulations compared with the BCSD methods.

\begin{figure*}[!h]
    \centering
    \begin{overpic}[width=\textwidth]{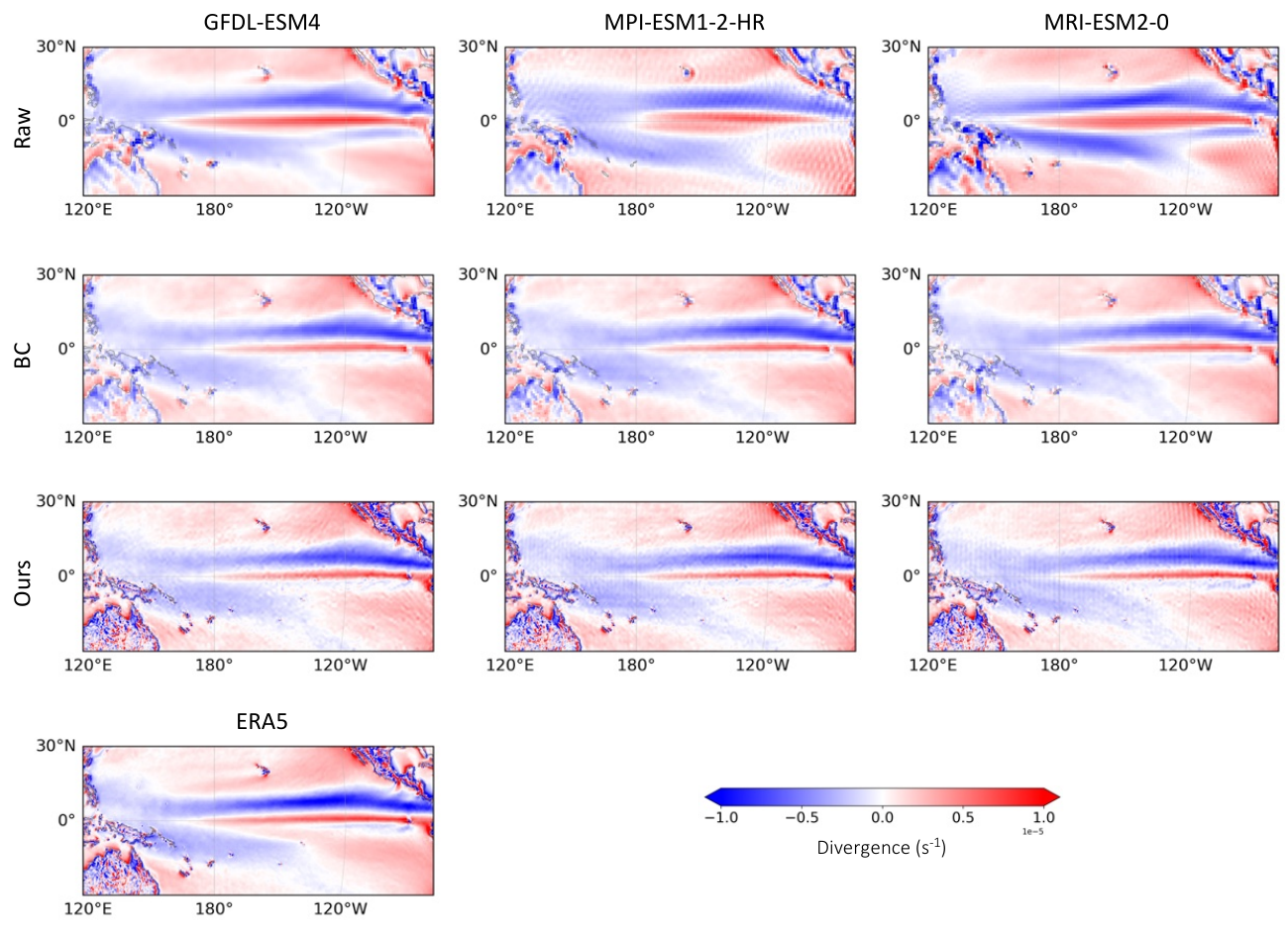}
        \put(1,71){\scriptsize{$\textbf{a}$}}
        \put(1,53.5){\scriptsize{$\textbf{b}$}}
        \put(1,35.5){\scriptsize{$\textbf{c}$}}
        \put(1,18){\scriptsize{$\textbf{d}$}}
    \end{overpic}
    \caption{
        \textbf{Double-ITCZ bias reduction performance of bias-correction preprocessing (BC) and neural downscaling framework for different ESMs.}
        \textbf{a}, The divergence of the near-surface wind field for the ESM simulations (GFDL-ESM4~\cite{Dunne2020GFDL}, MPI-ESM1-2-HR~\cite{Mauritsen2019developments}, and MRI-ESM2-0~\cite{Seiji2019meteorological}).
        \textbf{b}, The divergence of the wind field for the bias-corrected ESM simulations.
        \textbf{c}, The divergence of the wind field for the downscaled fields from our framework.
        \textbf{d}, The divergence of the wind field for the ground truth ERA5 reanalysis data.
    }
    \label{fig:a_bias_correction_double}
\end{figure*}

The Double-Intertropical Convergence Zone (ITCZ) bias, a persistent error in many ESMs, is typically associated with an unrealistic meridional splitting of tropical precipitation~\cite{Hess2025fast}. 
It is also expressed in the near-surface wind field through a robust dynamical signature: the long-term mean tropical wind divergence displays a pronounced three-band pattern, with enhanced divergence over the equator and two quasi-symmetric convergence maxima on either side. 
The divergence fields of the original ESM simulations (Fig.~\ref{fig:a_bias_correction_double}a) exhibit the same characteristic structure, indicating that this systematic bias is also present in the driving ESMs considered in this study. 
We find that bias correction substantially mitigates the Double-ITCZ bias (Fig.~\ref{fig:a_bias_correction_double}b). The neural operator further mitigates the Double-ITCZ bias (Fig.~\ref{fig:a_bias_correction_double}c), with the spatial distribution of near-surface wind divergence in the downscaled fields over the tropical Pacific Ocean becoming much more consistent with that in ERA5 (Fig.~\ref{fig:a_bias_correction_double}d).

The statistical summary of the performance comparison for the MPI-ESM1-2-HR and MRI-ESM2-0 simulations is reported in Table~\ref{tab:a_bias_correction_1}. Overall, our framework achieves errors in the long-term mean that are comparable to those of BCSD baselines, while substantially reducing the mean absolute error at both the \textit{95}$^{\text{th}}$ and \textit{5}$^{\text{th}}$ wind speed percentiles by more than 5\%. Our framework also better preserves the large-scale spatial structure of the original ESM simulations after downscaling. We further find that the magnitude of improvement achieved through downscaling is related to the severity of the original ESM biases, with simulations exhibiting larger initial biases generally benefiting more from bias correction. This behavior is consistently observed for both our framework and BCSD methods, indicating that stronger systematic errors provide greater potential for correction. Importantly, these results suggest that the framework can adapt to ESMs with varying levels of intrinsic accuracy, demonstrating a robust and broadly applicable bias-correction capability rather than one tailored to a specific model performance regime.

In terms of specific kinetic energy decomposition (Table~\ref{tab:a_bias_correction_2}), our framework exhibits performance comparable to BCSD (Clim.) for MKE, indicating a similar capability to represent the climatological mean flow. However, for EKE, which characterizes small-scale and high-frequency variability, our framework outperforms BCSD methods. This suggests that while both methods can correct large-scale mean-state biases to a comparable extent, our framework is more effective at recovering unresolved variability and preserving small-scale dynamical energy that is largely suppressed by climatology adjustment downscaling methods.

\begin{table}[!t]
    \setlength{\tabcolsep}{1.8pt}
    \centering
    \caption{
        \textbf{Statistical comparison in CMIP6 (MPI-ESM1-2-HR~\cite{Mauritsen2019developments} and MRI-ESM2-0~\cite{Seiji2019meteorological}) modeling of ERA5 performance across the neural downscaling framework and BCSD baselines.}
    }
    \footnotesize
    \label{tab:a_bias_correction_1}
        \begin{tabular}{llccccccccccccc}
        \toprule
        \multirow{2}{*}{ESMs}
        & \multirow{2}{*}{Methods}
        & \multicolumn{1}{c}{\makecell{MAE \\ (pool)}}
        & \multicolumn{2}{c}{\makecell{PCC}}
        & \multicolumn{4}{c}{ME}
        & \multicolumn{4}{c}{\textit{95}$^{\text{th}}$ PE}
        & \multicolumn{2}{c}{\textit{5}$^{\text{th}}$ PE} \\
        \cmidrule(lr){3-3}\cmidrule(lr){4-5}\cmidrule(lr){6-9}\cmidrule(lr){10-13}\cmidrule(lr){14-15}
        & & $UV$
        & $UV$ & Speed
        & $UV$ & \% & Speed & \%
        & $UV$ & \% & Speed & \%
        & $UV$ & \% \\
        \midrule

        \multirow{3}{*}{\makecell{MPI- \\ ESM1-2- \\ HR}}
        & BCSD (Interp.)
        & 0.115
        & 0.993 & 0.986
        & 0.250 & 60.13 & 0.289 & 60.82
        & 0.421 & 58.77 & 0.456 & 62.47
        & 0.401 & 56.94 \\

        & BCSD (Clim.)
        & 0.097
        & \textbf{0.996} & 0.992
        & \textbf{0.198} & \textbf{68.42} & 0.241 & 67.27
        & 0.388 & 61.91 & 0.396 & 67.42
        & 0.370 & 60.36 \\

        & Ours
        & \textbf{0.041}
        & 0.996 & \textbf{0.996}
        & 0.202 & 67.79 & \textbf{0.189} & \textbf{74.34}
        & \textbf{0.327} & \textbf{67.91} & \textbf{0.294} & \textbf{75.75}
        & \textbf{0.299} & \textbf{67.96} \\

        \midrule
        \multirow{3}{*}{\makecell{MRI- \\ ESM2-0}}
        & BCSD (Interp.)
        & 0.147
        & 0.991 & 0.983
        & 0.275 & 62.00 & 0.309 & 69.93
        & 0.440 & 68.75 & 0.509 & 72.12
        & 0.449 & 65.18 \\

        & BCSD (Clim.)
        & 0.126
        & \textbf{0.995} & 0.991
        & \textbf{0.214} & \textbf{70.45} & 0.251 & 75.55
        & 0.403 & 71.42 & 0.435 & 76.21
        & 0.411 & 68.14 \\

        & Ours
        & \textbf{0.066}
        & 0.995 & \textbf{0.996}
        & 0.218 & 69.94 & \textbf{0.185} & \textbf{81.99}
        & \textbf{0.323} & \textbf{77.08} & \textbf{0.296} & \textbf{83.82}
        & \textbf{0.313} & \textbf{75.75} \\

        \bottomrule
    \end{tabular}
    \newline
    
    \footnotesize 
    \justifying
    \noindent The pooled global mean absolute error, MAE (pool), for wind components (third column) with respect to LR bias-corrected ESM simulations. 
    The Pearson correlation coefficient, PCC, for wind components (fourth column) and wind speed (fifth column) with respect to ERA5.
    The global long-term mean error, ME, for wind components (sixth column) and wind speed (eighth column) with respect to ERA5. 
    The global absolute error in \textit{95}$^{\text{th}}$ wind component percentile (\textit{95}$^{\text{th}}$ PE, tenth column), \textit{95}$^{\text{th}}$ wind speed percentile (twelfth column), and global absolute error in \textit{5}$^{\text{th}}$ wind component percentile (\textit{5}$^{\text{th}}$ PE, fourteenth column) with respect to ERA5.
\end{table}

\begin{table}[!h]
    \setlength{\tabcolsep}{8pt}
    \centering
    \caption{
        \textbf{Statistical comparison in specific kinetic energy decomposition (MKE and EKE) for CMIP6 (GFDL-ESM4~\cite{Dunne2020GFDL}, MPI-ESM1-2-HR~\cite{Mauritsen2019developments}, and MRI-ESM2-0~\cite{Seiji2019meteorological}) modeling of ERA5 across the neural downscaling framework and BCSD baselines.}
    }
    \label{tab:a_bias_correction_2}
    \small
    \begin{tabular}{llcccc}
        \toprule
        ESMs & Methods & MAE (MKE) & \% & MAE (EKE) & \% \\
        \midrule

        \multirow{3}{*}{GFDL-ESM4}
        & BCSD (Interp.)
        & 1.081 & 57.05
        & 1.289 & 55.33 \\
        
        & BCSD (Clim.)
        & \textbf{0.855} & \textbf{66.03}
        & 1.243 & 56.92 \\
        
        & Ours
        & 0.856 & 65.98
        & \textbf{0.816} & \textbf{71.73} \\

        \midrule
        \multirow{3}{*}{MPI-ESM1-2-HR}
        & BCSD (Interp.)
        & 1.090 & 62.85
        & 1.184 & 58.49 \\
        
        & BCSD (Clim.)
        & \textbf{0.885} & \textbf{69.84}
        & 1.144 & 59.89 \\
        
        & Ours
        & 0.894 & 69.55
        & \textbf{0.833} & \textbf{70.81} \\
        
        \midrule
        \multirow{3}{*}{MRI-ESM2-0}
        & BCSD (Interp.)
        & 1.209 & 65.99
        & 1.391 & 74.81 \\
        
        & BCSD (Clim.)
        & \textbf{0.966} & \textbf{72.82}
        & 1.338 & 75.78 \\
        
        & Ours
        & 0.974 & 72.60
        & \textbf{0.862} & \textbf{84.40} \\
        
        \bottomrule
    \end{tabular}
    \newline
    
    \footnotesize 
    \justifying
    \noindent The global long-term mean error of MKE (third column) and EKE (fifth column) with respect to ERA5. 
\end{table}

\clearpage

\subsection{Temporal consistency}\label{subsec:temporal_consistency}

\begin{figure*}[!h]
    \centering
    \begin{overpic}[width=\textwidth,percent]{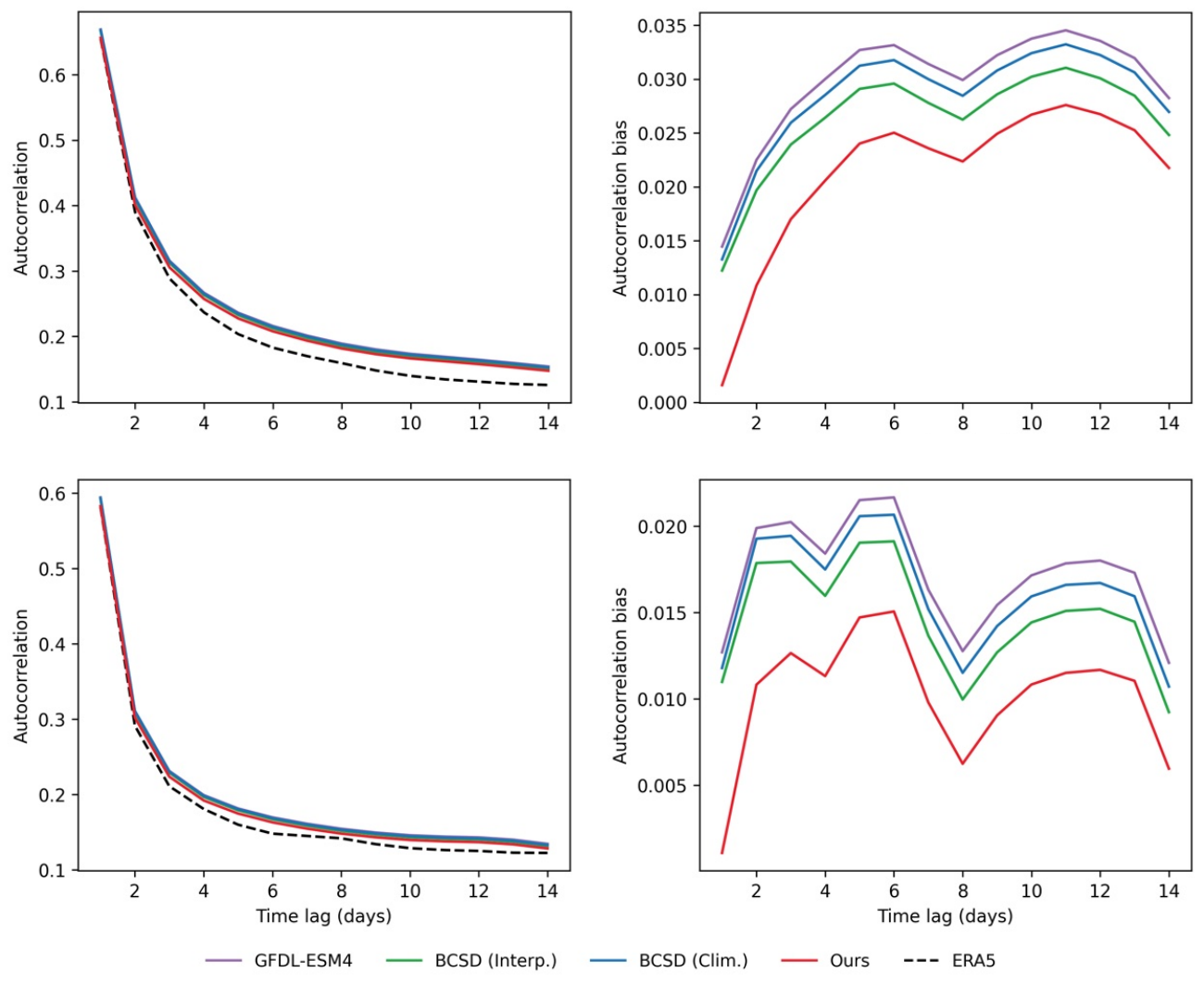}
        \put(1,81){\scriptsize{$\textbf{a}$}}
        \put(1,44){\scriptsize{$\textbf{b}$}}
    \end{overpic}
    \caption{
        \textbf{Comparison of temporal consistency across the neural downscaling framework and BCSD baselines.}
        \textbf{a}, Temporal autocorrelation and biases with respect to ERA5 of the zonal wind component $U$ for the our framework and the BCSD baselines.
        \textbf{b}, Temporal autocorrelation and biases of the meridional wind component $V$.
    }
    \label{fig:a_temporal_consistency}
\end{figure*}

We compute the temporal autocorrelation for the downscaled fields from our neural downscaling framework and the Bias Correction and Spatial Disaggregation (BCSD) baselines at each spatial grid point and then obtain a global-mean autocorrelation by spatial averaging (Fig.~\ref{fig:a_temporal_consistency}). 

Specifically, for each spatial grid point $x$, the lag-$t$ temporal autocorrelation is defined as
\begin{equation}
    \rho(x, t) =
    \frac{\sum_{i=1}^{N-t} \left( s_i(x) - \overline{s}(x) \right)
    \left( s_{i+t}(x) - \overline{s} \right)}
    {\sum_{i=1}^{N} \left( s_i(x) - \overline{s}(x) \right)^2},
\end{equation}
where $s_i(x)$ denotes the wind component speed, $\overline{s}$ is the temporal mean, and $N$ is the total number of time steps. The global-mean temporal autocorrelation is then computed by spatially averaging $\rho(x, t)$ over all spatial grid points.

\clearpage

\subsection{Uncertainty quantification}\label{subsec:uncertainty_quantification}

\begin{figure*}[!h]
    \centering
    \begin{overpic}[width=\textwidth]{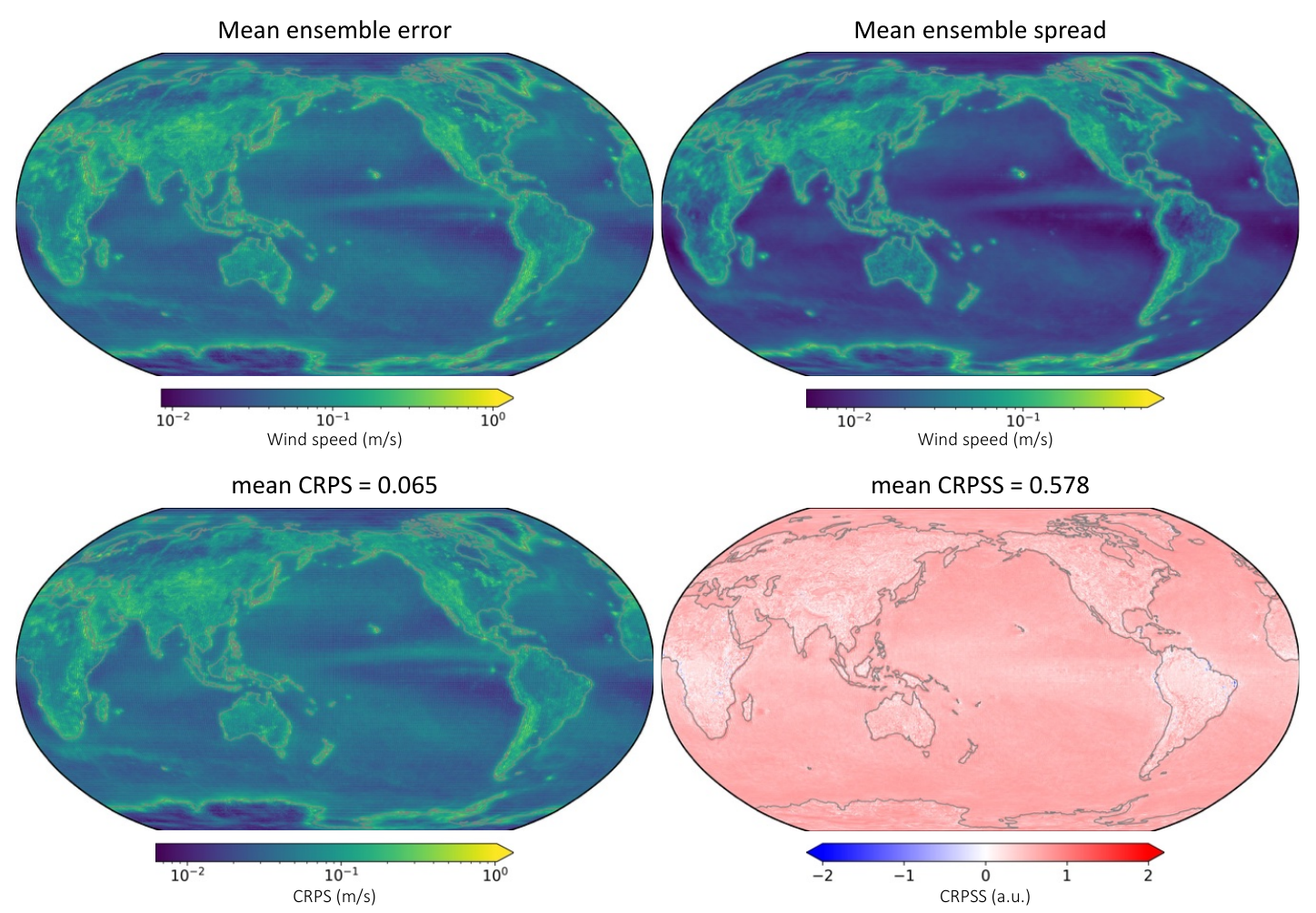}
        \put(2,66){\scriptsize{$\textbf{a}$}}
        \put(50,66){\scriptsize{$\textbf{b}$}}
        \put(2,32){\scriptsize{$\textbf{c}$}}
        \put(50,32){\scriptsize{$\textbf{d}$}}
    \end{overpic}
    \caption{
        \textbf{Uncertainty quantification of neural downscaling framework for the downscaled fields from GFDL-ESM4~\cite{Dunne2020GFDL} simulations.}
        \textbf{a,b}, The ensemble mean (\textbf{a}) and standard deviation (\textbf{b}) of the downscaled fields from our framework with stochastic forward passes for the wind speed.
        \textbf{c}, The CRPS for the ensemble of downscaled fields from our framework with stochastic forward passes for the wind speed.
        \textbf{d}, The CRPSS for the ensemble of downscaled fields from our framework with stochastic forward passes relative to BCSD (Clim.) baseline ensemble.
    }
    \label{fig:a_uncertainty}
\end{figure*}

Figure~\ref{fig:a_uncertainty}a,b illustrates the global distribution of mean ensemble error and standard deviation of the downscaled fields produced by our neural downscaling framework with Monte Carlo (MC) dropout~\cite{Gal2016Dropout}.
We quantitatively evaluate the probabilistic performance of the downscaled ensembles using the Continuous Ranked Probability Score (CRPS) and the Continuous Ranked Probability Skill Score (CRPSS) for wind speed, with the BCSD (Clim.) baseline ensemble as the reference (Fig.~\ref{fig:a_uncertainty}c,d). BCSD (Clim.) originally produces deterministic downscaled results by applying the climatological bias corresponding to the exact day of year (DOY) of each target time step. To construct a probabilistic reference for comparison, we generate 100 ensemble members for each day by randomly sampling DOY climatological biases from all 366 days of the year.

\clearpage

\subsection{Spatial downscaling of different future scenarios}\label{app:spatial_downscaling_different_scenarios}

\begin{figure*}[!h]
    \centering
    \begin{overpic}[width=\textwidth]{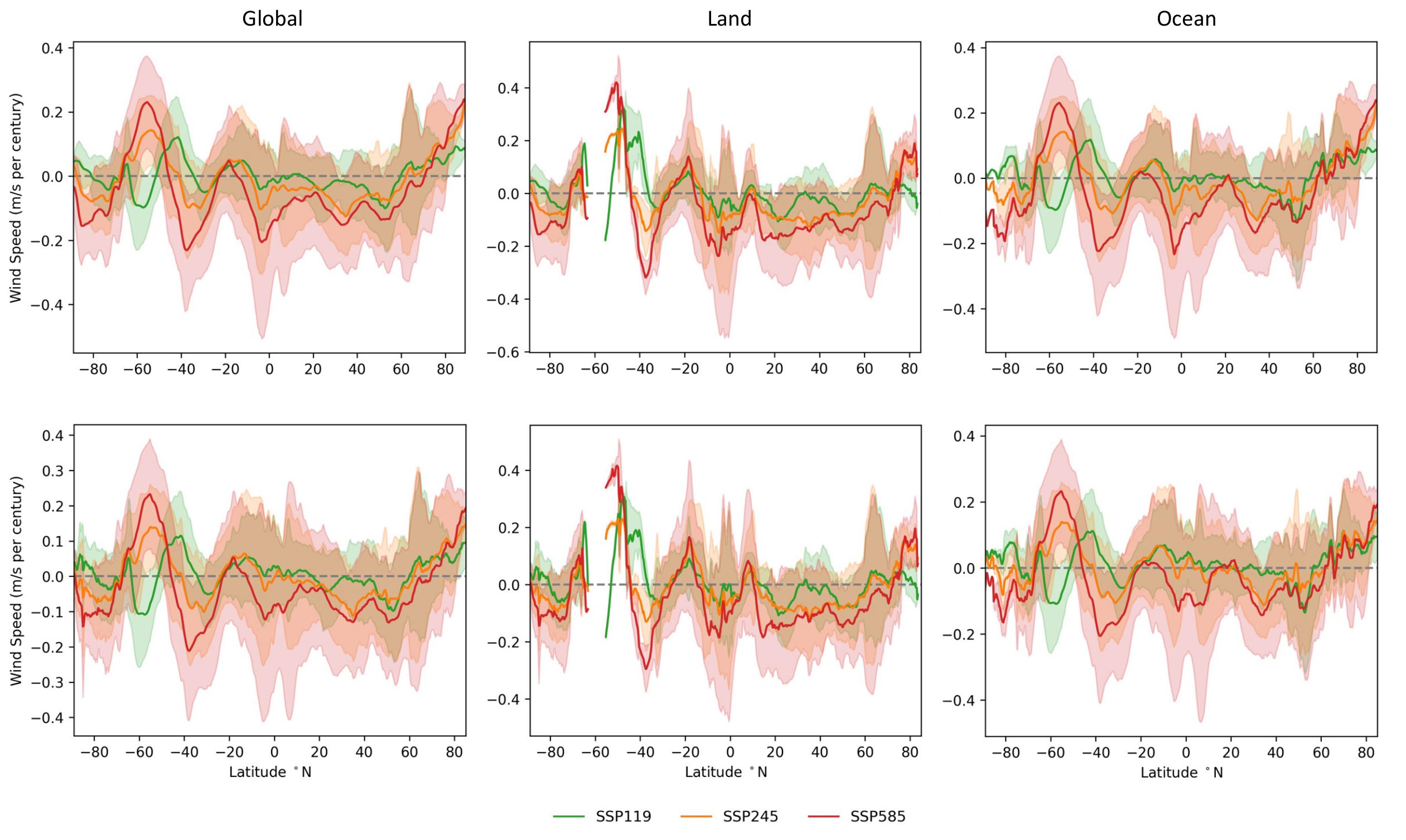}
        \put(1,59){\scriptsize{$\textbf{a}$}}
        \put(1,31){\scriptsize{$\textbf{b}$}}
    \end{overpic}
    \caption{
        \textbf{Future climate scenario projection trends under the SSP119, SSP245, and SSP585 scenarios over the global, land and ocean domains.}
        \textbf{a}, Latitudinally averaged wind speed trends of the GFDL-ESM4 simulations without bias correction over the global, land, and ocean domains.
        \textbf{b}, Latitudinally averaged wind speed trends of the downscaled fields obtained with our framework over the global, land, and ocean domains for three different future scenarios.
    }
    \label{fig:a_future_gen}
\end{figure*}

We report the projected future climate trends of the downscaled fields obtained with our neural downscaling framework and the GFDL-ESM4~\cite{Dunne2020GFDL} simulations under the SSP119, SSP245 and SSP585 scenarios over the entire global, land and ocean domains (Fig.~\ref{fig:a_future_gen}).

Based on GFDL-ESM4~\cite{Dunne2020GFDL} simulations, a coherent weakening of near-surface wind speed is projected over the equatorial region, affecting both land and ocean domains. A similar decreasing tendency is evident over large parts of the Northern Hemisphere land areas. In contrast, pronounced wind speed intensification is projected at high northern latitudes, particularly over the Arctic Ocean. These large-scale responses exhibit a clear dependence on emission intensity, becoming progressively more pronounced from SSP119 to SSP585.

Regions exhibiting substantial inter-scenario differences are primarily located in the Southern Hemisphere, most notably within the mid-latitude westerly belt. In this region, the low-emission SSP119 and high-emission SSP585 scenarios show opposite responses: under SSP119, wind speed increases around 40$^{\text{°}}$S and weakens near 60$^{\text{°}}$S, whereas SSP585 displays an inverse pattern, with reduced wind speeds near 40$^{\text{°}}$S and enhanced winds toward higher southern latitudes. This contrast highlights a strong scenario dependence of Southern Hemisphere mid-latitude circulation responses to radiative forcing.

We provide a quantitative comparison of SSP245 future projection trends between our framework and BCSD baselines for GFDL-ESM4~\cite{Dunne2020GFDL} simulations (Table~\ref{tab:a_future_1} and Table~\ref{tab:a_future_2}). The downscaling performance of our framework on near-future climate projections remains consistent with that of the historical evaluation, demonstrating its capacity to robustly correct projected climate signals rather than merely reproducing historical conditions.

\begin{table}[!h]
    \setlength{\tabcolsep}{2pt}
    \centering
    \caption{
        \textbf{Statistical comparison in SSP245 future climate scenario projections across the neural downscaling framework and BCSD baselines from GFDL-ESM4~\cite{Dunne2020GFDL} simulations.}
    }
    \footnotesize
    \label{tab:a_future_1}
    \begin{tabular}{llccccccccccccc}
    \toprule
    \multirow{2}{*}{ESMs}
    & \multirow{2}{*}{Methods}
    & \multicolumn{1}{c}{\makecell[c]{\shortstack{MAE \\ (pool)}}}
    & \multicolumn{2}{c}{\makecell[c]{PCC}}
    & \multicolumn{4}{c}{\makecell[c]{ME}} 
    & \multicolumn{4}{c}{\makecell[c]{\textit{95}$^{\text{th}}$ PE}} 
    & \multicolumn{2}{c}{\makecell[c]{\textit{5}$^{\text{th}}$ PE}}  \\
    \cmidrule(lr){3-3}\cmidrule(lr){4-5}\cmidrule(lr){6-9}\cmidrule(lr){10-13}\cmidrule(lr){14-15}
    & & $UV$ & $UV$ & Speed & $UV$ & \% & Speed & \%
    & $UV$ & \% & Speed & \%
    & $UV$ & \% \\
    \midrule

    \multirow{3}{*}{\makecell{GFDL- \\ ESM4}}
    & BCSD (Interp.)
        & 0.114
        & 0.993 & 0.987
        & 0.254 & 54.59
        & 0.300 & 55.57
        & 0.411 & 57.37
        & 0.510 & 57.68
        & 0.427 & 54.87  \\
    & BCSD (Clim.)
        & 0.099
        & \textbf{0.996} & 0.994
        & \textbf{0.198} & \textbf{64.54} 
        & 0.246 & 63.47 
        & 0.374 & 61.26 
        & 0.441 & 63.35 
        & 0.391 & 58.72 \\
    & Ours   
        & \textbf{0.048}
        & 0.996 & \textbf{0.997}
        & 0.202 & 63.96 
        & \textbf{0.180} & \textbf{73.28} 
        & \textbf{0.315} & \textbf{67.39} 
        & \textbf{0.332} & \textbf{72.46} 
        & \textbf{0.318} & \textbf{66.39} \\
    \bottomrule
    \end{tabular}
    \newline

    \footnotesize 
    \justifying
    \noindent The pooled global mean absolute error, MAE (pool), for wind components (third column) with respect to LR bias-corrected GFDL-ESM4 simulations.
    The Pearson correlation coefficient, PCC, for wind components (fourth column) and wind speed (fifth column) with respect to ERA5.
    The global long-term mean error, ME, for wind components (sixth column) and wind speed (eighth column) with respect to ERA5. 
    The global absolute error in \textit{95}$^{\text{th}}$ wind component percentile (\textit{95}$^{\text{th}}$ PE, tenth column), \textit{95}$^{\text{th}}$ wind speed percentile (twelfth column), and global absolute error in \textit{5}$^{\text{th}}$ wind component percentile (\textit{5}$^{\text{th}}$ PE, fourteenth column) with respect to ERA5.

\end{table}

\begin{table}[!h]
    \setlength{\tabcolsep}{8pt}
    \centering
    \caption{
        \textbf{Statistical comparison in specific kinetic energy decomposition (MKE and EKE) for SSP245 future climate scenario projections across the neural downscaling framework and BCSD baselines from GFDL-ESM4~\cite{Dunne2020GFDL} simulations.}
    }
    \small
    \label{tab:a_future_2}
    \begin{tabular}{llcccc}
        \toprule
        ESMs & Methods & MAE (MKE) & \% & MAE (EKE) & \% \\
        \midrule
        \multirow{3}{*}{GFDL-ESM4}
        & BCSD (Interp.)
        & 1.077 & 56.23
        & 1.239 & 58.57 \\
        
        & BCSD (Clim.)     
        & 0.848 & 65.56
        & 1.191 & 60.17 \\
        
        & Ours
        & \textbf{0.846} & \textbf{65.63}
        & \textbf{0.785} & \textbf{73.74} \\
        \bottomrule
    \end{tabular}
    \newline
    
    \footnotesize 
    \justifying
    \noindent The global long-term mean absolute error of MKE (third column) and EKE (fifth column) with respect to ERA5. 
\end{table}

\clearpage

\section{Additional ablation studies}\label{app:additional_ablation_studies}

\begin{figure*}[!h]
    \centering
    \begin{overpic}[width=\textwidth]{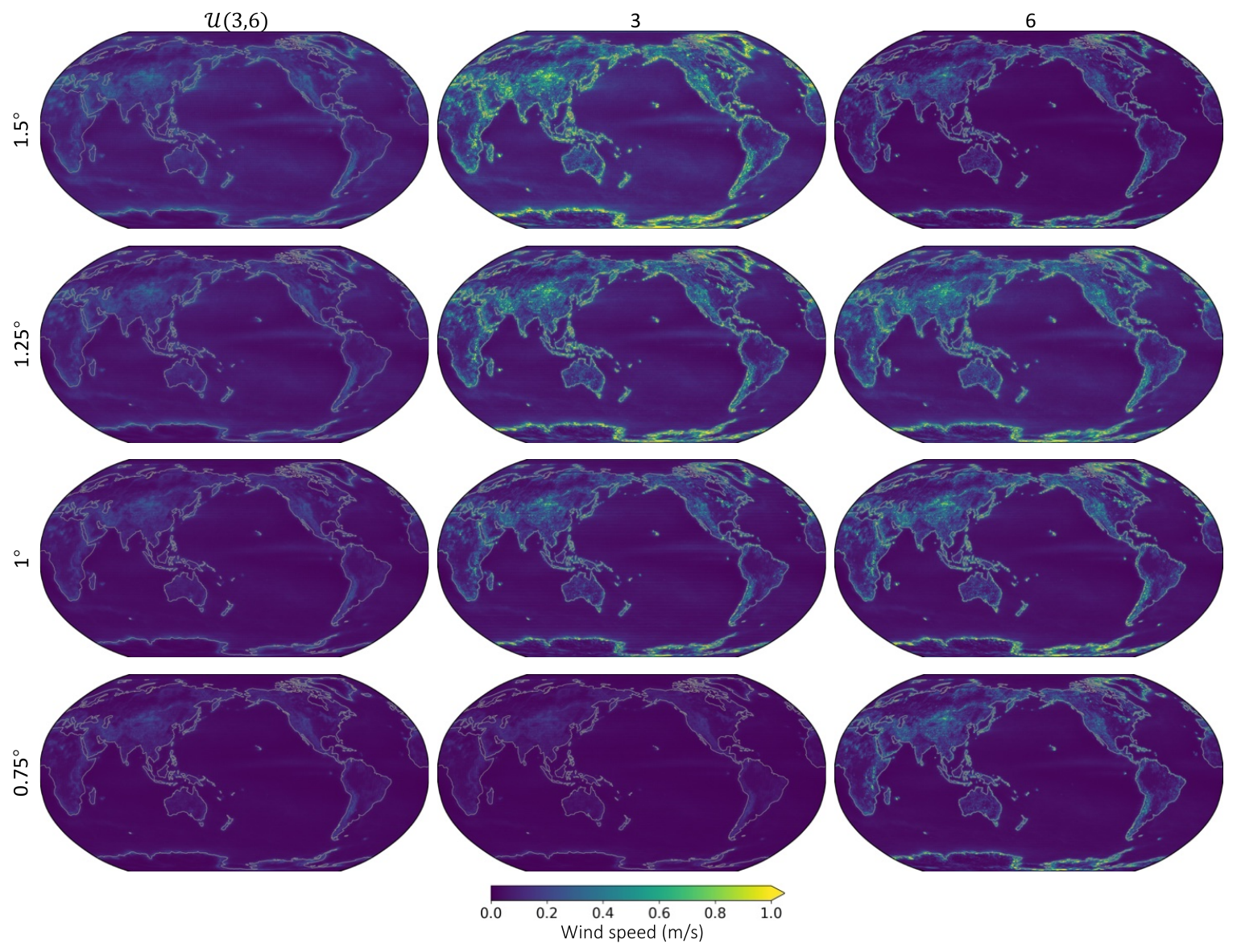}
    \end{overpic}
    \caption{
        \textbf{Comparison of global MAE distribution for state mapping with different training scale factor strategies.}
        From left to right, MAE of our framework with continuous sampling range $s\sim\mathcal{U}(3,6)$, and with discrete sampling range (3 and 6), respectively.
        From top to bottom, MAE of the downscaled fields from coarsened ERA5 reanalysis data with spatial resolution of 1.5$^\circ\times$1.5$^\circ$, 1.25$^\circ\times$1.25$^\circ$, 1$^\circ\times$1$^\circ$, and 0.75$^\circ\times$0.75$^\circ$ to HR ERA5 reanalysis with spatial resolution of 0.25$^\circ$.
    }
    \label{fig:a_ablation}
\end{figure*}

The ablation study on training scale factor strategies (Table~\ref{tab:a_ablation_2}) reveals that our downscaling neural operator trained with a continuous sampling range $s\sim\mathcal{U}(3,6)$ generally outperforms its counterpart trained with discrete scale factors ($s\in\{3,6\}$). This shows that the continuous multi-scale learning strategy is essential for enhancing the generalization capability of our framework. We additionally display the global MAE distribution for the different strategies in Fig.~\ref{fig:a_ablation}.

\begin{table}[!h]
    \setlength{\tabcolsep}{4.8pt}
    \centering
    \caption{
        \textbf{Ablation study on the downscaling neural operator with different training scale factor strategies.}
    }
    \label{tab:a_ablation_2}
    \begin{tabular}{lc|cc|cc|cc|cc}
        \toprule
        \multirow{2}{*}{Model} & \multirow{2}{*}{$s$} 
        & \multicolumn{2}{c}{$\times$3} & \multicolumn{2}{c}{$\times$4} 
        & \multicolumn{2}{c}{$\times$5} & \multicolumn{2}{c}{$\times$6} \\
        \cmidrule(lr){3-4} \cmidrule(lr){5-6} \cmidrule(lr){7-8} \cmidrule(lr){9-10}
         &  & MAE & RMSE & MAE & RMSE & MAE & RMSE & MAE & RMSE \\
        \midrule

        SD (Interp.) & -
        & 0.1602 & 0.3694 & 0.2359 & 0.4775 & 0.2937 & 0.5618 & 0.3586 & 0.6492 \\
        
        SD (Clim.) & -
        & 0.1256 & 0.2927 & 0.1883 & 0.3794 & 0.2372 & 0.4519 & 0.2929 & 0.5277 \\
        
        Ours & 3 
        & \textbf{0.0376} & \textbf{0.0663} & 0.1227 & 0.2442 & 0.1847 & 0.3728 & 0.2480 & 0.5665 \\
        
        Ours & 6 
        & 0.1110 & 0.3077 & 0.1477 & 0.3804 & 0.1819 & 0.3758 & \textbf{0.1131} & \textbf{0.1794} \\
        
        Ours & $\mathcal{U}(3,6)$ 
        & \textit{0.0484} & \textit{0.0840} & \textbf{0.0664} & \textbf{0.1115} & \textbf{0.0936} & \textbf{0.1505} & \textit{0.1167} & \textit{0.1859} \\
        
        \bottomrule
    \end{tabular}
    \newline

    \footnotesize 
    \justifying
    \noindent $s$ is the distribution of downscaling scale factors during training. The best performance highlighted in \textbf{bold}, the second best performance highlighted in \textit{italic}.
\end{table}

We evaluate the state-mapping capability of our downscaling neural operator by comparing it against alternative model architectures and varying gradient loss weights, with results summarized in Table~\ref{tab:a_ablation_1}. Our method consistently outperforms all baselines across every downscaling factor, and its performance is further enhanced by incorporating the gradient loss term.

\begin{table}[!h]
    \setlength{\tabcolsep}{5pt}
    \centering
    \caption{
        \textbf{State mapping capability of our downscaling neural operator with different model architectures and gradient loss weights.}
    }
    \label{tab:a_ablation_1}

    \begin{tabular}{lc|cc|cc|cc|cc}
        \toprule
        \multirow{2}{*}{Model} & \multirow{2}{*}{$\lambda_2$} 
        & \multicolumn{2}{c}{$\times$3} & \multicolumn{2}{c}{$\times$4} 
        & \multicolumn{2}{c}{$\times$5} & \multicolumn{2}{c}{$\times$6} \\
        \cmidrule(lr){3-4} \cmidrule(lr){5-6} \cmidrule(lr){7-8} \cmidrule(lr){9-10}
         &  & MAE & RMSE & MAE & RMSE & MAE & RMSE & MAE & RMSE \\
        \midrule

        SD (Interp.) & -
        & 0.1602 & 0.3694 & 0.2359 & 0.4775 & 0.2937 & 0.5618 & 0.3586 & 0.6492 \\
        
        SD (Clim.)~\cite{Wood2004hydrologic} & -
        & 0.1256 & 0.2927 & 0.1883 & 0.3794 & 0.2372 & 0.4519 & 0.2929 & 0.5277 \\
        
        CNN~\cite{Lim2017Enhanced} & 0 & 0.0639 & 0.1289 & 0.0941 & 0.1795 & 0.1301 & 0.2395 & 0.1695 & 0.3050 \\
        
        SRNO~\cite{Wei2023SuperResolution} & 0 & 0.0583 & 0.1155 & 0.0837 & 0.1530 & 0.1150 & 0.2019 & 0.1521 & 0.2631 \\

        DFNO~\cite{yang2024fourier} & 0 & 0.0545 & 0.0946 & 0.0699 & 0.1173 & 0.0968 & 0.1556 & 0.1192 & 0.1899 \\
        
        Ours & 0 & 0.0484 & 0.0840 & 0.0664 & 0.1115 & 0.0936 & 0.1505 & 0.1167 & 0.1859 \\
        
        Ours & 0.1 & \textit{0.0474} & \textit{0.0821} & \textbf{0.0660} & \textbf{0.1103} & \textbf{0.0933} & \textbf{0.1498} & \textbf{0.1164} & \textbf{0.1852} \\

        Ours & 0.2 & \textbf{0.0474} & \textbf{0.0816} & \textit{0.0661} & \textit{0.1103} & \textit{0.0935} & \textit{0.1500} & \textit{0.1168} & \textit{0.1858} \\

        \bottomrule
    \end{tabular}
    \newline
        
    \footnotesize 
    \justifying
    \noindent $\lambda_2$ is the gradient loss weight for the loss term in loss function. The best performance highlighted in \textbf{bold}, the second best performance highlighted in \textit{italic}.

\end{table}

\clearpage

\clearpage
\renewcommand\refname{References}
\bibliographystyle{unsrt}
\footnotesize
\bibliography{references}

@article{zhang2020new,
  title   = {A New Prediction Method Based on {{VMD-PRBF-ARMA-E}} Model Considering Wind Speed Characteristic},
  author  = {Zhang, Yagang and Zhao, Yuan and Kong, Chunhui and Chen, Bing},
  year    = 2020,
  journal = {Energy Conversion and Management},
  volume  = {203},
  pages   = {112254},
  issn    = {0196-8904},
  doi     = {10.1016/j.enconman.2019.112254}
}

@article{leung2012Wind,
  title   = {Wind Energy Development and Its Environmental Impact: {{A}} Review},
  author  = {Leung, Dennis Y. C. and Yang, Yuan},
  year    = 2012,
  journal = {Renewable and Sustainable Energy Reviews},
  volume  = {16},
  number  = {1},
  pages   = {1031--1039},
  issn    = {1364-0321},
  doi     = {10.1016/j.rser.2011.09.024}
}

@article{kim2015Recent,
  title   = {Recent Recovery of Surface Wind Speed after Decadal Decrease: A Focus on {{South Korea}}},
  author  = {Kim, JongChun and Paik, Kyungrock},
  year    = 2015,
  journal = {Climate Dynamics},
  volume  = {45},
  number  = {5},
  pages   = {1699--1712},
  issn    = {1432-0894},
  doi     = {10.1007/s00382-015-2546-9}
}

@article{mcvicar2012Global,
  title   = {Global Review and Synthesis of Trends in Observed Terrestrial Near-Surface Wind Speeds: {{Implications}} for Evaporation},
  author  = {McVicar, Tim R. and Roderick, Michael L. and Donohue, Randall J. and Li, Ling Tao and Van Niel, Thomas G. and Thomas, Axel and Grieser, J{\"u}rgen and Jhajharia, Deepak and Himri, Youcef and Mahowald, Natalie M. and Mescherskaya, Anna V. and Kruger, Andries C. and Rehman, Shafiqur and Dinpashoh, Yagob},
  year    = 2012,
  journal = {Journal of Hydrology},
  volume  = {416--417},
  pages   = {182--205},
  issn    = {0022-1694},
  doi     = {10.1016/j.jhydrol.2011.10.024}
}

@article{zhang2020Decreasing,
  title   = {Decreasing Atmospheric Visibility Associated with Weakening Winds from 1980 to 2017 over {{China}}},
  author  = {Zhang, Yong and Gao, Lina and Cao, Lijuan and Yan, Zhongwei and Wu, Yongxue},
  year    = 2020,
  journal = {Atmospheric Environment},
  volume  = {224},
  pages   = {117314},
  issn    = {1352-2310},
  doi     = {10.1016/j.atmosenv.2020.117314}
}

@article{pryor2012Analyses,
  title   = {Analyses of Possible Changes in Intense and Extreme Wind Speeds over Northern {{Europe}} under Climate Change Scenarios},
  author  = {Pryor, S. C. and Barthelmie, R. J. and Clausen, N. E. and Drews, M. and MacKellar, N. and Kjellstr{\"o}m, E.},
  year    = 2012,
  journal = {Climate Dynamics},
  volume  = {38},
  number  = {1},
  pages   = {189--208},
  issn    = {1432-0894},
  doi     = {10.1007/s00382-010-0955-3}
}

@article{zha2022Effects,
  title   = {Effects of {{Northern Hemisphere Annular Mode}} on Terrestrial Near-Surface Wind Speed over Eastern {{China}} from 1979 to 2017},
  author  = {Zha, Jin-Lin and Shen, Cheng and Wu, Jian and Zhao, De-Ming and Fan, Wen-Xuan and Jiang, Hui-Ping and {Azorin-molina}, Cesar and Chen, Deliang},
  year    = 2022,
  journal = {Advances in Climate Change Research},
  series  = {Special Topic on Explaining {{Chinas}} Climate in 2021},
  volume  = {13},
  number  = {6},
  pages   = {875--883},
  issn    = {1674-9278},
  doi     = {10.1016/j.accre.2022.10.005}
}

@article{lyu2026Reconstruction,
  title     = {Reconstruction and {{Projection}} of {{Near-Surface Wind Speed}} in {{China Based}} on {{Transformer Neural Network}}},
  author    = {Lyu, Yanjun and Zhang, Hao and Lu, Wenxi and Zhou, Bofan and Duo, Yunfeng},
  year      = 2026,
  journal   = {Advances in Meteorology},
  volume    = {2026},
  number    = {1},
  pages     = {9986477},
  issn      = {1687-9317},
  doi       = {10.1155/adme/9986477},
  copyright = {Copyright \copyright{} 2026 Yanjun Lyu et al. Advances in Meteorology published by John Wiley \& Sons Ltd.}
}

@article{haarsma2016High,
  title     = {High {{Resolution Model Intercomparison Project}} ({{HighResMIP}} v1.0) for {{CMIP6}}},
  author    = {Haarsma, Reindert J. and Roberts, Malcolm J. and Vidale, Pier Luigi and Senior, Catherine A. and Bellucci, Alessio and Bao, Qing and Chang, Ping and Corti, Susanna and Fu{\v c}kar, Neven S. and Guemas, Virginie and {von Hardenberg}, Jost and Hazeleger, Wilco and Kodama, Chihiro and Koenigk, Torben and Leung, L. Ruby and Lu, Jian and Luo, Jing-Jia and Mao, Jiafu and Mizielinski, Matthew S. and Mizuta, Ryo and Nobre, Paulo and Satoh, Masaki and Scoccimarro, Enrico and Semmler, Tido and Small, Justin and {von Storch}, Jin-Song},
  year      = 2016,
  journal   = {Geoscientific Model Development},
  volume    = {9},
  number    = {11},
  pages     = {4185--4208},
  publisher = {Copernicus GmbH},
  issn      = {1991-959X},
  doi       = {10.5194/gmd-9-4185-2016}
}

@article{Daines2016ModelBased,
  title   = {Model-Based Projections and Uncertainties of Near-Surface Wind Climate in Western Canada},
  author  = {Daines, Jeffrey T. and Monahan, Adam H. and Curry, Charles L.},
  year    = 2016,
  journal = {Journal of Applied Meteorology and Climatology},
  volume  = {55},
  number  = {10},
  pages   = {2229--2245},
  issn    = {1558-8424, 1558-8432},
  doi     = {10.1175/JAMC-D-16-0091.1}
}

@article{Maraun2016Bias,
  title   = {Bias Correcting Climate Change Simulations - a Critical Review},
  author  = {Maraun, Douglas},
  year    = 2016,
  journal = {Current Climate Change Reports},
  volume  = {2},
  number  = {4},
  pages   = {211--220},
  issn    = {2198-6061},
  doi     = {10.1007/s40641-016-0050-x}
}

@article{tang2016Statistical,
  title     = {Statistical Downscaling and Dynamical Downscaling of Regional Climate in {{China}}: {{Present}} Climate Evaluations and Future Climate Projections},
  author    = {Tang, Jianping and Niu, Xiaorui and Wang, Shuyu and Gao, Hongxia and Wang, Xueyuan and Wu, Jian},
  year      = 2016,
  journal   = {Journal of Geophysical Research: Atmospheres},
  volume    = {121},
  number    = {5},
  pages     = {2110--2129},
  issn      = {2169-8996},
  doi       = {10.1002/2015JD023977},
  copyright = {\copyright 2016. American Geophysical Union. All Rights Reserved.}
}

@article{rummukainen2010Stateofthearta,
  title     = {State-of-the-Art with Regional Climate Models},
  author    = {Rummukainen, Markku},
  year      = 2010,
  journal   = {Wiley Interdisciplinary Reviews: Climate Change},
  volume    = {1},
  number    = {1},
  pages     = {82--96},
  publisher = {John Wiley \& Sons, Ltd},
  issn      = {1757-7799},
  doi       = {10.1002/wcc.8}
}

@article{Xu2019Dynamical,
  title   = {Dynamical Downscaling of Regional Climate: A Review of Methods and Limitations},
  author  = {Xu, Zhongfeng and Han, Ying and Yang, Zongliang},
  year    = 2019,
  journal = {Science China Earth Sciences},
  volume  = {62},
  number  = {2},
  pages   = {365--375},
  issn    = {1674-7313, 1869-1897},
  doi     = {10.1007/s11430-018-9261-5}
}

@article{Tapiador2020Regional,
  title   = {Regional Climate Models: 30 Years of Dynamical Downscaling},
  author  = {Tapiador, Francisco J. and Navarro, Andr{\'e}s and Moreno, Ra{\'u}l and S{\'a}nchez, Jos{\'e} Luis and {Garc{\'i}a-Ortega}, Eduardo},
  year    = 2020,
  journal = {Atmospheric Research},
  volume  = {235},
  pages   = {104785},
  issn    = {01698095},
  doi     = {10.1016/j.atmosres.2019.104785}
}

@article{sun2024Deepa,
  title   = {Deep Learning in Statistical Downscaling for Deriving High Spatial Resolution Gridded Meteorological Data: {{A}} Systematic Review},
  author  = {Sun, Yongjian and Deng, Kefeng and Ren, Kaijun and Liu, Jia and Deng, Chongjiu and Jin, Yongjun},
  year    = 2024,
  journal = {ISPRS Journal of Photogrammetry and Remote Sensing},
  volume  = {208},
  pages   = {14--38},
  issn    = {0924-2716},
  doi     = {10.1016/j.isprsjprs.2023.12.011}
}

@article{soares2024Highresolution,
  title     = {High-Resolution Downscaling of {{CMIP6 Earth}} System and Global Climate Models Using Deep Learning for {{Iberia}}},
  author    = {Soares, Pedro M. M. and Johannsen, Frederico and Lima, Daniela C. A. and Lemos, Gil and Bento, Virg{\'i}lio A. and Bushenkova, Angelina},
  year      = 2024,
  journal   = {Geoscientific Model Development},
  volume    = {17},
  number    = {1},
  pages     = {229--259},
  publisher = {Copernicus GmbH},
  issn      = {1991-959X},
  doi       = {10.5194/gmd-17-229-2024}
}

@article{Rampal2024Enhancing,
  title     = {Enhancing Regional Climate Downscaling through Advances in Machine Learning},
  author    = {Rampal, Neelesh and Hobeichi, Sanaa and Gibson, Peter B. and {Ba{\~n}o-Medina}, Jorge and Abramowitz, Gab and Beucler, Tom and {Gonz{\'a}lez-Abad}, Jose and Chapman, William and Harder, Paula and Guti{\'e}rrez, Jos{\'e} Manuel},
  year      = 2024,
  journal   = {Artificial Intelligence for the Earth Systems},
  volume    = {3},
  number    = {2},
  pages     = {230066},
  issn      = {2769-7525},
  doi       = {10.1175/AIES-D-23-0066.1},
  copyright = {http://www.ametsoc.org/PUBSReuseLicenses}
}

@article{Schoof2013Statistical,
  title   = {Statistical Downscaling in Climatology},
  author  = {Schoof, Justin T.},
  year    = 2013,
  journal = {Geography Compass},
  volume  = {7},
  number  = {4},
  pages   = {249--265},
  issn    = {1749-8198, 1749-8198},
  doi     = {10.1111/gec3.12036}
}

@article{Reder2025Estimating,
  title   = {Estimating Pros and Cons of Statistical Downscaling Based on EQM Bias Adjustment as a Complementary Method to Dynamical Downscaling},
  author  = {Reder, Alfredo and Fedele, Giusy and Manco, Ilenia and Mercogliano, Paola},
  year    = 2025,
  journal = {Scientific Reports},
  volume  = {15},
  number  = {1},
  pages   = {621},
  issn    = {2045-2322},
  doi     = {10.1038/s41598-024-84527-5}
}

@inproceedings{Vandal2017DeepSD,
  author    = {Vandal, Thomas and Kodra, Evan and Ganguly, Sangram and Michaelis, Andrew and Nemani, Ramakrishna and Ganguly, Auroop R.},
  title     = {DeepSD: Generating High Resolution Climate Change Projections through Single Image Super-Resolution},
  year      = {2017},
  isbn      = {9781450348874},
  publisher = {Association for Computing Machinery},
  address   = {New York, NY, USA},
  url       = {https://doi.org/10.1145/3097983.3098004},
  doi       = {10.1145/3097983.3098004},
  pages     = {1663--1672},
  numpages  = {10},
  location  = {Halifax, NS, Canada},
  booktitle = {KDD '17},
  editor    = {Zaimis, E.}
}

@article{Neelesh2022high,
  title   = {High-resolution downscaling with interpretable deep learning: Rainfall extremes over New Zealand},
  journal = {Weather and Climate Extremes},
  volume  = {38},
  pages   = {100525},
  year    = {2022},
  issn    = {2212-0947},
  doi     = {https://doi.org/10.1016/j.wace.2022.100525},
  url     = {https://www.sciencedirect.com/science/article/pii/S2212094722001049},
  author  = {Neelesh Rampal and Peter B. Gibson and Abha Sood and Stephen Stuart and Nicolas C. Fauchereau and Chris Brandolino and Ben Noll and Tristan Meyers}
}

@article{Zhong2024investigating,
  author  = {Zhong, Xiaohui and Du, Fei and Chen, Lei and Wang, Zhibin and Li, Hao},
  title   = {Investigating transformer-based models for spatial downscaling and correcting biases of near-surface temperature and wind-speed forecasts},
  journal = {Quarterly Journal of the Royal Meteorological Society},
  volume  = {150},
  number  = {758},
  pages   = {275--289},
  doi     = {https://doi.org/10.1002/qj.4596},
  url     = {https://rmets.onlinelibrary.wiley.com/doi/abs/10.1002/qj.4596},
  eprint  = {https://rmets.onlinelibrary.wiley.com/doi/pdf/10.1002/qj.4596},
  year    = {2024}
}

@article{AmirHossein2025PrecipNet,
  title   = {PrecipNet: A transformer-based downscaling framework for improved precipitation prediction in San Diego County},
  journal = {Journal of Hydrology: Regional Studies},
  volume  = {62},
  pages   = {102738},
  year    = {2025},
  issn    = {2214-5818},
  doi     = {https://doi.org/10.1016/j.ejrh.2025.102738},
  url     = {https://www.sciencedirect.com/science/article/pii/S2214581825005671},
  author  = {AmirHossein Adibfar and Hassan Davani}
}

@article{Oyama2023deep,
  author  = {Oyama, Norihiro
             and Ishizaki, Noriko N.
             and Koide, Satoshi
             and Yoshida, Hiroaki},
  title   = {Deep generative model super-resolves spatially correlated multiregional climate data},
  journal = {Scientific Reports},
  year    = {2023},
  month   = {Apr},
  day     = {25},
  volume  = {13},
  number  = {1},
  pages   = {5992},
  issn    = {2045-2322},
  doi     = {10.1038/s41598-023-32947-0},
  url     = {https://doi.org/10.1038/s41598-023-32947-0}
}

@article{Glawion2025global,
  author  = {Glawion, Luca
             and Polz, Julius
             and Kunstmann, Harald
             and Fersch, Benjamin
             and Chwala, Christian},
  title   = {Global spatio-temporal ERA5 precipitation downscaling to km and sub-hourly scale using generative AI},
  journal = {npj Climate and Atmospheric Science},
  year    = {2025},
  month   = {Jun},
  day     = {15},
  volume  = {8},
  number  = {1},
  pages   = {219},
  issn    = {2397-3722},
  doi     = {10.1038/s41612-025-01103-y},
  url     = {https://doi.org/10.1038/s41612-025-01103-y}
}

@article{Hess2025fast,
  author  = {Hess, Philipp
             and Aich, Michael
             and Pan, Baoxiang
             and Boers, Niklas},
  title   = {Fast, scale-adaptive and uncertainty-aware downscaling of Earth system model fields with generative machine learning},
  journal = {Nature Machine Intelligence},
  year    = {2025},
  month   = {Mar},
  day     = {01},
  volume  = {7},
  number  = {3},
  pages   = {363--373},
  issn    = {2522-5839},
  doi     = {10.1038/s42256-025-00980-5},
  url     = {https://doi.org/10.1038/s42256-025-00980-5}
}

@article{Schmidt2025agenerative,
  author  = {Schmidt, Jonathan
             and Schmidt, Luca
             and Strnad, Felix M.
             and Ludwig, Nicole
             and Hennig, Philipp},
  title   = {A Generative Framework for Probabilistic, Spatiotemporally Coherent Downscaling of Climate Simulation},
  journal = {npj Climate and Atmospheric Science},
  year    = {2025},
  month   = {Jul},
  day     = {18},
  volume  = {8},
  number  = {1},
  pages   = {270},
  issn    = {2397-3722},
  doi     = {10.1038/s41612-025-01157-y},
  url     = {https://doi.org/10.1038/s41612-025-01157-y}
}

@article{Ignacio2025dynamical,
  author  = {Ignacio Lopez-Gomez  and Zhong Yi Wan  and Leonardo Zepeda-Núñez  and Tapio Schneider  and John Anderson  and Fei Sha },
  title   = {Dynamical-generative downscaling of climate model ensembles},
  journal = {Proceedings of the National Academy of Sciences},
  volume  = {122},
  number  = {17},
  pages   = {e2420288122},
  year    = {2025},
  doi     = {10.1073/pnas.2420288122},
  url     = {https://www.pnas.org/doi/abs/10.1073/pnas.2420288122},
  eprint  = {https://www.pnas.org/doi/pdf/10.1073/pnas.2420288122}
}

@article{nguyen2024Selecting,
  title     = {Selecting {{CMIP6}} Global Climate Models ({{GCMs}}) for {{Coordinated Regional Climate Downscaling Experiment}} ({{CORDEX}}) Dynamical Downscaling over {{Southeast Asia}} Using a Standardised Benchmarking Framework},
  author    = {Nguyen, Phuong Loan and Alexander, Lisa V. and Thatcher, Marcus J. and Truong, Son C. H. and Isphording, Rachael N. and McGregor, John L.},
  year      = 2024,
  journal   = {Geoscientific Model Development},
  volume    = {17},
  number    = {19},
  pages     = {7285--7315},
  publisher = {Copernicus GmbH},
  issn      = {1991-959X},
  doi       = {10.5194/gmd-17-7285-2024}
}

@article{zanchettin2022Effects,
  title     = {Effects of Forcing Differences and Initial Conditions on Inter-Model Agreement in the {{VolMIP}} Volc-Pinatubo-Full Experiment},
  author    = {Zanchettin, Davide and Timmreck, Claudia and Khodri, Myriam and Schmidt, Anja and Toohey, Matthew and Abe, Manabu and Bekki, Slimane and Cole, Jason and Fang, Shih-Wei and Feng, Wuhu and Hegerl, Gabriele and Johnson, Ben and Lebas, Nicolas and LeGrande, Allegra N. and Mann, Graham W. and Marshall, Lauren and Rieger, Landon and Robock, Alan and Rubinetti, Sara and Tsigaridis, Kostas and Weierbach, Helen},
  year      = 2022,
  journal   = {Geoscientific Model Development},
  volume    = {15},
  number    = {5},
  pages     = {2265--2292},
  publisher = {Copernicus GmbH},
  issn      = {1991-959X},
  doi       = {10.5194/gmd-15-2265-2022}
}

@article{Sinha2025on,
  title   = {On the effectiveness of neural operators at zero-shot weather downscaling},
  volume  = {4},
  doi     = {10.1017/eds.2025.11},
  journal = {Environmental Data Science},
  author  = {Sinha, Saumya and Benton, Brandon and Emami, Patrick},
  year    = {2025},
  pages   = {e21}
}

@article{Aich2025conditional,
  author  = {Aich, M. and Hess, P. and Pan, B. and Bathiany, S. and Huang, Y. and Boers, N.},
  title   = {Conditional diffusion models for downscaling \& bias correction of Earth system model precipitation},
  journal = {EGUsphere},
  volume  = {2025},
  year    = {2025},
  pages   = {1--21},
  url     = {https://egusphere.copernicus.org/preprints/2025/egusphere-2025-2646/},
  doi     = {10.5194/egusphere-2025-2646}
}

@article{Wood2004hydrologic,
  author  = {Wood, A. W.
             and Leung, L. R.
             and Sridhar, V.
             and Lettenmaier, D. P.},
  title   = {Hydrologic Implications of Dynamical and Statistical Approaches to Downscaling Climate Model Outputs},
  journal = {Climatic Change},
  year    = {2004},
  month   = {Jan},
  day     = {01},
  volume  = {62},
  number  = {1},
  pages   = {189--216},
  issn    = {1573-1480},
  doi     = {10.1023/B:CLIM.0000013685.99609.9e},
  url     = {https://doi.org/10.1023/B:CLIM.0000013685.99609.9e}
}

@article{Hersbach2020era5,
  author  = {Hersbach, Hans and Bell, Bill and Berrisford, Paul and Hirahara, Shoji and Horányi, András and Muñoz-Sabater, Joaquín and Nicolas, Julien and Peubey, Carole and Radu, Raluca and Schepers, Dinand and Simmons, Adrian and Soci, Cornel and Abdalla, Saleh and Abellan, Xavier and Balsamo, Gianpaolo and Bechtold, Peter and Biavati, Gionata and Bidlot, Jean and Bonavita, Massimo and De Chiara, Giovanna and Dahlgren, Per and Dee, Dick and Diamantakis, Michail and Dragani, Rossana and Flemming, Johannes and Forbes, Richard and Fuentes, Manuel and Geer, Alan and Haimberger, Leo and Healy, Sean and Hogan, Robin J. and Hólm, Elías and Janisková, Marta and Keeley, Sarah and Laloyaux, Patrick and Lopez, Philippe and Lupu, Cristina and Radnoti, Gabor and de Rosnay, Patricia and Rozum, Iryna and Vamborg, Freja and Villaume, Sebastien and Thépaut, Jean-Noël},
  title   = {The ERA5 global reanalysis},
  journal = {Quarterly Journal of the Royal Meteorological Society},
  volume  = {146},
  number  = {730},
  pages   = {1999--2049},
  doi     = {https://doi.org/10.1002/qj.3803},
  year    = {2020}
}

@article{Eyring2016overview,
  author  = {Eyring, V. and Bony, S. and Meehl, G. A. and Senior, C. A. and Stevens, B. and Stouffer, R. J. and Taylor, K. E.},
  title   = {Overview of the Coupled Model Intercomparison Project Phase 6 (CMIP6)
             experimental design and organization},
  journal = {Geoscientific Model Development},
  volume  = {9},
  year    = {2016},
  number  = {5},
  pages   = {1937--1958},
  doi     = {10.5194/gmd-9-1937-2016}
}

@article{Mauritsen2019developments,
  author  = {Mauritsen, Thorsten and Bader, Jürgen and Becker, Tobias and Behrens, Jörg and Bittner, Matthias and Brokopf, Renate and Brovkin, Victor and Claussen, Martin and Crueger, Traute and Esch, Monika and Fast, Irina and Fiedler, Stephanie and Fläschner, Dagmar and Gayler, Veronika and Giorgetta, Marco and Goll, Daniel S. and Haak, Helmuth and Hagemann, Stefan and Hedemann, Christopher and Hohenegger, Cathy and Ilyina, Tatiana and Jahns, Thomas and Jimenéz-de-la-Cuesta, Diego and Jungclaus, Johann and Kleinen, Thomas and Kloster, Silvia and Kracher, Daniela and Kinne, Stefan and Kleberg, Deike and Lasslop, Gitta and Kornblueh, Luis and Marotzke, Jochem and Matei, Daniela and Meraner, Katharina and Mikolajewicz, Uwe and Modali, Kameswarrao and Möbis, Benjamin and Müller, Wolfgang A. and Nabel, Julia E. M. S. and Nam, Christine C. W. and Notz, Dirk and Nyawira, Sarah-Sylvia and Paulsen, Hanna and Peters, Karsten and Pincus, Robert and Pohlmann, Holger and Pongratz, Julia and Popp, Max and Raddatz, Thomas Jürgen and Rast, Sebastian and Redler, Rene and Reick, Christian H. and Rohrschneider, Tim and Schemann, Vera and Schmidt, Hauke and Schnur, Reiner and Schulzweida, Uwe and Six, Katharina D. and Stein, Lukas and Stemmler, Irene and Stevens, Bjorn and von Storch, Jin-Song and Tian, Fangxing and Voigt, Aiko and Vrese, Philipp and Wieners, Karl-Hermann and Wilkenskjeld, Stiig and Winkler, Alexander and Roeckner, Erich},
  title   = {Developments in the MPI-M Earth System Model version 1.2 (MPI-ESM1.2) and Its Response to Increasing CO2},
  journal = {Journal of Advances in Modeling Earth Systems},
  volume  = {11},
  number  = {4},
  pages   = {998--1038},
  doi     = {https://doi.org/10.1029/2018MS001400},
  year    = {2019}
}

@article{Dunne2020GFDL,
  author  = {Dunne, J. P. and Horowitz, L. W. and Adcroft, A. J. and Ginoux, P. and Held, I. M. and John, J. G. and Krasting, J. P. and Malyshev, S. and Naik, V. and Paulot, F. and Shevliakova, E. and Stock, C. A. and Zadeh, N. and Balaji, V. and Blanton, C. and Dunne, K. A. and Dupuis, C. and Durachta, J. and Dussin, R. and Gauthier, P. P. G. and Griffies, S. M. and Guo, H. and Hallberg, R. W. and Harrison, M. and He, J. and Hurlin, W. and McHugh, C. and Menzel, R. and Milly, P. C. D. and Nikonov, S. and Paynter, D. J. and Ploshay, J. and Radhakrishnan, A. and Rand, K. and Reichl, B. G. and Robinson, T. and Schwarzkopf, D. M. and Sentman, L. T. and Underwood, S. and Vahlenkamp, H. and Winton, M. and Wittenberg, A. T. and Wyman, B. and Zeng, Y. and Zhao, M.},
  title   = {The GFDL Earth System Model Version 4.1 (GFDL-ESM 4.1): Overall Coupled Model Description and Simulation Characteristics},
  journal = {Journal of Advances in Modeling Earth Systems},
  volume  = {12},
  number  = {11},
  pages   = {e2019MS002015},
  doi     = {https://doi.org/10.1029/2019MS002015},
  year    = {2020}
}

@article{Seiji2019meteorological,
  title   = {The Meteorological Research Institute Earth System Model Version 2.0, MRI-ESM2.0: Description and Basic Evaluation of the Physical Component},
  author  = {Seiji YUKIMOTO and Hideaki KAWAI and Tsuyoshi KOSHIRO and Naga OSHIMA and Kohei YOSHIDA and Shogo URAKAWA and Hiroyuki TSUJINO and Makoto DEUSHI and Taichu TANAKA and Masahiro HOSAKA and Shokichi YABU and Hiromasa YOSHIMURA and Eiki SHINDO and Ryo MIZUTA and Atsushi OBATA and Yukimasa ADACHI and Masayoshi ISHII},
  journal = {Journal of the Meteorological Society of Japan. Ser. II},
  volume  = {97},
  number  = {5},
  pages   = {931--965},
  year    = {2019},
  doi     = {10.2151/jmsj.2019-051}
}

@article{Douglas2013bias,
  author    = {Douglas Maraun},
  title     = {Bias Correction, Quantile Mapping, and Downscaling: Revisiting the Inflation Issue},
  journal   = {Journal of Climate},
  year      = {2013},
  publisher = {American Meteorological Society},
  address   = {Boston MA, USA},
  volume    = {26},
  number    = {6},
  doi       = {10.1175/JCLI-D-12-00821.1},
  pages     = {2137--2143}
}

@article{Azizzadenesheli2024neural,
  author  = {Azizzadenesheli, Kamyar and Kovachki, Nikola and Li, Zongyi and Liu-Schiaffini, Miguel and Kossaifi, Jean and Anandkumar, Anima},
  title   = {Neural operators for accelerating scientific simulations and design},
  journal = {Nature Reviews Physics},
  year    = {2024},
  volume  = {6},
  number  = {5},
  pages   = {320--328},
  doi     = {10.1038/s42254-024-00712-5}
}

@article{kovachki2023neural,
  title   = {Neural operator: Learning maps between function spaces with applications to pdes},
  author  = {Kovachki, Nikola and Li, Zongyi and Liu, Burigede and Azizzadenesheli, Kamyar and Bhattacharya, Kaushik and Stuart, Andrew and Anandkumar, Anima},
  journal = {Journal of Machine Learning Research},
  volume  = {24},
  number  = {89},
  pages   = {1--97},
  year    = {2023}
}

@inproceedings{Li2021Fourier,
  title     = {Fourier Neural Operator for Parametric Partial Differential Equations},
  booktitle = {International Conference on Learning Representations},
  author    = {Li, Zongyi and Kovachki, Nikola Borislavov and Azizzadenesheli, Kamyar and Liu, Burigede and Bhattacharya, Kaushik and Stuart, Andrew and Anandkumar, Anima},
  year      = 2021
}

@article{yang2024fourier,
  title   = {Fourier neural operators for arbitrary resolution climate data downscaling},
  author  = {Yang, Qidong and Hernandez-Garcia, Alex and Harder, Paula and Ramesh, Venkatesh and Sattigeri, Prasanna and Szwarcman, Daniela and Watson, Campbell D and Rolnick, David},
  journal = {Journal of Machine Learning Research},
  volume  = {25},
  number  = {420},
  pages   = {1--30},
  year    = {2024}
}

@inproceedings{Lim2017Enhanced,
  title     = {Enhanced Deep Residual Networks for Single Image Super-Resolution},
  booktitle = {2017 IEEE Conference on Computer Vision and Pattern Recognition Workshops (CVPRW)},
  author    = {Lim, B. and Son, S. and Kim, H. and Nah, S. and Lee, K. M.},
  year      = 2017,
  pages     = {1132--1140},
  doi       = {10.1109/CVPRW.2017.151},
  isbn      = {2160-7516}
}

@article{Kang2017Evaluation,
  title   = {On the Evaluation of Seasonal Variability of the Ocean Kinetic Energy},
  author  = {Kang, Dujuan and Curchitser, Enrique N.},
  year    = 2017,
  journal = {Journal of Physical Oceanography},
  volume  = {47},
  number  = {7},
  pages   = {1675--1683},
  issn    = {0022-3670, 1520-0485},
  doi     = {10.1175/JPO-D-17-0063.1}
}

@article{Riahi2017Shared,
  title   = {The Shared Socioeconomic Pathways and Their Energy, Land Use, and Greenhouse Gas Emissions Implications: An Overview},
  author  = {Riahi, Keywan and {van Vuuren}, Detlef P. and Kriegler, Elmar and Edmonds, Jae and O'Neill, Brian C. and Fujimori, Shinichiro and Bauer, Nico and Calvin, Katherine and Dellink, Rob and Fricko, Oliver and Lutz, Wolfgang and Popp, Alexander and Cuaresma, Jesus Crespo and KC, Samir and Leimbach, Marian and Jiang, Leiwen and Kram, Tom and Rao, Shilpa and Emmerling, Johannes and Ebi, Kristie and Hasegawa, Tomoko and Havlik, Petr and Humpen{\"o}der, Florian and Da Silva, Lara Aleluia and Smith, Steve and Stehfest, Elke and Bosetti, Valentina and Eom, Jiyong and Gernaat, David and Masui, Toshihiko and Rogelj, Joeri and Strefler, Jessica and Drouet, Laurent and Krey, Volker and Luderer, Gunnar and Harmsen, Mathijs and Takahashi, Kiyoshi and Baumstark, Lavinia and Doelman, Jonathan C. and Kainuma, Mikiko and Klimont, Zbigniew and Marangoni, Giacomo and {Lotze-Campen}, Hermann and Obersteiner, Michael and Tabeau, Andrzej and Tavoni, Massimo},
  year    = 2017,
  journal = {Global Environmental Change},
  volume  = {42},
  pages   = {153--168},
  issn    = {0959-3780},
  doi     = {10.1016/j.gloenvcha.2016.05.009}
}

@article{Hernanz2024limitations,
  title     = {On the Limitations of Deep Learning for Statistical Downscaling of Climate Change Projections: The Transferability and the Extrapolation Issues},
  author    = {Hernanz, Alfonso and Correa, Carlos and {S{\'a}nchez-Perrino}, Juan-Carlos and {Prieto-Rico}, Ignacio and {Rodr{\'i}guez-Guisado}, Esteban and Dom{\'i}nguez, Marta and {Rodr{\'i}guez-Camino}, Ernesto},
  year      = 2024,
  journal   = {Atmospheric Science Letters},
  volume    = {25},
  number    = {2},
  pages     = {e1195},
  publisher = {John Wiley \& Sons, Ltd},
  issn      = {1530-261X},
  doi       = {10.1002/asl.1195}
}

@inproceedings{Wei2023SuperResolution,
  title     = {Super-Resolution Neural Operator},
  booktitle = {2023 IEEE/CVF Conference on Computer Vision and Pattern Recognition (CVPR)},
  author    = {Wei, M. and Zhang, X.},
  year      = 2023,
  pages     = {18247--18256},
  doi       = {10.1109/CVPR52729.2023.01750},
  isbn      = {2575-7075}
}

@article{Bano-Medina2022Downscaling,
  title     = {Downscaling Multi-Model Climate Projection Ensembles with Deep Learning (DeepESD): Contribution to CORDEX EUR-44},
  author    = {{Ba{\~n}o-Medina}, Jorge and Manzanas, Rodrigo and Cimadevilla, Ezequiel and Fern{\'a}ndez, Jes{\'u}s and {Gonz{\'a}lez-Abad}, Jose and Cofi{\~n}o, Antonio S. and Guti{\'e}rrez, Jos{\'e} Manuel},
  year      = 2022,
  journal   = {Geoscientific Model Development},
  volume    = {15},
  number    = {17},
  pages     = {6747--6758},
  issn      = {1991-9603},
  doi       = {10.5194/gmd-15-6747-2022},
  copyright = {https://creativecommons.org/licenses/by/4.0/}
}

@article{Yoshikane2023downscaling,
  title     = {A Downscaling and Bias Correction Method for Climate Model Ensemble Simulations of Local-Scale Hourly Precipitation},
  author    = {Yoshikane, Takao and Yoshimura, Kei},
  year      = 2023,
  journal   = {Scientific Reports},
  volume    = {13},
  number    = {1},
  pages     = {9412},
  publisher = {Nature Publishing Group},
  issn      = {2045-2322},
  doi       = {10.1038/s41598-023-36489-3},
  copyright = {2023 The Author(s)}
}

@article{Yuan2024HiCPC,
  title     = {HiCPC: A New 10-Km CMIP6 Downscaled Daily Climate Projections over China},
  author    = {Yuan, Huihui and Ning, Like and Zhou, Jiewei and Shi, Wen and Huang, Jianbin and Luo, Yong},
  year      = 2024,
  journal   = {Scientific Data},
  volume    = {11},
  number    = {1},
  pages     = {1167},
  publisher = {Nature Publishing Group},
  issn      = {2052-4463},
  doi       = {10.1038/s41597-024-03982-x},
  copyright = {2024 The Author(s)}
}

@inproceedings{Gal2016Dropout,
  title     = {Dropout as a Bayesian Approximation: Representing Model Uncertainty in Deep Learning},
  booktitle = {Proceedings of The 33rd International Conference on Machine Learning},
  author    = {Gal, Yarin and Ghahramani, Zoubin},
  year      = 2016,
  pages     = {1050--1059},
  publisher = {PMLR},
  issn      = {1938-7228},
  address   = {New York, USA}
}

@article{Ding2025Superresolution,
  title     = {Super-Resolution Wind Mapping with Deep Learning for Scalable Renewable Energy Planning},
  author    = {Ding, Jun-Wei and Hsieh, I.-Yun Lisa},
  year      = 2025,
  journal   = {Communications Earth \& Environment},
  volume    = {7},
  number    = {1},
  pages     = {51},
  publisher = {Nature Publishing Group},
  issn      = {2662-4435},
  doi       = {10.1038/s43247-025-03072-9},
  copyright = {2025 The Author(s)}
}

@article{Bevacqua2025year,
  title     = {A Year above 1.5 {$^\circ$}C Signals That Earth Is Most Probably within the 20-Year Period That Will Reach the Paris Agreement Limit},
  author    = {Bevacqua, Emanuele and Schleussner, Carl-Friedrich and Zscheischler, Jakob},
  year      = 2025,
  journal   = {Nature Climate Change},
  volume    = {15},
  number    = {3},
  pages     = {262--265},
  publisher = {Nature Publishing Group},
  issn      = {1758-6798},
  doi       = {10.1038/s41558-025-02246-9},
  copyright = {2025 The Author(s)}
}

@article{Lin2024GFDL,
  title   = {The GFDL Variable-Resolution Global Chemistry-Climate Model for Research at the Nexus of US Climate and Air Quality Extremes},
  author  = {Lin, Meiyun and Horowitz, Larry W. and Zhao, Ming and Harris, Lucas and Ginoux, Paul and Dunne, John and Malyshev, Sergey and Shevliakova, Elena and Ahsan, Hamza and Garner, Steve and Paulot, Fabien and Pouyaei, Arman and Smith, Steven J. and Xie, Yuanyu and Zadeh, Niki and Zhou, Linjiong},
  year    = 2024,
  journal = {Journal of Advances in Modeling Earth Systems},
  volume  = {16},
  number  = {4},
  pages   = {e2023MS003984},
  issn    = {1942-2466},
  doi     = {10.1029/2023MS003984}
}

@article{Shen2022Evaluation,
  title     = {Evaluation of Global Terrestrial Near-Surface Wind Speed Simulated by CMIP6 Models and Their Future Projections},
  author    = {Shen, Cheng and Zha, Jinlin and Li, Zhibo and {Azorin-Molina}, Cesar and Deng, Kaiqiang and Minola, Lorenzo and Chen, Deliang},
  year      = 2022,
  journal   = {Annals of the New York Academy of Sciences},
  volume    = {1518},
  number    = {1},
  pages     = {249--263},
  issn      = {1749-6632},
  doi       = {10.1111/nyas.14910},
  copyright = {\copyright{} 2022 The Authors. Annals of the New York Academy of Sciences published by Wiley Periodicals LLC on behalf of New York Academy of Sciences.}
}

@article{Liu2024Rapid,
  title     = {Rapid Acceleration of Arctic Near-Surface Wind Speed in a Warming Climate},
  author    = {Liu, Wanlei and Yang, Song and Chen, Deliang and Zha, Jinlin and Zhang, Gangfeng and Zhang, Zhengtai and Zhang, Tuantuan and Xu, Lianlian and Hu, Xiaoming and Deng, Kaiqiang},
  year      = 2024,
  journal   = {Geophysical Research Letters},
  volume    = {51},
  number    = {8},
  pages     = {e2024GL109385},
  publisher = {John Wiley \& Sons, Ltd},
  issn      = {0094-8276},
  doi       = {10.1029/2024GL109385}
}

@article{daust2025Capturinga,
  title     = {Capturing {{Climatic Variability}}: {{Using Deep Learning}} for {{Stochastic Downscaling}}},
  author    = {Daust, Kiri and Monahan, Adam},
  year      = 2025,
  journal   = {Artificial Intelligence for the Earth Systems},
  volume    = {4},
  number    = {2},
  pages     = {e240044},
  publisher = {American Meteorological Society},
  address   = {Boston MA, USA},
  doi       = {10.1175/AIES-D-24-0044.1}
}

@article{xu2019Application,
  title     = {Application of the {{Bias Correction}} and {{Spatial Downscaling Algorithm}} on the {{Temperature Extremes From CMIP5 Multimodel Ensembles}} in {{China}}},
  author    = {Xu, Lianlian and Wang, Aihui},
  year      = 2019,
  journal   = {Earth and Space Science},
  volume    = {6},
  number    = {12},
  pages     = {2508--2524},
  issn      = {2333-5084},
  doi       = {10.1029/2019EA000995},
  copyright = {\copyright 2019. The Authors.}
}

@article{delimamoraes2024Climate,
  title     = {The {{Climate Data}} for {{Adaptation}} and {{Vulnerability Assessments}} and the {{Spatial Interactions Downscaling Method}}},
  author    = {{de Lima Moraes}, Andre Geraldo and Khoshnood Motlagh, Sajad},
  year      = 2024,
  journal   = {Scientific Data},
  volume    = {11},
  number    = {1},
  pages     = {1157},
  publisher = {Nature Publishing Group},
  issn      = {2052-4463},
  doi       = {10.1038/s41597-024-03995-6},
  copyright = {2024 The Author(s)}
}

@article{nahar2017rankbased,
  title   = {A Rank-Based Approach for Correcting Systematic Biases in Spatial Disaggregation of Coarse-Scale Climate Simulations},
  author  = {Nahar, Jannatun and Johnson, Fiona and Sharma, Ashish},
  year    = 2017,
  journal = {Journal of Hydrology},
  volume  = {550},
  pages   = {716--725},
  issn    = {0022-1694},
  doi     = {10.1016/j.jhydrol.2017.05.045}
}

@article{wu2022Quantifying,
  title     = {Quantifying the {{Uncertainty Sources}} of {{Future Climate Projections}} and {{Narrowing Uncertainties With Bias Correction Techniques}}},
  author    = {Wu, Yi and Miao, Chiyuan and Fan, Xuewei and Gou, Jiaojiao and Zhang, Qi and Zheng, Haiyan},
  year      = 2022,
  journal   = {Earth's Future},
  volume    = {10},
  number    = {11},
  pages     = {e2022EF002963},
  publisher = {John Wiley \& Sons, Ltd},
  issn      = {2328-4277},
  doi       = {10.1029/2022EF002963}
}

@article{lafferty2023Downscaling,
  title     = {Downscaling and Bias-Correction Contribute Considerable Uncertainty to Local Climate Projections in {{CMIP6}}},
  author    = {Lafferty, David C. and Sriver, Ryan L.},
  year      = 2023,
  journal   = {npj Climate and Atmospheric Science},
  volume    = {6},
  number    = {1},
  pages     = {158},
  publisher = {Nature Publishing Group},
  issn      = {2397-3722},
  doi       = {10.1038/s41612-023-00486-0},
  copyright = {2023 The Author(s)}
}

@article{Zhang2019double,
  title   = {The Double ITCZ Syndrome in GCMs: A Coupled Feedback Problem among Convection, Clouds, Atmospheric and Ocean Circulations},
  author  = {Zhang, Guang J. and Song, Xiaoliang and Wang, Yong},
  year    = 2019,
  journal = {Atmospheric Research},
  volume  = {229},
  pages   = {255--268},
  issn    = {0169-8095},
  doi     = {10.1016/j.atmosres.2019.06.023}
}

@article{Liu2002Double,
  title     = {Double Intertropical Convergence Zones---a New Look Using Scatterometer},
  author    = {Liu, W. Timothy and Xie, X.},
  year      = 2002,
  journal   = {Geophysical Research Letters},
  volume    = {29},
  number    = {22},
  pages     = {29-1-29-4},
  issn      = {1944-8007},
  doi       = {10.1029/2002GL015431},
  copyright = {Copyright 2002 by the American Geophysical Union.}
}

@inproceedings{kendall2017What,
  title     = {What {{Uncertainties Do We Need}} in {{Bayesian Deep Learning}} for {{Computer Vision}}?},
  booktitle = {Advances in {{Neural Information Processing Systems}}},
  author    = {Kendall, Alex and Gal, Yarin},
  year      = 2017,
  volume    = {30},
  publisher = {Curran Associates, Inc.}
}

@article{Hess2022Physically,
  title     = {Physically Constrained Generative Adversarial Networks for Improving Precipitation Fields from Earth System Models},
  author    = {Hess, Philipp and Dr{\"u}ke, Markus and Petri, Stefan and Strnad, Felix M. and Boers, Niklas},
  year      = 2022,
  journal   = {Nature Machine Intelligence},
  volume    = {4},
  number    = {10},
  pages     = {828--839},
  publisher = {Nature Publishing Group},
  issn      = {2522-5839},
  doi       = {10.1038/s42256-022-00540-1},
  copyright = {2022 The Author(s), under exclusive licence to Springer Nature Limited}
}

\end{document}